\documentclass[aps,prd,superscriptaddress,showkeys,showpacs,nofootinbib,10pt,onecolumn,longbibliography,floatfix]{revtex4-2}

\usepackage[english]{babel}

\usepackage{color}
\usepackage{graphicx}
\usepackage{amsmath,amssymb,empheq}
\graphicspath{{./figures/}}
\usepackage{appendix}

\allowdisplaybreaks

\usepackage[breaklinks=true]{hyperref}
\hypersetup{
	colorlinks=true,
	linkcolor=blue,
	filecolor=magenta,      
	urlcolor=blue,
	citecolor=red,
}

\usepackage{orcidlink}

\usepackage{mathrsfs}
\usepackage[normalem]{ulem}

\usepackage{placeins}
\usepackage[utf8]{inputenc}

\input epsf
\usepackage{mathtools}
\usepackage{amsfonts}
\usepackage{upgreek}
\usepackage{latexsym}
\usepackage{stfloats}

\usepackage{afterpage}
\usepackage{bm}

\usepackage{siunitx}
\usepackage{multirow}

\usepackage{stackengine}
\usepackage{scalerel}
\usepackage{xcolor}

\newcommand{\syml}{(\!\!(}
\newcommand{\symr}{)\!\!)}

\begin{document}
{\hfill MS-TP-24-10}
\title{Scalar perturbations from inflation in the presence of gauge fields}

\author{R.~Durrer\,\orcidlink{0000-0001-9833-2086}}
\email{ruth.durrer@unige.ch}
\affiliation{D\'{e}partement de Physique Th\'{e}orique and Center for Astroparticle Physics, \href{https://ror.org/01swzsf04}{Universit\'{e} de Gen\`{e}ve},  24 quai Ernest Ansermet, 1211 Gen\`{e}ve 4, Switzerland}

\author{R.~von~Eckardstein\,\orcidlink{0009-0006-9176-2343}}
\affiliation{Institute for Theoretical Physics, \href{https://ror.org/00pd74e08}{University of M\"{u}nster}, Wilhelm-Klemm-Stra{\ss}e 9, 48149 M\"{u}nster, Germany}

\author{Deepen~Garg\,\orcidlink{0000-0001-5226-1913}}
\affiliation{D\'{e}partement de Physique Th\'{e}orique and Center for Astroparticle Physics, \href{https://ror.org/01swzsf04}{Universit\'{e} de Gen\`{e}ve},  24 quai Ernest Ansermet, 1211 Gen\`{e}ve 4, Switzerland}

\author{K.~Schmitz\,\orcidlink{0000-0003-2807-6472}}
\affiliation{Institute for Theoretical Physics, \href{https://ror.org/00pd74e08}{University of M\"{u}nster}, Wilhelm-Klemm-Stra{\ss}e 9, 48149 M\"{u}nster, Germany}

\author{O.~Sobol\,\orcidlink{0000-0002-6300-3079}}
\email{oleksandr.sobol@knu.ua}
\affiliation{Institute for Theoretical Physics, \href{https://ror.org/00pd74e08}{University of M\"{u}nster}, Wilhelm-Klemm-Stra{\ss}e 9, 48149 M\"{u}nster, Germany}
\affiliation{Physics Faculty, \href{https://ror.org/02aaqv166}{Taras Shevchenko National University of Kyiv}, 64/13, Volodymyrska Street, 01601 Kyiv, Ukraine}
	
\author{S.~Vilchinskii\,\orcidlink{0000-0002-9294-9939}}
\affiliation{D\'{e}partement de Physique Th\'{e}orique and Center for Astroparticle Physics, \href{https://ror.org/01swzsf04}{Universit\'{e} de Gen\`{e}ve},  24 quai Ernest Ansermet, 1211 Gen\`{e}ve 4, Switzerland}
\affiliation{Physics Faculty, \href{https://ror.org/02aaqv166}{Taras Shevchenko National University of Kyiv}, 64/13, Volodymyrska Street, 01601 Kyiv, Ukraine}	

\date{\today}

\begin{abstract}
We study how Abelian-gauge-field production during inflation affects  scalar perturbations in the case when the gauge field interacts with the inflaton directly (by means of generic kinetic and axial couplings) and via gravity. The homogeneous background solution is defined by self-consistently taking into account the backreaction of the gauge field on the evolution of the inflaton and the scale factor. For the perturbations on top of this background, all possible scalar contributions coming from the inflaton, the metric, and the gauge field are considered.
We derive a second-order differential equation for the curvature perturbation, $\zeta$, capturing the impact of the gauge field, both on the background dynamics and on the evolution of scalar perturbations. The latter is described by a source term in the $\zeta$ equation, which is quadratic in the gauge-field operators and leads to non-Gaussianities in the curvature perturbations. We derive general expressions for the induced scalar power spectrum and bispectrum. Finally, we apply our formalism to the well-known case of axion inflation without backreaction. 
Numerical results show that, in this example, the effect of including metric perturbations is small for values of the gauge-field production parameter $\xi> 3$. This is in agreement with the previous results in the literature. However, in the region of smaller values, $\xi\lesssim 2$, our new results exhibit order-of-unity deviations when compared to previous results.
\end{abstract}

\maketitle

%%%%%%%%%%%%%%%%%%%%%%%%%%%%%%%%%%%%%%%%%%%%%%%%%%%%%%%%%
%%%%%%%%%%%%%%%%%%%%%%%%%%%%%%%%%%%%%%%%%%%%%%%%%%%%%%%%%
\section{Introduction}
\label{sec:introduction}
%%%%%%%%%%%%%%%%%%%%%%%%%%%%%%%%%%%%%%%%%%%%%%%%%%%%%%%%%
%%%%%%%%%%%%%%%%%%%%%%%%%%%%%%%%%%%%%%%%%%%%%%%%%%%%%%%%%

Inflation is an important pillar of the standard model of cosmology, $\Lambda$CDM. The main role played by this period of very rapid expansion before the hot big bang~\cite{Starobinsky:1980te,Guth:1980zm,Linde:1981mu,Starobinsky:1982ee,Albrecht:1982wi,Linde:1983gd} is that it leads to the amplification of vacuum fluctuations of the inflaton field and of the metric~\cite{Starobinsky:1979ty,Mukhanov:1981xt,Mukhanov:1982nu,Guth:1982ec,Hawking:1982cz,Bardeen:1983qw}. This  mechanism for the generation of cosmological perturbations is supported by observations of the cosmic microwave background (CMB)~\cite{Planck:2013jfk,Planck:2015sxf,Planck:2018jri} and of the cosmological large-scale structure (LSS)~\cite{Reid:2009xm,Cabass:2022wjy}; for reviews, see Refs.~\cite{Martin:2013tda,Chluba:2015bqa,Martin:2018ycu,Durrer:2020fza,Martin:2024qnn}. In the simplest models of inflation, the scalar fluctuations generated during inflation, which are the seeds of the small density fluctuations in the hot cosmic plasma after inflation, are very close to Gaussian, which is in good agreement with observations~\cite{Bartolo:2004if,Planck:2015zfm,Planck:2019kim}. Thus far, no primordial non-Gaussianities (PNG) have been observed, and inflation is characterized purely by the primordial power spectrum of scalar perturbations. Clearly, the detection of PNG would open a new window to study the physics of inflation, making it an extremely important observable. Processes during inflation that may lead to measurable PNG are, therefore, of great interest.

Another subject of great interest is the origin of magnetic fields in cosmology. Magnetic fields are ubiquitous in the Universe~\cite{Kronberg:1993vk,Ryu:2011hu,Vallee:2011zz,Vachaspati:2020blt} and have been observed in galaxies, even at high redshifts~\cite{Bernet:2008qp}, in clusters~\cite{Bonafede:2010wg,Feretti:2012vk}, in filaments~\cite{Vacca:2018rta,Carretti:2022fqk}, and, albeit indirectly, through the observation of blazar spectra, even in voids~\cite{Neronov:2010gir,Tavecchio:2010ja,Tavecchio:2010mk,Ando:2010rb,Dermer:2010mm,Dolag:2010ni,Vovk:2011aa,Huan:2011kp,Taylor:2011bn,Neronov:2013zka,Caprini:2015gga,AlvesBatista:2020oio} (for a review, see Ref.~\cite{AlvesBatista:2021sln}). In Ref.~\cite{MAGIC:2022piy}, a lower limit of about $10^{-17}\,$Gauss was derived for the amplitude of large-scale magnetic fields in voids. Such fields are difficult to generate in the late stages of the evolution of the Universe. Therefore, processes in the early Universe leading to the generation of magnetic fields seem more plausible. The authors of Ref.~\cite{Tjemsland:2023hmj} recently argued that blazar observations require the intergalactic magnetic field to fill a fraction $f \gtrsim 0.67$ of space, which again points to a primordial origin, through processes during inflation or cosmological phase transitions.

To generate magnetic fields during inflation, the conformal symmetry of the electromagnetic field has to be broken~\cite{Parker:1968mv}. This can be done through a coupling of the gauge field to the inflaton, to the Ricci scalar, or to the Riemann tensor~\cite{Turner:1987bw}. During a phase of nearly de Sitter inflation, all these couplings are in fact equivalent. The amplification of gauge-field vacuum fluctuations coupled to the inflaton during inflation has been studied intensively in the past; see the seminal works in Refs.~\cite{Turner:1987bw,Ratra:1991bn,Garretson:1992vt,Martin:2007ue,Anber:2006xt} and the review articles in Refs.~\cite{Giovannini:2003yn,Subramanian:2009fu,Kandus:2010nw,Durrer:2013pga,Subramanian:2015lua}. The homogeneous dynamics of the inflaton leads to the production of gauge fields due to the coupling between them. On the other hand, these gauge fields, through their interaction with the inflaton and gravity, affect the inflaton and the metric fluctuations. A systematic perturbative study of this effect is the goal of the present paper.

There is a big body of literature in which the production of curvature perturbations in the presence of vector fields during inflation is studied. They consider the kinetic coupling $\propto I(\phi) F_{\mu\nu}F^{\mu\nu}$~\cite{Seery:2008ms,Yokoyama:2008xw,Dimastrogiovanni:2010sm,Watanabe:2010fh,Caldwell:2011ra,Bonvin:2011dt,Bartolo:2012sd,Motta:2012rn,Jain:2012ga,Jain:2012vm,Shiraishi:2012xt,Soda:2012zm,Suyama:2012wh,Abolhasani:2013zya,Nurmi:2013gpa,Thorsrud:2013kya,Giovannini:2013rme,Fujita:2013qxa,Choi:2015wva,Chua:2018dqh,Chen:2023bcz} or the axial coupling $\propto I(\phi) F_{\mu\nu}\tilde{F}^{\mu\nu}$~\cite{Anber:2009ua,Barnaby:2010vf,Barnaby:2011vw,Barnaby:2011qe,Barnaby:2012xt,Linde:2012bt,Meerburg:2012id,Bamba:2014vda,Ferreira:2014zia,Ferreira:2015omg,Cheng:2015oqa,Fujita:2015iga,Namba:2015gja,Dimastrogiovanni:2016fuu,Domcke:2016bkh,Garcia-Bellido:2016dkw,Papageorgiou:2018rfx,Papageorgiou:2019ecb,Domcke:2020zez,Dimastrogiovanni:2023oid,Domcke:2023tnn,Corba:2024tfz} of the gauge field to the (pseudo)scalar inflaton or spectator field. In most cases, Abelian gauge fields were studied, except for Refs.~\cite{Dimastrogiovanni:2010sm,Dimastrogiovanni:2016fuu,Papageorgiou:2018rfx,Papageorgiou:2019ecb,Dimastrogiovanni:2023oid}, which deal with non-Abelian fields. References~\cite{Valenzuela-Toledo:2009bzd,Ringeval:2013hfa,Fujita:2014sna,Kamada:2020bmb} derive model-independent results for scalar perturbations in the presence of gauge fields. In addition to perturbative studies, numerical simulations on a lattice have been performed in Refs.~\cite{Deskins:2013dwa,Caravano:2022epk,Figueroa:2023oxc}.

In a previous paper~\cite{Durrer:2023rhc}, some of us found that the interaction of the gauge field with the inflaton leads to considerable backreaction onto the inflaton dynamics, which actually prolongs inflation.  This is in agreement with similar analyses~\cite{Domcke:2020zez,Gorbar:2021rlt,vonEckardstein:2023gwk} and fully numerical studies~\cite{Figueroa:2023oxc}.  For this reason, we expect that the influence of gauge-field perturbations on the scalar perturbations of the metric and the inflaton, which seed the density fluctuations observed in the Universe today, can be considerable\,---\,especially, as we shall see, because these fluctuations are non-Gaussian and may lead to observable PNG. However, to the best of our knowledge, a self-consistent treatment of the impact of gauge fields on both the background evolution and the scalar perturbations (including those of the metric) is still missing in the literature. This provides the main motivation for the present work.

We treat the gauge fields as sources of perturbations in the curvature and neglect the backreaction of metric perturbations on the evolution of the gauge-field sources. This formalism has been first used in~\cite{Durrer:1990mk} and later generalized in Ref.~\cite{Durrer:1993tti}. In the present situation, however, the sources also interact with the inflaton field via nongravitational interactions and thus backreact on the evolution of the universe. Therefore, metric perturbations have to be taken into account also in the evolution equations of the source terms. We extend the formalism derived in the above references to take these modifications consistently into account.

The remainder of this paper is organized as follows: In the next section, we present the Lagrangian for the type of models studied in this paper and the resulting equations of motion for all background quantities. In Sec.~\ref{sec:eqs-for-perturbations}, which is the heart of the present work, we derive the full set of equations describing the evolution of scalar perturbations, taking into account (i)~linear perturbations of the inflaton, (ii)~scalar linear perturbations of the metric, as well as (iii)~perturbations up to second order in the gauge field (which are first-order perturbations in its energy momentum tensor). We then derive a master equation for the gauge-invariant curvature fluctuation $\zeta$, which accounts for all sources of scalar perturbations in our model, including the contributions from the gauge fields. 
In Sec.~\ref{sec:correlators}, we derive general expressions for the two- and three-point correlation functions of $\zeta$ in Fourier space, i.e., the power spectrum and the bispectrum. We write these in terms of the gauge-field mode functions and the vacuum inflaton perturbations.
In Sec.~\ref{sec:application}, we illustrate our formalism with a simple example and compare our results to the literature. Finally, in Sec.~\ref{sec:conclusion}, we conclude and present an outlook on the future applications of our formalism. Some lengthy derivations of important equations and explicit expressions for the constituents of the differential equation for the variable $\zeta$ are deferred to several appendixes.

\vspace{0.2cm}
\textit{Notation.---}We work with a mostly negative metric signature, $\left(+,-,-,-\right)$, and denote conformal time by $\eta$. A prime on a quantity denotes the derivative of this quantity with respect to $\eta$. For functions depending on a different argument, though, a prime denotes a derivative with respect to this argument. In these cases, the argument is always shown explicitly. For instance, $V'(\phi)$, $I_1'(\phi)$, and $I_2'(\phi)$ denote derivatives of theses functions with respect to the scalar field $\phi$. The conformal Hubble rate is denoted by $\mathcal{H} =a'/a=aH$, where $H$ is the physical Hubble rate. Spacetime indices are denoted by lowercase Greek letters, while spatial indices are lowercase Latin letters. Spatial vectors are written in boldface. The three-dimensional Levi-Civita symbol is denoted by  $\varepsilon_{ijk}=\varepsilon^{ijk}$. Throughout the work, we use natural units and set $\hbar=c=1$. We use the notation $M_{\mathrm{P}}=(8\pi G)^{-1/2}\approx 2.43\times 10^{18}\,{\rm GeV}$ for the reduced Planck mass.

%%%%%%%%%%%%%%%%%%%%%%%%%%%%%%%%%%%%%%%%%%%%%%%%%%%%%%%%%
%%%%%%%%%%%%%%%%%%%%%%%%%%%%%%%%%%%%%%%%%%%%%%%%%%%%%%%%%
\section{Model and background evolution}
\label{sec:background}
%%%%%%%%%%%%%%%%%%%%%%%%%%%%%%%%%%%%%%%%%%%%%%%%%%%%%%%%%
%%%%%%%%%%%%%%%%%%%%%%%%%%%%%%%%%%%%%%%%%%%%%%%%%%%%%%%%%

We consider a model of inflation with a real scalar inflaton field, $\phi$, and an Abelian gauge field, $A_{\mu}$, which can have both kinetic and axial couplings to the inflaton. Such a general model breaks spatial parity invariance, $P$, \textit{explicitly}.%
\footnote{However, in particular cases of a purely kinetic coupling or a purely axial coupling in terms of an odd function of the pseudoscalar (axial) inflaton, parity is still preserved.}
The corresponding action has the form
%
%%ONE-COLUMN MODE
\begin{equation}
    S[g_{\mu\nu},\phi, A_{\mu}] =\!\int\!\! d^4 x \sqrt{-g}\Big[-\frac{M_{\mathrm{P}}^2}{2} R +\frac{1}{2} g^{\mu\nu}\partial_\mu \phi\partial_\nu \phi - V(\phi) -\frac{1}{4}I_1(\phi)\syml F_{\mu\nu}F^{\mu\nu}\symr-\frac{1}{4}I_2(\phi)\syml F_{\mu\nu}\tilde{F}^{\mu\nu}\symr\Big]\, ,
\label{action-1}
\end{equation}
%%TWO-COLUMN MODE
% \begin{align}
%     S[g_{\mu\nu},\phi, A_{\mu}] &=\!\int\!\! d^4 x \sqrt{-g}\Big[-\frac{M_{\mathrm{P}}^2}{2} R \nonumber \\
%     &\ \   +\frac{1}{2} g^{\mu\nu}\partial_\mu \phi\partial_\nu \phi - V(\phi) 
%     \label{action-1} \\
%     &\ \  -\frac{1}{4}I_1(\phi)\syml F_{\mu\nu}F^{\mu\nu}\symr\!-\!\frac{1}{4}I_2(\phi)\syml F_{\mu\nu}\tilde{F}^{\mu\nu}\symr\Big]\, , \nonumber
% \end{align}
%
where $g_{\mu\nu}$ is the spacetime metric, $g=\mathrm{det}\,g_{\mu\nu}$ is its determinant,  $R$ is the Ricci curvature scalar, $V(\phi)$ is the inflaton potential, $F_{\mu\nu}=\partial_{\mu}A_{\nu}-\partial_{\nu}A_{\mu}$ is the field-strength tensor, and $\tilde F^{\mu\nu}=1/2\,\epsilon^{\mu\nu\alpha\beta}F_{\alpha\beta}/\sqrt{-g}$ is its dual. Here, $\varepsilon^{\mu\nu\alpha\beta}$ (with $\varepsilon^{0123}=+1$) is the totally antisymmetric Levi-Civita symbol in four dimensions. 
In the last two terms in the action, we introduced the notation $\syml \cdots\symr$ for the symmetrized product of two operators
\begin{equation*}
    \syml PQ \symr \equiv \frac{1}{2}(PQ+QP)\, ,
\end{equation*}
which, after quantization, may no longer commute. Introduction of this symmetrization is prescribed by the correspondence principle if we write down the action in terms of quantum fields. This symmetrization, although not strictly necessary yet, is introduced here for convenience in deriving later equations like, for example, Eq.~\eqref{EMT}.

The functions $I_{1,2}(\phi)$ in the action~\eqref{action-1} are, respectively, the kinetic and axial couplings of the gauge field to the inflaton. For the sake of generality, we will not specify explicit forms of the inflaton potential $V(\phi)$ and the coupling functions $I_{1,2}(\phi)$. However, in any case, the kinetic coupling function $I_1(\phi)$ must be (i)~positive to ensure the positive definiteness of the gauge-field energy density and (ii)~never much smaller than unity during inflation to avoid the strong coupling problem. On the other hand, the axial coupling function $I_2(\phi)$ is completely arbitrary, because it does not enter the energy--momentum tensor and does not have any impact on the coupling of other matter to the gauge field.

The energy--momentum tensor is obtained by varying the matter part (inflaton and gauge field) of the action \eqref{action-1} with respect to the inverse metric $g^{\mu\nu}$,
%
%%ONE-COLUMN MODE
\begin{equation}
    T_{\mu\nu} = \partial_{\mu}\phi\partial_{\nu}\phi-
    I_1(\phi)g^{\alpha\beta}\syml F_{\mu\alpha}F_{\nu\beta}\symr -g_{\mu\nu} \left[\frac{1}{2}\partial_{\alpha}\phi
    \partial^{\alpha}\phi-V(\phi)-\frac{1}{4}I_1(\phi)\syml F_{\alpha\beta}F^{\alpha\beta}\symr\right]\, .
\label{EMT}
\end{equation}
%%TWO-COLUMN MODE
% \begin{align}
%     T_{\mu\nu}\! &=\! \partial_{\mu}\phi\partial_{\nu}\phi-
%     I_1(\phi)g^{\alpha\beta}\syml F_{\mu\alpha}F_{\nu\beta}\symr 
%     \label{EMT} \\    
%     &\quad-g_{\mu\nu} \left[\frac{1}{2}\partial_{\alpha}\phi
%     \partial^{\alpha}\phi-V(\phi)-\frac{1}{4}I_1(\phi)\syml F_{\alpha\beta}F^{\alpha\beta}\symr\right]\, . \nonumber
% \end{align}
%
The Einstein equations read as
\begin{equation}
    G_{\mu\nu}=\frac{1}{M_{\mathrm{P}}^2}T_{\mu\nu} \,,
\end{equation}
where $G_{\mu\nu}$ is the Einstein tensor.
The modified Maxwell equations are obtained by varying the action \eqref{action-1} with respect to the gauge field $A_{\mu}$,
\begin{equation}
\label{Maxwell}
    \frac{1}{\sqrt{-g}}\partial_{\mu}\left[\sqrt{-g}I_1(\phi)F^{\mu\nu}\right]=-\tilde{F}^{\mu\nu}\partial_\mu I_2(\phi) \,.
\end{equation}
They are supplemented by the Bianchi identity
\begin{equation}
\label{e:const}
    \frac{1}{\sqrt{-g}}\partial_{\mu}\left[\sqrt{-g}\tilde{F}^{\mu\nu}\right]=0\, .
\end{equation}
Finally, varying the action \eqref{action-1} with respect to the inflaton field, we obtain the inflaton equation of motion,
%
%%ONE-COLUMN MODE
\begin{equation}
\label{KGF}
  \frac{1}{\sqrt{-g}}\partial_{\mu}\left[\sqrt{-g}g^{\mu\nu}\partial_{\nu}\phi\right]+\frac{dV}{d\phi} =-\frac{1}{4}\frac{dI_1}{d\phi} \syml F_{\mu\nu}F^{\mu\nu}\symr-\frac{1}{4}\frac{dI_2}{d\phi}\syml F_{\mu\nu}\tilde{F}^{\mu\nu}\symr\, .
\end{equation}
%%TWO-COLUMN MODE
% \begin{multline}
%     \frac{1}{\sqrt{-g}}\partial_{\mu}\left[\sqrt{-g}g^{\mu\nu}\partial_{\nu}\phi\right]+\frac{dV}{d\phi}  =-\frac{1}{4}\frac{dI_1}{d\phi}\syml F_{\mu\nu}F^{\mu\nu}\symr
%     \\-\frac{1}{4}\frac{dI_2}{d\phi}\syml F_{\mu\nu}\tilde{F}^{\mu\nu}\symr\, .
% \label{KGF}
% \end{multline}
%
Let us first describe the solutions to these equations in the case of a spatially flat homogeneous and isotropic Friedmann--Lema\^{\i}tre--Robertson--Walker (FLRW) geometry. The corresponding metric is given by
\begin{equation}
    g^{(0)}_{\mu\nu}\mathrm{d}x^\mu \mathrm{d}x^\nu = a^2(\eta)[\mathrm{d}\eta^2 - \delta_{ij}\mathrm{d}x^{i}\mathrm{d}x^{j}]\, ,
\end{equation}
where $a(\eta)$ is the scale factor. The background inflaton field, which is considered as a spatially homogeneous classical field, is denoted by $\phi_c\left(\eta\right) = \left<\phi\left(\eta,\bm{x}\right)\right>$. We define the electric- and magnetic-field three-vectors from the components of the field-strength tensor,
\begin{equation}
\label{E-B-def}
    F_{0i} = a^2 E^{i}\, , \quad  F_{ij} = -a^{2}\varepsilon_{ijk}B^{k}\, .
\end{equation}
These correspond to the electric and magnetic fields as measured by a comoving observer in a background FLRW universe.

In this case, the Einstein equations simply become the Friedmann equations, which, for our matter content, read
%
%%ONE-COLUMN MODE
\begin{align}
    \frac{a^{\prime 2}}{a^2}\equiv \mathcal{H}^2 &= 
    \frac{1}{3M_{\mathrm{P}}^{2}}\bigg\{
    \frac{\phi_{c}^{\prime 2}}{2} +  a^2V(\phi_c)  + \frac{a^2}{2}I_1(\phi_c) \Big[\langle\boldsymbol{E}^2\rangle+\langle\boldsymbol{B}^2\rangle\Big]\bigg\}\, ,
    \label{Friedmann} \\
    2\mathcal{H}'+\mathcal{H}^2 &=  -\frac{1}{M_{\mathrm{P}}^2}
    \bigg\{\frac{\phi_{c}^{\prime 2}}{2} - a^2 V(\phi_c) + \frac{a^2}{6}I_1(\phi_c) \Big[\langle\boldsymbol{E}^2\rangle+\langle\boldsymbol{B}^2\rangle\Big]\bigg\}\, .
    \label{Friedmann2}
\end{align}
%
%%TWO-COLUMN MODE 
% \begin{align}
%     \frac{a^{\prime 2}}{a^2}\equiv \mathcal{H}^2 &= 
%     \frac{1}{3M_{\mathrm{P}}^{2}}\bigg\{
%     \frac{\phi_{c}^{\prime 2}}{2} +  a^2V(\phi_c)   \nonumber \\
%     &\quad+ \frac{a^2}{2}I_1(\phi_c) \Big[\langle\boldsymbol{E}^2\rangle+\langle\boldsymbol{B}^2\rangle\Big]\bigg\}\, ,
%     \label{Friedmann} \\
%     2\mathcal{H}'+\mathcal{H}^2 &=  -\frac{1}{M_{\mathrm{P}}^2}
%     \bigg\{\frac{\phi_{c}^{\prime 2}}{2} - a^2 V(\phi_c) \nonumber \\ 
%     &\quad+ \frac{a^2}{6}I_1(\phi_c) \Big[\langle\boldsymbol{E}^2\rangle+\langle\boldsymbol{B}^2\rangle\Big]\bigg\}\, .
%     \label{Friedmann2}
% \end{align}
%
Here, $\langle\boldsymbol{E}^2\rangle$ and $\langle\boldsymbol{B}^2\rangle$ are the vacuum expectation values of the gauge-field contributions to the total energy density and pressure.

The Klein--Gordon equation for the scalar field in this background becomes
%
%%ONE-COLUMN MODE
\begin{equation}
   \phi^{\prime\prime}_c+2\mathcal{H}\phi'_c+a^2 V'(\phi_{c}) =\frac{1}{2}a^2 I'_1(\phi_{c})\Big[\langle\boldsymbol{E}^2\rangle-\langle\boldsymbol{B}^2\rangle\Big] + a^2 I'_2(\phi_{c}) \langle \syml\boldsymbol{E}\cdot\boldsymbol{B}\symr \rangle\, . \label{KGF-c}
\end{equation}
%
%%TWO-COLUMN MODE 
% \begin{align}
%    &\phi^{\prime\prime}_c+2\mathcal{H}\phi'_c+a^2 V'(\phi_{c})  \label{KGF-c} \\
%    &\quad=\frac{1}{2}a^2 I'_1(\phi_{c})\Big[\langle\boldsymbol{E}^2\rangle-\langle\boldsymbol{B}^2\rangle\Big] + a^2 I'_2(\phi_{c}) \langle \syml\boldsymbol{E}\cdot\boldsymbol{B} \symr\rangle\, . \nonumber
% \end{align}
%
Note that we apply the symmetrization operation explicitly only for the products of the electric and magnetic fields, $\syml E_j B_k\symr$, as the corresponding field operators do not commute; see Refs.~\cite{Jordan:1928zkl,Heisenberg:1929xj,Bohr:1933b,Stewart:2010by}.%
\footnote{In some cases, e.g., in $\langle \syml\boldsymbol{E}\cdot\boldsymbol{B}\symr \rangle$, the symmetrization is redundant as the commutator $[E_i, B_i]$ vanishes. Nevertheless, we keep the symmetrization here for the consistency of our notation.}
In cases where the operators obviously commute, e.g., for $\boldsymbol{E}^2$, this operation is omitted for the sake of brevity.

The Maxwell equations read
%
%%ONE-COLUMN MODE
\begin{align}
    \operatorname{div}\boldsymbol{B}&=0\, ,\label{div-B-bckgr}\\
    {\boldsymbol{B}}'+2\mathcal{H}\boldsymbol{B}+\operatorname{rot}\boldsymbol{E}&=0\, ,\label{Maxwell-second-equation}\\
    \operatorname{div}\boldsymbol{E}&=0\, ,\label{div-E-bckgr}\\
    \boldsymbol{E}'+2\mathcal{H}\boldsymbol{E}-\operatorname{rot}\boldsymbol{B}+\frac{I'_{1}(\phi_c)}{I_{1}(\phi_c)}\phi'_c\boldsymbol{E}+\frac{I'_{2}(\phi_c)}{I_{1}(\phi_c)}\phi'_c\boldsymbol{B}&=0\, .\label{Maxwell-2}
\end{align}
%%TWO-COLUMN MODE
% \begin{align}
%     \operatorname{div}\boldsymbol{B}&=0\, ,\label{div-B-bckgr}\\
%     {\boldsymbol{B}}'+2\mathcal{H}\boldsymbol{B}+\operatorname{rot}\boldsymbol{E}&=0\, ,\label{Maxwell-second-equation}\\
%     \operatorname{div}\boldsymbol{E}&=0\, ,\label{div-E-bckgr}\\
%     \boldsymbol{E}'\!+\!2\mathcal{H}\boldsymbol{E}\!-\!\operatorname{rot}\boldsymbol{B}\!+\!\frac{I'_{1}(\phi_c)}{I_{1}(\phi_c)}\phi'_c\boldsymbol{E}
%     %\quad& \nonumber\\
%     \!+\!\frac{I'_{2}(\phi_c)}{I_{1}(\phi_c)}\phi'_c\boldsymbol{B}&=0\, .\label{Maxwell-2}
% \end{align}
%
With these, we can derive equations for the time derivatives of the vacuum expectation values to which only the background quantities contribute:
\begin{align}
    \langle \boldsymbol{E}^2 \rangle' &= -\left[4\mathcal{H}+2\frac{I'_1(\phi_c)}{I_1(\phi_c)}\phi'_c\right]\langle \boldsymbol{E}^2 \rangle 
   % \nonumber \\ &  \hspace{-0.26cm} %%COMMENT FOR 1-COLUMN
    + 2\langle \syml\boldsymbol{E}\cdot \operatorname{rot} \boldsymbol{B}\symr \rangle 
    -2\frac{I'_2(\phi_c)}{I_1(\phi_c)}\phi'_c \langle \syml\boldsymbol{E}\cdot\boldsymbol{B}\symr \rangle\, ,
    \label{EE-prime} \\
\label{BB-prime}
    \langle \boldsymbol{B}^2 \rangle'&=  -4\mathcal{H}\langle \boldsymbol{B}^2 \rangle - 2\langle\syml \boldsymbol{E}\cdot \operatorname{rot} \boldsymbol{B}\symr \rangle\, ,\\
    \langle \syml\boldsymbol{E}\cdot\boldsymbol{B}\symr \rangle'&= -\left[4\mathcal{H}+\frac{I'_1(\phi_c)}{I_1(\phi_c)}\phi'_c\right]\langle \syml\boldsymbol{E}\cdot\boldsymbol{B}\symr \rangle
  %  \nonumber\\&  \hspace{-1.0cm}   %%COMMENT FOR 1-COLUMN
    -\frac{I'_2(\phi_c)}{I_1(\phi_c)}\phi'_c \langle \boldsymbol{B}^2 \rangle - \langle \boldsymbol{E}\cdot \operatorname{rot} \boldsymbol{E} \rangle + \langle \boldsymbol{B}\cdot \operatorname{rot} \boldsymbol{B} \rangle\, .
    \label{EB-prime}
\end{align}
These equations for the expectation values are not closed as new terms with a larger number of spatial curls appear on their right-hand side. Writing down equations of motion for those new terms leads to an infinite chain of differential equations. However, there is a way to truncate this system at a certain finite order, which allows one to solve it numerically. This method, known as the gradient-expansion formalism, was developed in Refs.~\cite{Sobol:2019xls,Sobol:2020lec,Gorbar:2021rlt,Durrer:2023rhc} and can be used to obtain the background solution in the presence of gauge-field backreaction.

%%%%%%%%%%%%%%%%%%%%%%%%%%%%%%%%%%%%%%%%%%%%%%%%%%%%%%%%%
%%%%%%%%%%%%%%%%%%%%%%%%%%%%%%%%%%%%%%%%%%%%%%%%%%%%%%%%%
\section{Scalar perturbations}
\label{sec:eqs-for-perturbations}
%%%%%%%%%%%%%%%%%%%%%%%%%%%%%%%%%%%%%%%%%%%%%%%%%%%%%%%%%
%%%%%%%%%%%%%%%%%%%%%%%%%%%%%%%%%%%%%%%%%%%%%%%%%%%%%%%%%

We now switch to a perturbed FLRW universe with spatially flat sections. We consider only scalar perturbations and will work in the longitudinal gauge. However, our final master equation will be gauge invariant. Of course, gauge fields also generate vector and tensor fluctuations, but we do not study these here.

The situation we have in mind is a phase of slow roll of the scalar field with its energy density dominated by the scalar-field potential leading to nearly de Sitter expansion. The slow-roll parameters
\begin{equation}
\label{e:slowroll}
    \epsilon_{H} \equiv -\frac{\dot H}{H^2}=1-\frac{\mathcal{H}'}{\mathcal{H}^2} \,, \quad \eta_{\phi}\equiv \frac{\ddot\phi_c}{H\dot\phi_c} = \frac{\phi''_c}{\mathcal{H}\phi'_c}-1
\end{equation}
are assumed to be much smaller than unity in absolute value. An overdot is a derivative with respect to cosmic time $t$.

At early times, subhorizon modes with $|k\eta|\gg 1$ (where $k=|\boldsymbol{k}|$ is the absolute value of the comoving momentum $\boldsymbol{k}$ carried by the perturbation mode) are in the vacuum state. Scalar-field perturbations are then excited by the rapid expansion, while gauge modes are excited due to their coupling to the background inflaton field. How this happens in detail is governed by the perturbation equations that we derive in this section. The quantum vacuum simply determines the initial state for each mode. 

The metric is given by
\begin{equation}
\label{metric}
    g_{\mu\nu}\mathrm{d}x^\mu \mathrm{d}x^\nu = (1+2\Psi)a^2\mathrm{d}\eta^2 - (1-2\Phi)a^{2}\delta_{ij}\mathrm{d}x^i\mathrm{d}x^j \, ,
\end{equation}
where $\Psi(\eta,\boldsymbol{x})$ and $\Phi(\eta,\boldsymbol{x})$ are the Bardeen potentials describing small perturbations of the metric, $|\Phi|,\,|\Psi|\ll 1$. The square root of the determinant of the metric up to linear order in the Bardeen potentials reads 
$$
    \sqrt{-g} \approx a^4(1+\Psi-3\Phi).
$$

The definitions of the electric- and magnetic-field three-vectors remain unchanged and are given by Eq.~\eqref{E-B-def}. The contravariant components of the field-strength tensor and the components of the dual tensor can be expressed in terms of $\boldsymbol{E}$, $\boldsymbol{B}$, and the metric perturbations, given by Eqs.~\eqref{EM-tensor-0i-low}--\eqref{EM-tensor-dual-ij-up}. While we work to first order in the metric perturbations in these expressions, we have not made any assumptions about the magnitude of the gauge-field amplitude so far.

In addition to the Bardeen potentials, we define the following perturbation variables:
\begin{equation}
\label{d-phi-def}
    \delta \varphi(\eta,\,\boldsymbol{x}) \equiv \phi(\eta,\,\boldsymbol{x})-\phi_{c}(\eta)
\end{equation}
for the inflaton perturbation and
\begin{equation}
\label{d-elmag-gen-def}
    \delta_{P_i Q_j} = \syml P_i Q_j \symr - \frac{1}{3}\delta_{ij} \langle \syml \boldsymbol{P}\cdot \boldsymbol{Q}\symr \rangle
\end{equation}
for the product of any two gauge-field vector quantities, where $\boldsymbol{P}$ and $\boldsymbol{Q}$ can each be equal to any spatial vectors out of the set $\{\boldsymbol{E}$, $\boldsymbol{B}$, $\operatorname{rot}\boldsymbol{E}$, $\operatorname{rot}\boldsymbol{B},  \operatorname{rot}^2\boldsymbol{E}, \dots\}$. Here, again, $\langle\cdots\rangle$ denotes the vacuum expectation value which, as a consequence of the isotropy of the background, is independent of position. In all equations of motion, only the terms up to linear order in these perturbations are retained. 

It may appear confusing that we consider terms only to the first order in the inflaton perturbations but to the second order in the gauge fields. However, this is necessary if we want to study the backreaction of the gauge fields onto the inflaton perturbations.
As we shall see, only the perturbation variables defined above enter in the inflaton-perturbation equations. To determine the gauge fields generated during inflation, we will, however, have to solve Eqs.~\eqref{Maxwell} and \eqref{e:const}, which are linear in the gauge fields. 

%%%%%%%%%%%%%%%%%%%%%%%%%%%%%%%%%%%%%%%%%%%%%%%%%%%%%%%%%
\subsection{Perturbed Einstein equations}
\label{subsec:Einstein-eqs}
%%%%%%%%%%%%%%%%%%%%%%%%%%%%%%%%%%%%%%%%%%%%%%%%%%%%%%%%%

In Appendix~\ref{app:Einstein}, we explicitly write down the perturbed Einstein and energy--momentum tensors. While the contributions from inflaton perturbations are always scalar, this is not the case for the gauge field. To isolate the scalar perturbations from the gauge-field energy--momentum tensor, we act on the latter with the corresponding projection operators and introduce the following set of sources:
%
%%ONE-COLUMN MODE
\begin{align}
    F_{\rho} &= \frac{a^2}{2M_{\mathrm{P}}^2}\delta T^{\ 0 \,\mathrm{(EM)}}_{0}= \frac{a^2 I_1(\phi_{c})}{4M_{\mathrm{P}}^2} \Big[\delta_{\boldsymbol{E}^2}+\delta_{\boldsymbol{B}^2}\Big]\, , \label{f-rho-def}\\
    F_v &= \frac{a^2}{2M_{\mathrm{P}}^2} \frac{\partial_i}{\triangle}\delta T^{\ 0 \,\mathrm{(EM)}}_{i}= -\frac{a^2 I_1(\phi_{c})}{2M_{\mathrm{P}}^2}\frac{\partial_i}{\triangle} \varepsilon_{ijk} \delta_{E_j B_k}\, , \label{f-v-def}\\
    F_p &= -\frac{a^2}{6M_{\mathrm{P}}^2}\delta^{i}_{j}\, \delta T^{\ j \,\mathrm{(EM)}}_{i} = \frac{a^2 I_1(\phi_{c})}{12M_{\mathrm{P}}^2} \big[\delta_{\boldsymbol{E}^2}+\delta_{\boldsymbol{B}^2}\big]
    = \frac{F_\rho}{3} \, , \label{f-p-def}\\
    F_\pi &= \frac{a^2}{4M_{\mathrm{P}}^2}\frac{1}{\triangle}\Big(\delta^{i}_{j}-3\frac{\partial_i \partial_j}{\triangle} \Big) \delta T^{\ j \,\mathrm{(EM)}}_{i}
    =\frac{a^2 I_1(\phi_{c})}{4M_{\mathrm{P}}^2}\frac{1}{\triangle} \big[\delta_{\boldsymbol{E}^2} + \delta_{\boldsymbol{B}^2}\big] - \frac{3a^2 I_1(\phi_{c})}{4M_{\mathrm{P}}^2}\frac{\partial_i \partial_j}{\triangle^2}\big[\delta_{E_i E_j} + \delta_{B_i B_j}\big]\, .
\label{f-pi-def}
\end{align}
%%TWO-COLUMN-MODE
% \begin{align}
%     &\hspace{-1.7mm}F_{\rho} \!= \!\frac{a^2}{2M_{\mathrm{P}}^2}\delta T^{\ 0 \,\mathrm{(EM)}}_{0}\!=\!\frac{a^2 \! I_1 \!(\phi_{c})}{4M_{\mathrm{P}}^2} \Big[\delta_{\boldsymbol{E}^2}+\delta_{\boldsymbol{B}^2}\Big]\, , \label{f-rho-def}\\
%     &\hspace{-1.7mm}F_v \!=\! \frac{a^2}{2M_{\mathrm{P}}^2} \frac{\partial_i}{\triangle}\delta T^{\ 0 \,\mathrm{(EM)}}_{i}= -\frac{a^2 \! I_1\!(\phi_{c})}{2M_{\mathrm{P}}^2}\frac{\partial_i}{\triangle} \varepsilon_{ijk} \delta_{E_j B_k}\, , \label{f-v-def}\\
%     &\hspace{-1.7mm}F_p \!=\! -\frac{a^2}{6M_{\mathrm{P}}^2}\delta^{i}_{j}\, \delta T^{\ j \,\mathrm{(EM)}}_{i}\!\! =\! \frac{a^2\! I_1\!(\phi_{c})}{12M_{\mathrm{P}}^2}\!\big[\delta_{\boldsymbol{E}^2}\!+\!\delta_{\boldsymbol{B}^2}\big]
%     \!\!=\! \frac{F_\rho}{3} \, , \label{f-p-def}\\
%     &\hspace{-1.7mm}F_\pi\! = \!\frac{a^2}{4M_{\mathrm{P}}^2}\frac{1}{\triangle}\Big(\delta^{i}_{j}-3\frac{\partial_i \partial_j}{\triangle} \Big) \delta T^{\ j \,\mathrm{(EM)}}_{i} 
% \label{f-pi-def}\\
%     &\hspace{-1.7mm}=\!\frac{a^2\! I_1\!(\phi_{c})}{4M_{\mathrm{P}}^2}\frac{1}{\triangle}\! \big[\delta_{\boldsymbol{E}^2}\! +\! \delta_{\boldsymbol{B}^2}\big]\!\! -\! \frac{3a^2\! I_1\!(\phi_{c})}{4M_{\mathrm{P}}^2}\frac{\partial_i \partial_j}{\triangle^2}\!\big[\delta_{E_i E_j}\! + \!\delta_{B_i B_j}\big]\, . \nonumber
% \end{align}
%
Here, the factors $a^2/(2M_{\mathrm{P}}^2)$ have been introduced for further convenience; all spatial indices $i, j, k$ are raised and lowered by the Euclidean metric. The triangle $\triangle$ denotes the spatial Laplacian operator. The inverse Laplacian operators are easily interpreted in Fourier space. Their appearance explicitly shows that the decomposition into scalar, vector, and tensor contributions is nonlocal.

With these definitions, the first-order perturbed Einstein equations become
%
%\begin{widetext}    %%COMMENT FOR 1-COLUMN
\begin{align}
    \triangle\Phi -3\mathcal{H}\Phi' -3\mathcal{H}^2\Psi &=\frac{a^2}{2M_{\mathrm{P}}^2}
    \bigg\{ \frac{\phi'_c}{a^2}\delta \varphi'+\Big[V'(\phi_{c})\!+\!\frac{1}{2}I'_1(\phi_{c})\Big(\langle\boldsymbol{E}^2\rangle\! +\!\langle\boldsymbol{B}^2\rangle\Big)\Big] \delta \varphi \nonumber\\ \label{Einstein-laplace-Phi-1}
    &\quad-\frac{\phi^{\prime 2}_c}{a^2}\Psi
    + I_1(\phi_{c})\Big[(\Phi-\Psi)\langle\boldsymbol{E}^2\rangle+2\Phi\langle\boldsymbol{B}^2\rangle \Big]\bigg\} + F_\rho \, ,\\
    \label{Einstein-Phi-prime}
    \Phi' +\mathcal{H}\Psi &= \frac{\phi'_c}{2M_{\mathrm{P}}^2}\delta \varphi + F_v\, ,\\
    \Phi'' +\mathcal{H}(\Psi'+2\Phi') +(2\mathcal{H}'+\mathcal{H}^2)\Psi &= \frac{a^2}{2M_{\mathrm{P}}^2} \bigg\{ 
    \frac{\phi'_c}{a^2}\delta \varphi'-\Big[V'(\phi_{c})\!-\!\frac{1}{6}I'_1(\phi_{c})\Big(\langle\boldsymbol{E}^2\rangle\! +\!\langle\boldsymbol{B}^2\rangle\Big)\Big] \delta \varphi \nonumber\\ \label{Einstein-Phi-prime-prime}
    &\quad-\frac{\phi^{\prime 2}_c}{a^2}\Psi
    + \frac{1}{3}I_1(\phi_{c})\Big[(\Phi-\Psi)\langle\boldsymbol{E}^2\rangle +2\Phi\langle\boldsymbol{B}^2\rangle \Big]\bigg\} + \frac{1}{3} (F_{\rho}+2\triangle F_\pi)\,, \\
    \label{Einstein-Psi-Phi}
    \Phi-\Psi &= 2F_\pi\,.
\end{align}
The first two equations are the $00$ and $0i$ constraint equations while the last two are the $ii$ and $i\ne j$ evolution equations. The gauge field enters here in two ways. First, through their coupling to the inflaton field, the background terms $\langle\boldsymbol{B}^2\rangle $ and $\langle\boldsymbol{E}^2\rangle$ couple to $\delta \varphi$. Second, the energy--momentum tensor generates metric perturbations via the functions $F_\rho$, $F_v$, and $F_\pi$.

Using Eq.~\eqref{Einstein-Phi-prime}, we can rewrite Eq.~\eqref{Einstein-laplace-Phi-1} in a simpler form
\begin{align}
\label{Einstein-laplace-Phi}
    \triangle\Phi &= 
    \frac{a^2}{2M_{\mathrm{P}}^2}
    \bigg\{ \frac{\phi'_c}{a^2}(\delta \varphi'+3\mathcal{H}\delta \varphi)+\Big[V'(\phi_{c})+\frac{1}{2}I'_1(\phi_{c})\Big(\langle\boldsymbol{E}^2\rangle+\langle\boldsymbol{B}^2\rangle\Big)\Big] \delta \varphi \nonumber \\
    &-\frac{\phi^{\prime 2}_c}{a^2}\Psi
    + I_1(\phi_{c})\Big[(\Phi-\Psi)\langle\boldsymbol{E}^2\rangle +2\Phi\langle\boldsymbol{B}^2\rangle \Big]\bigg\} +F_\rho+3\mathcal{H}F_v \, .
\end{align}
With the help of Eq.~\eqref{Einstein-Psi-Phi}, $\Psi$ can be replaced by $\Phi$, and, with Eq.~\eqref{Einstein-Phi-prime}, $\delta \varphi$ can be eliminated. One thus obtains an equation for $\Phi$ alone, which, however, contains not only the source terms $F_\alpha$ but also their time derivatives. Since we prefer to work with the curvature perturbation variable $\zeta$, we do not write the resulting equation for $\Phi$ here.

%%%%%%%%%%%%%%%%%%%%%%%%%%%%%%%%%%%%%%%%%%%%%%%%%%%%%%%%%
\subsection{Maxwell's equations and source terms}
\label{subsec:Maxwell-eqs}
%%%%%%%%%%%%%%%%%%%%%%%%%%%%%%%%%%%%%%%%%%%%%%%%%%%%%%%%%

To determine the source terms $F_\alpha$ and their time derivatives, $F'_\alpha$, we must solve the Maxwell equations. Accounting for the scalar metric and inflaton perturbations in Maxwell's equations yields
\begin{align}
\label{divB}
    \operatorname{div}\boldsymbol{B} &=0\, , \\
\label{B-prime}
    \boldsymbol{B}'+2\mathcal{H}\boldsymbol{B}+\operatorname{rot}\boldsymbol{E} &=0\, ,\\
\label{divE}
    \operatorname{div}\boldsymbol{E}+\Big[\frac{I'_1(\phi_c)}{I_1(\phi_c)}\boldsymbol{E}+\frac{I'_2(\phi_c)}{I_1(\phi_c)}\boldsymbol{B}\Big]\cdot \boldsymbol{\nabla}\delta \varphi-\boldsymbol{E}\cdot\boldsymbol{\nabla}(\Phi+\Psi) &= 0\, , \\
    \boldsymbol{E}'+2\mathcal{H}\boldsymbol{E}-\operatorname{rot}\boldsymbol{B}+\Big[\frac{I'_1(\phi_c)}{I_1(\phi_c)}\boldsymbol{E}+\frac{I'_2(\phi_c)}{I_1(\phi_c)}\boldsymbol{B}\Big](\phi'_c+\delta \varphi')+\Big[\Big(\frac{I'_1(\phi_c)}{I_1(\phi_c)}\boldsymbol{B}-\frac{I'_2(\phi_c)}{I_1(\phi_c)}\boldsymbol{E}\Big)\!\times\!\boldsymbol{\nabla}\delta \varphi\Big] &  \nonumber \\
    +\bigg[\Big(\frac{I''_1(\phi_c)}{I_1(\phi_c)}-\frac{I'_1(\phi_c)^2}{I_1(\phi_c)^2}\Big)\boldsymbol{E}+\Big(\frac{I''_2(\phi_c)}{I_1(\phi_c)}-\frac{I'_1(\phi_c)I'_2(\phi_c)}{I_1(\phi_c)^2}\Big)\boldsymbol{B}\bigg]\phi'_c \delta \varphi &  \nonumber \\
    - (\Phi+\Psi)'\boldsymbol{E}-2(\Phi+\Psi)\operatorname{rot}\boldsymbol{B}+[\boldsymbol{B}\!\times\!\boldsymbol{\nabla}(\Phi+\Psi)]+\frac{I'_2(\phi_c)}{I_1(\phi_c)}\phi'_c (\Phi+\Psi)\boldsymbol{B} &= 0\, . \label{E-prime}
\end{align}
With the help of the perturbed Maxwell equations, \eqref{divB} to 
\eqref{E-prime}, we determine the time derivatives of the source terms:
\begin{align}
    F'_\rho &=\Big[-2\mathcal{H}+\frac{I'_1(\phi_c)}{I_1(\phi_c)}\phi'_c\Big] F_\rho + \triangle F_v - \frac{a^2 \phi'_c}{2M_{\mathrm{P}}^2} [I'_1(\phi_c) \delta_{\boldsymbol{E}^2}+I'_2(\phi_c)\delta_{\boldsymbol{E}\cdot\boldsymbol{B}}] \nonumber \\
    &\quad-\frac{a^2 I_1(\phi_c)}{2M_{\mathrm{P}}^2}\bigg\{\langle\boldsymbol{E}^2\rangle \Big[\frac{I'_1(\phi_c)}{I_1(\phi_c)}\delta \varphi'+\Big(\frac{I''_1(\phi_c)}{I_1(\phi_c)}-\frac{I^{\prime 2}_1(\phi_c)}{I^2_1(\phi_c)}\Big)\phi'_c\delta \varphi -(\Phi+\Psi)'\Big] \nonumber\\
    &\quad+\langle\syml\boldsymbol{E}\cdot\boldsymbol{B}\symr\rangle \Big[\frac{I'_2(\phi_c)}{I_1(\phi_c)}\delta \varphi'+\Big(\frac{I''_2(\phi_c)}{I_1(\phi_c)}-\frac{I^{\prime}_1(\phi_c)I^{\prime}_2(\phi_c)}{I^2_1(\phi_c)}\Big)\phi'_c\delta \varphi +\frac{I'_2(\phi_c)}{I_1(\phi_c)}\phi'_c(\Phi+\Psi)\Big] \nonumber \\
    &\quad+2\langle\syml\boldsymbol{E}\cdot\operatorname{rot}\boldsymbol{B}\symr\rangle (\Phi+\Psi)\bigg\}\, ,  \label{eq-F-rho}\\
    F'_v &= -2\mathcal{H} F_v + \frac{2}{3}\triangle F_\pi + \frac{1}{3}F_\rho -\frac{a^2}{2M_{\mathrm{P}}^2}\bigg\{\frac{1}{3}\langle\boldsymbol{E}^2\rangle \big[I'_1(\phi_c)\delta \varphi-I_1(\phi_c)(\Phi+\Psi)\big] + \langle\syml\boldsymbol{E}\cdot\boldsymbol{B}\symr\rangle I'_2(\phi_c) \delta \varphi \nonumber \\
    &\quad -\frac{2}{3}\langle\boldsymbol{B}^2\rangle \big[I'_1(\phi_c)\delta \varphi+I_1(\phi_c)(\Phi+\Psi)\big]\bigg\}\, ,
    \label{eq-F-v}\\
    F'_\pi &= \Big[-2\mathcal{H}+\frac{I'_1(\phi_c)}{I_1(\phi_c)}\phi'_c\Big] F_\pi \nonumber\\
    &\quad -\frac{a^2 I_1(\phi_c)}{2M_{\mathrm{P}}^2}\Big(\frac{\delta_{ij}}{\triangle}-3\frac{\partial_i \partial_j}{\triangle^2}\Big)\Big[\frac{I'_1(\phi_c)}{I_1(\phi_c)}\phi'_c \delta_{E_i E_j}+\frac{I'_2(\phi_c)}{I_1(\phi_c)}\phi'_c \delta_{E_i B_j}+\delta_{(\operatorname{rot}\boldsymbol{E})_i B_j}-\delta_{(\operatorname{rot}\boldsymbol{B})_i E_j}\Big]\, .
    \label{eq-F-pi}
\end{align}
Note that the equations of motion for $F_\rho$ and $F_v$ contain metric perturbations while the one for $F_\pi$ does not. This fact will be important below, where we derive a differential equation for the curvature-perturbation variable $\zeta$.

%%%%%%%%%%%%%%%%%%%%%%%%%%%%%%%%%%%%%%%%%%%%%%%%%%%%%%%%%
\subsection{Perturbed Klein--Gordon equation}
\label{subsec:KG-eq}
%%%%%%%%%%%%%%%%%%%%%%%%%%%%%%%%%%%%%%%%%%%%%%%%%%%%%%%%%

For completeness, we also write down the perturbed Klein--Gordon equation, which can be directly obtained from Eq.~\eqref{KGF} or by combining the Einstein equations:
\begin{align}
\label{KGF-pert}
\delta \varphi''+2\mathcal{H}\delta \varphi'-\triangle \delta \varphi + a^2& \Big[V''(\phi_{c})-\frac{I''_1(\phi_{c})}{2} \Big(\langle\boldsymbol{E}^2\rangle-\langle\boldsymbol{B}^2\rangle\Big)- I''_2(\phi_{c}) \langle\syml \boldsymbol{E}\cdot\boldsymbol{B}\symr \rangle \Big]\delta \varphi\nonumber
\\=\,&\phi'_c(\Psi'+3\Phi')+2(\phi^{\prime\prime}_c+2\mathcal{H}\phi'_c) \Psi + 
a^2 I'_1(\phi_{c})\Big[(\Phi-\Psi)\langle\boldsymbol{E}^2\rangle-2\Phi \langle\boldsymbol{B}^2\rangle\Big]   \nonumber   
\\&+a^2 I'_2(\phi_{c}) \langle\syml \boldsymbol{E}\cdot\boldsymbol{B}\symr \rangle (3\Phi-\Psi)+ \frac{a^2 I'_1(\phi_{c})}{2}\Big[\delta_{\boldsymbol{E}^2}-\delta_{\boldsymbol{B}^2}\Big] + a^2 I'_2(\phi_{c}) \delta_{\boldsymbol{E}\cdot\boldsymbol{B}}\, .
\end{align}
%\end{widetext}    %%COMMENT FOR 1-COLUMN
%
Note that this equation is not independent from the Einstein equations \eqref{Einstein-laplace-Phi-1} and \eqref{Einstein-Phi-prime}. We will use them interchangeably in our computations below.

%%%%%%%%%%%%%%%%%%%%%%%%%%%%%%%%%%%%%%%%%%%%%%%%%%%%%%%%%
\subsection{Curvature  perturbation variable \texorpdfstring{\boldmath{$\zeta$}}{zeta}}
\label{subsec:zeta-eq}
%%%%%%%%%%%%%%%%%%%%%%%%%%%%%%%%%%%%%%%%%%%%%%%%%%%%%%%%%

The curvature-perturbation variable, $\zeta$, can be expressed in terms of the Bardeen variables as~\cite{Durrer:2020fza}
\begin{align}
\zeta &= \Phi + \frac{\mathcal{H}}{\mathcal{H}^2-\mathcal{H}^\prime}(\Phi'+\mathcal{H}\Psi) \label{zeta-def-1}\\
    &= \Phi + \frac{\mathcal{H}}{\mathcal{H}^2-\mathcal{H}^\prime}\Big(\frac{\phi'_c}{2M_{\mathrm{P}}^2}\delta \varphi + F_v\Big)\, ,\label{zeta-def-2}
\end{align}
where we used Eq.~\eqref{Einstein-Phi-prime} for the second equality.
In  the absence of gauge fields on the hypersurfaces $\delta \varphi=0$ (unitary gauge), this variable describes the curvature of the spatial sections. It is very useful as it is simply related to the canonical perturbation variable that has to be quantized to find the scalar perturbations of the inflaton. In the absence of gauge fields, the scalar perturbations of the metric and the inflaton combine to 1 degree of freedom which is traditionally cast in terms of this $\zeta$ variable. The advantage of $\zeta$ is that it is constant on large, superhorizon scales in the absence of gauge fields. Therefore, the transition from inflation to the radiation-dominated era, at a time when all relevant modes are superhorizon, is especially simple in terms of this variable. Even though these nice properties are in general lost when gauge fields are present, and despite its rather contrived derivation, we nonetheless aim to find a second-order differential equation for $\zeta$ alone in which the source terms, $F_\alpha$, appear on the right-hand side. Standard inflationary predictions are then easily obtained by just setting the right-hand side to zero. This will also allow us to split $\zeta$ into a homogeneous solution, describing the inflaton perturbations, and a peculiar solution, due to the gauge-field source terms.

The derivation of this differential equation for the $\zeta$ variable is presented in full detail in Appendix~\ref{app:eq-for-zeta}. It can be summarized as follows:
we use Eqs.~\eqref{Einstein-Psi-Phi} and \eqref{Einstein-laplace-Phi} to express $\delta \varphi'$ as a function of $\Phi$, $\delta \varphi$, and the sources $F_v$, $F_\rho$, $F_\pi$. Then, we write $\Phi'$ in terms  of $\Phi$, $\delta \varphi$, and the sources  with the help of Eqs.~\eqref{Einstein-Phi-prime} and \eqref{Einstein-Psi-Phi}. Using Eqs.~\eqref{Einstein-Phi-prime-prime} and \eqref{Einstein-Psi-Phi}, we can now find $(\Phi'+\mathcal{H}\Psi)'$ as a function of $\Phi$, $\delta \varphi$, and the sources. Taking the time derivative of Eq.~\eqref{zeta-def-2}, we can also express $\zeta'$ as a function of $\Phi$, $\delta \varphi$, and the sources.
We then invert the relations $\zeta=\zeta(\Phi, \,\delta \varphi,\ F_\alpha)$ and $\zeta'=\zeta'(\Phi, \,\delta \varphi,\ F_\alpha)$ to determine $\Phi=\Phi(\zeta,\,\zeta',\,F_\alpha)$ and $\delta \varphi=\delta \varphi(\zeta,\,\zeta',\,F_\alpha)$.
Next, we take the time derivative of $\zeta'$ and express it as a function of $\Phi$, $\delta \varphi$, the sources, and their first time derivatives. Finally, we use the equations of motion for the sources derived in the previous section to express $\zeta''$ as a function of $\zeta'$, $\zeta$, the sources, and $F_\pi'$. This results in the desired second-order differential equation for $\zeta$. This derivation is best performed in Fourier space, where Laplacians simply generate a factor $(-k^2)$. The final equation takes the form
%
%%ONE-COLUMN MODE
\begin{equation}
    \zeta''_{\boldsymbol{k}}+p \zeta'_{\boldsymbol{k}} + q\zeta_{\boldsymbol{k}} = r^{(v)} F_v + r^{(\rho)} F_\rho + r^{(\pi)}_1 F_\pi  + r^{(\pi)}_2 F'_\pi
    + r^{(\boldsymbol{E}^2)} \delta_{\boldsymbol{E}^2} + r^{(\boldsymbol{E}\cdot\boldsymbol{B})} \delta_{\boldsymbol{E}\cdot\boldsymbol{B}}\, .
    \label{zeta-prime-prime-2}
\end{equation}
%
%%TWO-COLUMN MODE
% \begin{align}
%         \zeta''_{\boldsymbol{k}}+p \zeta'_{\boldsymbol{k}} + q\zeta_{\boldsymbol{k}} &= r^{(v)} F_v + r^{(\rho)} F_\rho + r^{(\pi)}_1 F_\pi  + r^{(\pi)}_2 F'_\pi
%         \nonumber \\ & \qquad
%         + r^{(\boldsymbol{E}^2)} \delta_{\boldsymbol{E}^2} + r^{(\boldsymbol{E}\cdot\boldsymbol{B})} \delta_{\boldsymbol{E}\cdot\boldsymbol{B}}\, .
%             \label{zeta-prime-prime-2}
% \end{align}
%
The coefficients $p$, $q$, and $r^{(\alpha)}$ are functions of time that depend only on the background evolution, including the zero modes, $\langle\boldsymbol{E}^2\rangle$, $\langle\boldsymbol{B}^2\rangle$, and
$\langle \syml\boldsymbol{E}\cdot \boldsymbol{B}\symr\rangle$. They are determined and presented in detail in Appendix~\ref{app:eq-for-zeta}. We emphasize that the decomposition of the source on the right-hand side of Eq.~\eqref{zeta-prime-prime-2} into six terms is performed merely for convenience. These terms are not independent, but rather depend on the same gauge-field mode functions.

In the case where we may neglect backreaction and set  $\langle\boldsymbol{E}^2\rangle = \langle\boldsymbol{B}^2\rangle
=\langle \syml\boldsymbol{E}\cdot \boldsymbol{B}\symr\rangle
=0$,
it is easy to check that the coefficients  $p$ and $q$ reduce to the usual form
\begin{align}
    q &= k^2\, ,\\
    p &= 2\frac{\mathcal{H}^2-\mathcal{H}'}{\mathcal{H}} +  \frac{\left(\mathcal{H}^2-\mathcal{H}'\right)'}{\mathcal{H}^2-\mathcal{H}'} = 2\frac{z'}{z} \, , \\
\intertext{where $z$ is the Mukhanov--Sasaki variable,}
    z &= \frac{a\, M_{\mathrm{P}}\sqrt{2(\mathcal{H}^2-\mathcal{H}')}}{\mathcal{H}} = \frac{a\phi_c'}{\mathcal{H}}\, .
\end{align}
As we have seen in our previous work~\cite{Durrer:2023rhc}, while the contribution of the gauge fields to the energy--momentum tensor remains small, they do significantly affect the inflaton evolution. Therefore, terms such as 
$I'_1(\phi_c) \langle\boldsymbol{E}^2\rangle$ or $I'_2(\phi_c) \langle\syml\boldsymbol{E}\cdot \boldsymbol{B}\symr\rangle$ cannot be neglected when compared to $\mathcal{H}\phi_c'$ or $V'$. These are exactly the kind of correction terms that enter $p$ and $q$; see Appendix~\ref{app:eq-for-zeta} for details.

Note that even in the absence of a direct gauge-field coupling to the inflaton, our $\zeta$ equation has a nontrivial source term on the right-hand side (see Appendix~\ref{subapp:no-direct-coupling}). This demonstrates the fact that the very presence of the gauge field during inflation already induces scalar perturbations through its gravity.

In general, one should treat the scalar variable $\zeta$ on the same footing as the gauge-field sources. For this, one would need to consider a coupled system of equations consisting of Eq.~\eqref{zeta-prime-prime-2}, the equations of motion for the sources, Eqs.~\eqref{eq-F-rho}--\eqref{eq-F-pi} for $F_\rho$, $F_v$, and $F_\pi$, and analogous equations for the other source terms. Then, this system of equations would have to be solved with the appropriate initial conditions in a similar way as is done for the adiabatic and entropy perturbations in multifield inflationary models; see, e.g., Refs.~\cite{Nakamura:1996da,Gordon:2000hv,Gong:2011uw}. This would allow one to take into account the impact of scalar inflaton and metric perturbations on the gauge-field modes. However, we expect this effect to be small compared to the mode amplification that comes from the background evolution via the kinetic and/or axial couplings. Therefore, in what follows, we compute the sources using the gauge-field mode functions determined on a spatially homogeneous background. Hence, we assume in Eq.~\eqref{zeta-prime-prime-2} that the sources on the right-hand side are known as soon as the background solution of the system is obtained.

To solve this inhomogeneous equation, we first have to determine its homogeneous solutions, $\zeta_{1,\boldsymbol{k}}$ and $\zeta_{2,\boldsymbol{k}}$, which obey
\begin{equation}
\label{eq-zeta-homogeneous}
    \zeta_{\boldsymbol{k}}''+p \zeta_{\boldsymbol{k}}' + q\zeta_{\boldsymbol{k}} \equiv \mathcal{D}\zeta_{\boldsymbol{k}} =0 \,.
\end{equation}
The Green function is then given by
\begin{equation}
\label{Green-function-zeta12}
    G_{\boldsymbol{k}}(\eta,\eta') \!= \!\frac{\zeta_{2,\boldsymbol{k}}(\eta)\zeta_{1,\boldsymbol{k}}(\eta')\! -\! \zeta_{1,\boldsymbol{k}}(\eta)\zeta_{2,\boldsymbol{k}}(\eta')}{W_{\boldsymbol{k}}(\eta')}\theta(\eta-\eta')\, ,
\end{equation}
where $W_{\boldsymbol{k}}$ is the Wronskian of the solutions $(\zeta_{1,\boldsymbol{k}},~\zeta_{2,\boldsymbol{k}})$,
\begin{equation}
W_{\boldsymbol{k}}(\eta') =\zeta_{1,\boldsymbol{k}}(\eta')\zeta'_{2,\boldsymbol{k}}(\eta')-\zeta_{2,\boldsymbol{k}}(\eta')\zeta'_{1,\boldsymbol{k}}(\eta')\, .
\end{equation}
It is straightforward to check that the Green function~\eqref{Green-function-zeta12} satisfies
\begin{equation}
\mathcal{D} G_{\boldsymbol{k}}(\eta,\eta') = \delta(\eta-\eta')
\end{equation}
together with the boundary conditions 
\begin{equation}
    G_{\boldsymbol{k}}(\eta,\eta'\geq \eta)\equiv 0\, , \quad \frac{\partial}{\partial \eta}G_{\boldsymbol{k}}(\eta,\eta')\Big\vert_{\eta'\to\eta} =1\, ,
\end{equation}
corresponding to the retarded Green function. Then, a particular solution $\bar{\zeta}_{\boldsymbol{k}}$ of \eqref{zeta-prime-prime-2}, with the source term $\mathcal{S}_{\boldsymbol{k}}(\eta)$ given by the right-hand side of \eqref{zeta-prime-prime-2}, can be written as
\begin{equation}
    \bar{\zeta}_{\boldsymbol{k}}(\eta) = \int\limits_{-\infty}^{\eta}d\eta'\, G_{\boldsymbol{k}}(\eta,\eta') \mathcal{S}_{\boldsymbol{k}}(\eta')\, .
\end{equation}
The general solution of Eq.~\eqref{zeta-prime-prime-2} is the sum of a homogeneous and this particular solution, and  has the form
\begin{equation}
\label{solution-zeta-formal}
    \zeta_{\boldsymbol{k}}(\eta)= \zeta^{(0)}_{\boldsymbol{k}}(\eta) + \int\limits_{-\infty}^{\eta}d\eta'\, G_{\boldsymbol{k}}(\eta,\eta') \mathcal{S}_{\boldsymbol{k}}(\eta')\, .
\end{equation}
In general, $\zeta^{(0)}_{\boldsymbol{k}}$ is an arbitrary linear combination of $\zeta_{1,\boldsymbol{k}}$ and $\zeta_{2,\boldsymbol{k}}$.
In our situation, the coefficients have to be chosen such that the homogeneous solution corresponds to the vacuum fluctuations of the canonical variable $v_{\boldsymbol{k}}=z\zeta_{\boldsymbol{k}}$ on subhorizon scales, $|k\eta|\gg 1$.

%%%%%%%%%%%%%%%%%%%%%%%%%%%%%%%%%%%%%%%%%%%%%%%%%%%%%%%%%
%%%%%%%%%%%%%%%%%%%%%%%%%%%%%%%%%%%%%%%%%%%%%%%%%%%%%%%%%
\section{Two-point and three-point correlation functions}
\label{sec:correlators}
%%%%%%%%%%%%%%%%%%%%%%%%%%%%%%%%%%%%%%%%%%%%%%%%%%%%%%%%%
%%%%%%%%%%%%%%%%%%%%%%%%%%%%%%%%%%%%%%%%%%%%%%%%%%%%%%%%%

The two contributions in Eq.~\eqref{solution-zeta-formal} are statistically independent since the first one can be interpreted as the vacuum fluctuations of the inflaton and the metric scalar degrees of freedom, i.e., the standard inflationary scalar perturbation, while the second one is induced by the gauge-field source $\mathcal{S}$. In practice, this means that there are no correlations between the two terms,
%
%%ONE-COLUMN MODE
\begin{equation}
\label{zeta-correlator}
    \langle \zeta_{\boldsymbol{k}}(\eta)\zeta_{\boldsymbol{k}'}(\eta)\rangle = \langle \zeta^{(0)}_{\boldsymbol{k}}(\eta)\zeta^{(0)}_{\boldsymbol{k}'}(\eta)\rangle
    +\int\limits_{-\infty}^{\eta}d\eta'\int\limits_{-\infty}^{\eta}d\eta''\, 
G_{\boldsymbol{k}}(\eta,\eta') G_{\boldsymbol{k}'}(\eta,\eta'') \langle \mathcal{S}_{\boldsymbol{k}}(\eta')\mathcal{S}_{\boldsymbol{k}'}(\eta'')\rangle\, .
\end{equation}
%
%%TWO-COLUMN MODE
% \begin{align}
% \label{zeta-correlator}
%     &\langle \zeta_{\boldsymbol{k}}(\eta)\zeta_{\boldsymbol{k}'}(\eta)\rangle = \langle \zeta^{(0)}_{\boldsymbol{k}}(\eta)\zeta^{(0)}_{\boldsymbol{k}'}(\eta)\rangle \nonumber\\
%     &\ +\!\!\int\limits_{-\infty}^{\eta}\!\!\!d\eta'\!\!\int\limits_{-\infty}^{\eta}\!\!\!d\eta''\, G_{\boldsymbol{k}}(\eta,\eta') G_{\boldsymbol{k}'}(\eta,\eta'') \langle \mathcal{S}_{\boldsymbol{k}}(\eta')\mathcal{S}_{\boldsymbol{k}'}(\eta'')\rangle\, .
% \end{align}
%
The part of the scalar power spectrum induced by the gauge fields is fully determined by the unequal-time correlation function of the source terms.%
\footnote{Note that the induced term is statistically independent from the vacuum fluctuations also in the case of a classical random source; see, e.g., Sec.~5.6 in Ref.~\cite{Ballesteros:2023dno}.}

In a similar way, we express the three-point correlation function of $\zeta$ Fourier modes, also called the \textit{bispectrum}. This non-Gaussian contribution is especially constraining since CMB observations have led to stringent limits on the bispectrum~\cite{Planck:2019kim}. Using Eq.~\eqref{solution-zeta-formal} and neglecting the three-point correlation function of the inflaton perturbations, which is slow-roll suppressed during inflation~\cite{Gangui:1993tt,Acquaviva:2002ud,Maldacena:2002vr}, we obtain
%
%%ONE-COLUMN MODE
\begin{equation}
    \label{bispectrum-general}
    \langle \zeta_{\boldsymbol{k}_1}(\eta) \zeta_{\boldsymbol{k}_2}(\eta) \zeta_{\boldsymbol{k}_3}(\eta) \rangle = \!\! \int\limits_{-\infty}^{\eta}\!\!d\eta_1 \!\!\int\limits_{-\infty}^{\eta}\!\!d\eta_2 \!\!\int\limits_{-\infty}^{\eta}\!\!d\eta_3\, G_{\boldsymbol{k}_1}(\eta,\eta_1) G_{\boldsymbol{k}_2}(\eta,\eta_2) G_{\boldsymbol{k}_3}(\eta,\eta_3) \langle \mathcal{S}_{\boldsymbol{k}_1}(\eta_1)\mathcal{S}_{\boldsymbol{k}_2}(\eta_2) \mathcal{S}_{\boldsymbol{k}_3}(\eta_3)\rangle\, ,
\end{equation}
%
%%TWO-COLUMN MODE
% \begin{multline}
%     \label{bispectrum-general}
%     \langle \zeta_{\boldsymbol{k}_1}(\eta) \zeta_{\boldsymbol{k}_2}(\eta) \zeta_{\boldsymbol{k}_3}(\eta) \rangle = \!\!\! \int\limits_{-\infty}^{\eta}\!\!d\eta_1 \!\!\int\limits_{-\infty}^{\eta}\!\!d\eta_2 \!\!\int\limits_{-\infty}^{\eta}\!\!d\eta_3\, G_{\boldsymbol{k}_1}(\eta,\eta_1)\\
%     \times G_{\boldsymbol{k}_2}(\eta,\eta_2) G_{\boldsymbol{k}_3}(\eta,\eta_3) \langle \mathcal{S}_{\boldsymbol{k}_1}(\eta_1)\mathcal{S}_{\boldsymbol{k}_2}(\eta_2) \mathcal{S}_{\boldsymbol{k}_3}(\eta_3)\rangle\, ,
% \end{multline}
%
which involves the unequal-time three-point correlation function of the source terms.

Below, we consider the source term in more detail and compute these unequal-time two- and three-point correlation functions.

%%%%%%%%%%%%%%%%%%%%%%%%%%%%%%%%%%%%%%%%%%%%%%%%%%%%%%%%%
\subsection{Structure of the source term}
\label{subsec:source-term}
%%%%%%%%%%%%%%%%%%%%%%%%%%%%%%%%%%%%%%%%%%%%%%%%%%%%%%%%%

In this subsection, we derive the explicit expression for the gauge-field source terms on the right-hand side of Eq.~\eqref{zeta-prime-prime-2}, needed to compute the scalar power spectrum and bispectrum. As discussed above Eq.~\eqref{eq-zeta-homogeneous}, one can compute the source using the gauge-field operators determined on a spatially homogeneous background. This implies that one can use the temporal gauge for the gauge field, $A_{\mu}=(0,\,-\boldsymbol{A})$, together with the Coulomb gauge condition, $\operatorname{div}\boldsymbol{A}=0$. The electric- and magnetic-field three-vectors are then given by
\begin{equation}
\label{fields-E-and-B}
    \boldsymbol{E}\equiv -\frac{1}{a^2}\boldsymbol{A}'\, , \qquad \boldsymbol{B}\equiv \frac{1}{a^{2}}{\rm rot\,}\boldsymbol{A}\, .
\end{equation}
We now decompose the quantized gauge-field vector potential into a set of annihilation and creation operators of Fourier modes with different momenta $\boldsymbol{p}$ and circular polarizations $\lambda$:
%
%%ONE-COLUMN MODE
\begin{equation}
\label{A-decomposition}
    \hat{\boldsymbol{A}}(\eta,\boldsymbol{x})=\int\frac{d^3 \boldsymbol{p}}{(2\pi)^{\frac{3}{2}}\sqrt{I_1(\phi_c)}}
     \sum_{\lambda=\pm} \Big[\boldsymbol{\epsilon}_{\lambda}(\boldsymbol{p}) \mathcal{A}_{\lambda,\boldsymbol{p}}(\eta)\hat{b}_{\lambda,\boldsymbol{p}}e^{i \boldsymbol{p}\cdot\boldsymbol{x}}
     +\boldsymbol{\epsilon}^{\ast}_{\lambda}(\boldsymbol{p}) \mathcal{A}^{\ast}_{\lambda,\boldsymbol{p}}(\eta)\hat{b}^{\dagger}_{\lambda,\boldsymbol{p}}e^{-i \boldsymbol{p}\cdot\boldsymbol{x}}\Big]\, ,
\end{equation}
%
%%TWO-COLUMN MODE
% \begin{multline}
% \label{A-decomposition}
%     \hat{\boldsymbol{A}}(\eta,\boldsymbol{x})\!=\!\!\int\!\!\frac{d^3 \boldsymbol{p}}{(2\pi)^{\frac{3}{2}}\sqrt{I_1(\phi_c)}}\!
%      \sum_{\lambda=\pm} \! \Big[\boldsymbol{\epsilon}_{\lambda}(\boldsymbol{p}) \mathcal{A}_{\lambda,\boldsymbol{p}}(\eta)\hat{b}_{\lambda,\boldsymbol{p}}e^{i \boldsymbol{p}\cdot\boldsymbol{x}}
%      \\+\boldsymbol{\epsilon}^{\ast}_{\lambda}(\boldsymbol{p}) \mathcal{A}^{\ast}_{\lambda,\boldsymbol{p}}(\eta)\hat{b}^{\dagger}_{\lambda,\boldsymbol{p}}e^{-i \boldsymbol{p}\cdot\boldsymbol{x}}\Big]\, ,
% \end{multline}
%
where $\mathcal{A}_{\lambda,\boldsymbol{p}}(\eta)$ is the mode function satisfying the Wronskian normalization condition
\begin{equation}
    \mathcal{A}^{\ast\prime}_{\lambda,\boldsymbol{p}}(\eta)\mathcal{A}_{\lambda,\boldsymbol{p}}(\eta)-\mathcal{A}^{\prime}_{\lambda,\boldsymbol{p}}(\eta)\mathcal{A}^{\ast}_{\lambda,\boldsymbol{p}}(\eta) = i\, ,
\end{equation}
$\boldsymbol{\epsilon}_{\lambda}(\boldsymbol{p})$ is the polarization vector with the properties
%
%%ONE-COLUMN MODE
\begin{equation}
    \boldsymbol{p}\cdot \boldsymbol{\epsilon}_{\lambda}(\boldsymbol{p})=0\, ,\quad \big(\boldsymbol{\epsilon}^{\ast}_{\lambda}(\boldsymbol{p})\cdot \boldsymbol{\epsilon}_{\lambda'}(\boldsymbol{p})\big)=\delta_{\lambda\lambda'}\, , \quad
    \boldsymbol{\epsilon}^{*}_{\lambda}(\boldsymbol{p})=\boldsymbol{\epsilon}_{-\lambda}(\boldsymbol{p})=\boldsymbol{\epsilon}_{\lambda}(-\boldsymbol{p})\, , \quad [i\boldsymbol{p}\times \boldsymbol{\epsilon}_{\lambda}(\boldsymbol{p})]=\lambda p \boldsymbol{\epsilon}_{\lambda}(\boldsymbol{p})\, ,
\end{equation}
%
%%TWO-COLUMN MODE
% \begin{subequations}
% \begin{align}
%     \boldsymbol{p}\cdot \boldsymbol{\epsilon}_{\lambda}(\boldsymbol{p})&=0\, ,\\ \big(\boldsymbol{\epsilon}^{\ast}_{\lambda}(\boldsymbol{p})\cdot \boldsymbol{\epsilon}_{\lambda'}(\boldsymbol{p})\big)&=\delta_{\lambda\lambda'}\, , \\
%     \boldsymbol{\epsilon}^{*}_{\lambda}(\boldsymbol{p})&=\boldsymbol{\epsilon}_{-\lambda}(\boldsymbol{p})=\boldsymbol{\epsilon}_{\lambda}(-\boldsymbol{p})\, , \\ [i\boldsymbol{p}\times \boldsymbol{\epsilon}_{\lambda}(\boldsymbol{p})]&=\lambda p \boldsymbol{\epsilon}_{\lambda}(\boldsymbol{p})\, ,
% \end{align}
% \end{subequations}
%
and $\hat{b}_{\lambda,\boldsymbol{p}}$ ($\hat{b}^{\dagger}_{\lambda,\boldsymbol{p}}$) are the annihilation (creation) operators satisfying the canonical commutation relations
\begin{equation}
    [\hat{b}_{\lambda,\boldsymbol{p}},\hat{b}^{\dagger}_{\lambda',\boldsymbol{p}'}]=
    \delta_{\lambda\lambda'} \delta(\boldsymbol{p}-\boldsymbol{p}')\, .
\end{equation}
Using the expressions for the electric- and magnetic-field operators in Eq.~\eqref{fields-E-and-B}, it can be readily shown that they are of the same form as  Eq.~\eqref{A-decomposition} with the mode functions $\mathcal{A}_{\lambda,\boldsymbol{p}}(\eta)$ replaced by
\begin{equation}
    \mathcal{E}_{\lambda,\boldsymbol{p}}(\eta)=-\frac{1}{a^2(\eta)}\Big[\mathcal{A}'_{\lambda,\boldsymbol{p}}(\eta)-\frac{I'_1(\phi_c)\phi'_c(\eta)}{2I_1(\phi_c)}\mathcal{A}_{\lambda,\boldsymbol{p}}(\eta)\Big]
\end{equation}
for the operator of the electric field, $\hat{\boldsymbol{E}}(\eta,\boldsymbol{x})$, or by
\begin{equation}
    \mathcal{B}_{\lambda,\boldsymbol{p}}(\eta)=\frac{\lambda p}{a^2(\eta)}\mathcal{A}_{\lambda,\boldsymbol{p}}(\eta)
\end{equation}
for the operator of the magnetic field, $\hat{\boldsymbol{B}}(\eta,\boldsymbol{x})$.

The right-hand side of Eq.~\eqref{zeta-prime-prime-2} contains six different terms. As already remarked upon previously, this decomposition is made for the sake of convenience and does not represent the expansion with respect to a particular basis. Upon realizing that all six terms are quadratic in the gauge fields, one can simply summarize them as one source-term operator that is quadratic in the gauge-field operators. This allows us to generically write the Fourier representation of the source as
%
%%ONE-COLUMN MODE
\begin{align}
\label{S-Fourier}
    \mathcal{S}_{\boldsymbol{k}}(\eta)=\int \frac{d^3\boldsymbol{p}}{(2\pi)^{\frac{3}{2}}}\,
    \sum_{\lambda,\lambda'=\pm} \Big[    &K_1(\lambda,\boldsymbol{p};\lambda',\boldsymbol{k}-\boldsymbol{p};\eta, \boldsymbol{k}) \hat{b}_{\lambda,\boldsymbol{p}}\hat{b}_{\lambda',\boldsymbol{k}-\boldsymbol{p}}+K_2(\lambda,\boldsymbol{p};\lambda',\boldsymbol{p}-\boldsymbol{k};\eta, \boldsymbol{k}) \hat{b}_{\lambda,\boldsymbol{p}}\hat{b}^{\dagger}_{\lambda',\boldsymbol{p}-\boldsymbol{k}}\nonumber\\
    +&K_3(\lambda,\boldsymbol{p};\lambda',\boldsymbol{k}+\boldsymbol{p};\eta, \boldsymbol{k}) \hat{b}^{\dagger}_{\lambda,\boldsymbol{p}}\hat{b}_{\lambda',\boldsymbol{k}+\boldsymbol{p}}
    +K_4(\lambda,\boldsymbol{p};\lambda',-\boldsymbol{k}-\boldsymbol{p};\eta, \boldsymbol{k}) \hat{b}^{\dagger}_{\lambda,\boldsymbol{p}}\hat{b}^{\dagger}_{\lambda',-\boldsymbol{k}-\boldsymbol{p}}\Big]\, ,
\end{align}
%
%%TWO-COLUMN MODE
% \begin{align}
%     \mathcal{S}_{\boldsymbol{k}}(\eta)&\!=\!\!\int\!\! \frac{d^3\boldsymbol{p}}{(2\pi)^{\!\frac{3}{2}}}\!\!
%     \sum_{\lambda,\lambda'=\pm}\!\! \Big[K_1(\lambda,\boldsymbol{p};\lambda',\boldsymbol{k}\!-\!\boldsymbol{p};\eta, \boldsymbol{k}) \hat{b}_{\lambda,\boldsymbol{p}}\hat{b}_{\lambda',\boldsymbol{k}-\boldsymbol{p}}\nonumber\\
%     &+K_2(\lambda,\boldsymbol{p};\lambda',\boldsymbol{p}-\boldsymbol{k};\eta, \boldsymbol{k}) \hat{b}_{\lambda,\boldsymbol{p}}\hat{b}^{\dagger}_{\lambda',\boldsymbol{p}-\boldsymbol{k}}\nonumber\\
%     &+K_3(\lambda,\boldsymbol{p};\lambda',\boldsymbol{k}+\boldsymbol{p};\eta, \boldsymbol{k}) \hat{b}^{\dagger}_{\lambda,\boldsymbol{p}}\hat{b}_{\lambda',\boldsymbol{k}+\boldsymbol{p}}\nonumber\\
%     &+K_4(\lambda,\boldsymbol{p};\lambda',-\boldsymbol{k}-\boldsymbol{p};\eta, \boldsymbol{k}) \hat{b}^{\dagger}_{\lambda,\boldsymbol{p}}\hat{b}^{\dagger}_{\lambda',-\boldsymbol{k}-\boldsymbol{p}}\Big]\, ,
% \label{S-Fourier}
% \end{align}
%
where the functions $K_1$, $K_2$, $K_3$, and $K_4$ are expressed in terms of the polarization vectors and the mode functions of the gauge field. Not all of these functions are independent as they satisfy the relations
\begin{align}
\label{K1-sym}
    K_{1}(\lambda,\boldsymbol{p}; \lambda', \boldsymbol{p}'; \eta, \boldsymbol{k}) &= K_{1}(\lambda',\boldsymbol{p}'; \lambda, \boldsymbol{p}; \eta, \boldsymbol{k}), \\
\label{K4-sym}
    K_{4}(\lambda,\boldsymbol{p}; \lambda', \boldsymbol{p}'; \eta, \boldsymbol{k}) &= K_{4}(\lambda',\boldsymbol{p}'; \lambda, \boldsymbol{p}; \eta, \boldsymbol{k}), \\
\label{K2-K3-sym}
    K_{2}(\lambda,\boldsymbol{p}; \lambda', \boldsymbol{p}'; \eta, \boldsymbol{k}) &= K_{3}(\lambda',\boldsymbol{p}'; \lambda, \boldsymbol{p}; \eta, \boldsymbol{k})\, ,
\end{align}
which follow directly from the definition \eqref{S-Fourier}, the fact that one can freely change the dummy integration and summation variables, and that the creation/annihilation operators in each pair in Eq.~\eqref{S-Fourier} commute if $\boldsymbol{k}\neq 0$. Moreover, the functions $K_1$ and $K_4$ are complex conjugates of each other,
\begin{equation}
\label{rel-K1-K4}
    K_4(\lambda,\boldsymbol{p};\lambda',\boldsymbol{p}';\eta,-\boldsymbol{k})=K^{\ast}_1(\lambda,\boldsymbol{p};\lambda',\boldsymbol{p}';\eta,\boldsymbol{k})\, .
\end{equation}
This property follows from the fact that the source $\mathcal{S}_{\boldsymbol{k}}$ is a Hermitian operator. Thus, it is sufficient to only find the expressions for two independent functions, e.g., $K_1$ and $K_3$.

Given our representation of the source term in Eq.~\eqref{zeta-prime-prime-2}, it is convenient to compute the functions $K_1$ and $K_3$ for each of the six terms separately. For this purpose, we introduce the additional upper index $L$ which runs from 1 to 6, enumerating the six terms in the source. The explicit expressions for the functions $K_1^L$ and $K_3^L$ are listed in Appendix~\ref{app:K-functions}.

%%%%%%%%%%%%%%%%%%%%%%%%%%%%%%%%%%%%%%%%%%%%%%%%%%%%%%%%%
\subsection{Scalar power spectrum}
\label{subsec:power-spectrum-general}
%%%%%%%%%%%%%%%%%%%%%%%%%%%%%%%%%%%%%%%%%%%%%%%%%%%%%%%%%

Taking the expression for the source in Eq.~\eqref{S-Fourier}, we can now compute the unequal-time two-point correlation function of the sources. Using Wick's theorem and $\hat{b}_{\lambda,\boldsymbol{p}}|0\rangle =\langle 0 |\hat{b}^{\dagger}_{\lambda,\boldsymbol{p}}=0$, we find the following expression for the correlator:
%
%%ONE-COLUMN MODE
\begin{equation}
\label{source-correlator}
    \langle \mathcal{S}_{\boldsymbol{k}}(\eta')\mathcal{S}_{\boldsymbol{k}'}(\eta'')\rangle=\delta(\boldsymbol{k}+\boldsymbol{k}') \int \frac{d^3\boldsymbol{p}}{(2\pi)^3}\sum_{\lambda,\lambda'=\pm} \Big[2K_1(\lambda,\boldsymbol{p};\lambda',\boldsymbol{k}-\boldsymbol{p};\eta', \boldsymbol{k}) K^{\ast}_1(\lambda,\boldsymbol{p};\lambda',\boldsymbol{k}-\boldsymbol{p};\eta'',\boldsymbol{k}) \Big]\, ,
\end{equation}
%
%%TWO-COLUMN MODE
% \begin{multline}
% \label{source-correlator}
%     \langle \mathcal{S}_{\boldsymbol{k}}(\eta')\mathcal{S}_{\boldsymbol{k}'}(\eta'')\rangle=\delta(\boldsymbol{k}+\boldsymbol{k}')
%     \\ \times \int \frac{d^3\boldsymbol{p}}{(2\pi)^3}\sum_{\lambda,\lambda'=\pm} \Big[
%     2K_1(\lambda,\boldsymbol{p};\lambda',\boldsymbol{k}-\boldsymbol{p};\eta', \boldsymbol{k})
%     \\ \times K^{\ast}_1(\lambda,\boldsymbol{p};\lambda',\boldsymbol{k}-\boldsymbol{p};\eta'',\boldsymbol{k}) \Big]\, ,
% \end{multline}
%
where we used the identity in Eq.~\eqref{rel-K1-K4} to eliminate $K_4$. Note that the two-point correlation function of the sources depends only on the function $K_1$. The terms proportional to $K_2$ and $K_3$ for nonvanishing momenta $\boldsymbol{k}$ always contain either at least one $\hat{b}_{\lambda,\boldsymbol{p}}$ on the right or a $\hat{b}^{\dagger}_{\lambda,\boldsymbol{p}}$ on the left and therefore do not contribute to the expectation value.

We can now exploit Eq.~\eqref{source-correlator} to derive the expression for the scalar power spectrum.  We substitute it into Eq.~\eqref{zeta-correlator}, the two-point correlation function of the $\zeta$ functions, and use the definition of the power spectrum,
\begin{equation}
    \langle \zeta_{\boldsymbol{k}}(\eta)\zeta_{\boldsymbol{k}'}(\eta)\rangle= \frac{2\pi^2}{k^3} P_{\zeta}(k,\eta) \delta(\boldsymbol{k}+\boldsymbol{k}'),
\end{equation}
to obtain
%
%%ONE-COLUMN MODE
\begin{equation}
\label{scalar-power-spectrum}
     P_{\zeta}(k,\eta)=P^{(0)}_{\zeta}(k,\eta)+\frac{k^3}{\pi^2}\int \frac{d^3\boldsymbol{p}}{(2\pi)^{3}}    \sum_{\lambda,\lambda'=\pm}\bigg|\int\limits_{-\infty}^{\eta}d\eta'\,G_{\boldsymbol{k}}(\eta,\eta') K_1(\lambda,\boldsymbol{p};\lambda',\boldsymbol{k}-\boldsymbol{p};\eta', \boldsymbol{k})\bigg|^2.
\end{equation}
%
%%TWO-COLUMN MODE
% \begin{multline}
% \label{scalar-power-spectrum}
%     P_{\zeta}(k,\eta)=P^{(0)}_{\zeta}(k,\eta)+\int \frac{d^3\boldsymbol{p}}{(2\pi)^{3}}    \sum_{\lambda,\lambda'=\pm} \frac{k^3}{\pi^2}\\
%     \times \bigg|\int\limits_{-\infty}^{\eta}d\eta'\,G_{\boldsymbol{k}}(\eta,\eta') K_1(\lambda,\boldsymbol{p};\lambda',\boldsymbol{k}-\boldsymbol{p};\eta', \boldsymbol{k})\bigg|^2.
% \end{multline}
%
Here, the first term originates from the vacuum fluctuations of $\zeta$, and the second one represents the induced contribution due to the presence of the gauge fields, which is sometimes referred to as ``the inverse decay'' term~\cite{Barnaby:2011vw}.

%%%%%%%%%%%%%%%%%%%%%%%%%%%%%%%%%%%%%%%%%%%%%%%%%%%%%%%%%
\subsection{Scalar bispectrum}
\label{subsec:bispectrum-general}
%%%%%%%%%%%%%%%%%%%%%%%%%%%%%%%%%%%%%%%%%%%%%%%%%%%%%%%%%

To compute the scalar bispectrum in Eq.~\eqref{bispectrum-general}, we need the unequal-time three-point correlation function of the sources. This correlation function can be found by using the representation of the source function obtained in Eq.~\eqref{S-Fourier} and computing the vacuum expectation value of the products of six annihilation and creation operators using Wick's theorem. The final expression reads as
%
%%ONE-COLUMN MODE
\begin{multline}
\label{source-three-point}
    \langle \mathcal{S}_{\boldsymbol{k}_1}(\eta_1)\mathcal{S}_{\boldsymbol{k}_2}(\eta_2) \mathcal{S}_{\boldsymbol{k}_3}(\eta_3)\rangle =\  \delta(\boldsymbol{k}_1 + \boldsymbol{k}_2 + \boldsymbol{k}_3)
    \int\frac{d^{3}\boldsymbol{p}}{(2\pi)^{9/2}} \sum\limits_{\{\lambda\}} \Big[ 8K_{1}(\lambda_1,\boldsymbol{p}; \lambda_2, \boldsymbol{k}_1-\boldsymbol{p}; \eta_1, \boldsymbol{k}_1)\\
    \times K_{3}(\lambda_2, \boldsymbol{k}_1-\boldsymbol{p}; \lambda_3,-\boldsymbol{k}_3-\boldsymbol{p};  \eta_2, \boldsymbol{k}_2) K_{4}( \lambda_3,-\boldsymbol{k}_3-\boldsymbol{p}; \lambda_1,\boldsymbol{p}; \eta_3, \boldsymbol{k}_3)\Big]\, ,
\end{multline}
%
%%TWO-COLUMN MODE
% \begin{multline}
% \label{source-three-point}
%     \langle \mathcal{S}_{\boldsymbol{k}_1}(\eta_1)\mathcal{S}_{\boldsymbol{k}_2}(\eta_2) \mathcal{S}_{\boldsymbol{k}_3}(\eta_3)\rangle =\  \delta(\boldsymbol{k}_1 + \boldsymbol{k}_2 + \boldsymbol{k}_3)
%     \\ \times \!\!\int\!\!\frac{d^{3}\boldsymbol{p}}{(2\pi)^{9/2}} \sum\limits_{\{\lambda\}} \Big[ 8K_{1}(\lambda_1,\boldsymbol{p}; \lambda_2, \boldsymbol{k}_1-\boldsymbol{p}; \eta_1, \boldsymbol{k}_1)\\
%     \times K_{3}(\lambda_2, \boldsymbol{k}_1-\boldsymbol{p}; \lambda_3,-\boldsymbol{k}_3-\boldsymbol{p};  \eta_2, \boldsymbol{k}_2) \\
%     \times K_{4}( \lambda_3,-\boldsymbol{k}_3-\boldsymbol{p}; \lambda_1,\boldsymbol{p}; \eta_3, \boldsymbol{k}_3)\Big]\, ,
% \end{multline}
%
where the shorthand notation $\{\lambda\}$ means that the summation is performed over all polarization indices (in this case, over three of them, $\lambda_1,\, \lambda_2,$ and $\lambda_3$). Note that we used the symmetry properties of the functions $K_1$ to $K_4$ given in Eqs.~\eqref{K1-sym}--\eqref{K2-K3-sym} to simplify the final result.

Substituting Eq.~\eqref{source-three-point} into the equal-time three-point correlation function for $\zeta$, Eq.~\eqref{bispectrum-general}, we obtain 
\begin{equation}
    \langle \zeta_{\boldsymbol{k}_1}(\eta) \zeta_{\boldsymbol{k}_2}(\eta) \zeta_{\boldsymbol{k}_3}(\eta) \rangle=\delta(\boldsymbol{k}_1 + \boldsymbol{k}_2 + \boldsymbol{k}_3) \mathscr{B}_{\zeta}(k_1,\,k_2,\,k_3)\, ,
\end{equation}
where the bispectrum $\mathscr{B}_{\zeta}$ is given by
%
%%ONE-COLUMN MODE
\begin{align}
      \mathscr{B}_{\zeta}(k_1,\,k_2,\,k_3) &= \Big(\frac{2}{\pi}\Big)^{3/2}\!\int\frac{d^{3}\boldsymbol{p}}{(2\pi)^{3}} \sum\limits_{\{\lambda\}} \int\limits_{-\infty}^{\eta}\!\! d\eta_1\,  G_{\boldsymbol{k}_1}(\eta,\eta_1) K_{1}(\lambda_1,\boldsymbol{p}; \lambda_2, \boldsymbol{k}_1-\boldsymbol{p}; \eta_1, \boldsymbol{k}_1) \nonumber\\
    &\hspace{-1.5cm}\times\!\! \int\limits_{-\infty}^{\eta}\!\! d\eta_2\, G_{\boldsymbol{k}_2}(\eta,\eta_2) K_{3}(\lambda_2, \boldsymbol{k}_1-\boldsymbol{p}; \lambda_3,-\boldsymbol{k}_3-\boldsymbol{p};  \eta_2, \boldsymbol{k}_2) \int\limits_{-\infty}^{\eta}\!\! d\eta_3\,  G_{\boldsymbol{k}_3}(\eta,\eta_3) K_{4}( \lambda_3,-\boldsymbol{k}_3-\boldsymbol{p}; \lambda_1,\boldsymbol{p}; \eta_3, \boldsymbol{k}_3)\, . \label{zeta-bispectrum}
\end{align}
%
%%TWO-COLUMN MODE
% \begin{align}
%     &\mathscr{B}_{\zeta}(k_1,\,k_2,\,k_3) = \!\int\frac{d^{3}\boldsymbol{p}}{(2\pi)^{3}} \sum\limits_{\{\lambda\}} \Big(\frac{2}{\pi}\Big)^{3/2} \nonumber\\
%     &\times\!\!\int\limits_{-\infty}^{\eta} \!\!d\eta_1 \, G_{\boldsymbol{k}_1}(\eta,\eta_1) K_{1}(\lambda_1,\boldsymbol{p}; \lambda_2, \boldsymbol{k}_1\!-\!\boldsymbol{p}; \eta_1, \boldsymbol{k}_1) \nonumber\\
%     &\times\!\! \int\limits_{-\infty}^{\eta} \!\!d\eta_2\,  G_{\boldsymbol{k}_2}(\eta,\eta_2) K_{3}(\lambda_2, \boldsymbol{k}_1\!-\!\boldsymbol{p}; \lambda_3,-\boldsymbol{k}_3\!-\!\boldsymbol{p};  \eta_2, \boldsymbol{k}_2)\nonumber\\
%     &\times\!\!\int\limits_{-\infty}^{\eta} \!\!d\eta_3 \, G_{\boldsymbol{k}_3}(\eta,\eta_3) K_{4}( \lambda_3,-\boldsymbol{k}_3\!-\!\boldsymbol{p}; \lambda_1,\boldsymbol{p}; \eta_3, \boldsymbol{k}_3)\, . \label{zeta-bispectrum}
% \end{align}
%
Note that this quantity also depends on time, $\eta$; however, for brevity, we do not write this dependence explicitly in the arguments. As a consequence of statistical homogeneity, the bispectrum is nonvanishing only for $\boldsymbol{k}_1 + \boldsymbol{k}_2 + \boldsymbol{k}_3=0$, i.e., for configurations of momenta that form a triangle. Because of the statistical spatial isotropy of the system under study, the orientation of this triangle in space does not matter, and the bispectrum depends only on the shape of the triangle, which can be parametrized, e.g., by the moduli of its sides $k_1$, $k_2$, and $k_3$. They satisfy the triangle inequality
\begin{equation}
    k_i + k_j \geq k_l\, ,
\end{equation}
where $(ijl)$ runs over all permutations of the indices 1, 2, and 3. Note that the integrand in Eq.~\eqref{zeta-bispectrum} depends on the three-vectors $\boldsymbol{k}_1$, $\boldsymbol{k}_2$, and $\boldsymbol{k}_3$; however, the dependence on the orientation will disappear after integrating over the momentum $\boldsymbol{p}$. Moreover, the bispectrum does not change under a permutation of the sides of the triangle since any of the resulting triangles can be brought back to the original one by means of a spatial rotation. The latter property allows us to consider only configurations of momenta for which $k_1\geq k_2 \geq k_3$.

Thus, the bispectrum has a more complicated structure than the scalar power spectrum in Eq.~\eqref{scalar-power-spectrum}. Similar to the latter case, there is a dependence on the overall momentum scale, which can be determined, for instance, from the equilateral configuration $k_1=k_2=k_3=k$,
\begin{equation}
    \mathscr{B}^{\mathrm{eq}}_{\zeta}(k) \equiv \mathscr{B}_{\zeta}(k,\,k,\,k)\, .
\end{equation}
In addition, the dependence on the shape of the triangle for a fixed overall momentum scale $k\equiv \max\{k_1,k_2,k_3\} = k_1$ can be described by a dimensionless shape function
\begin{equation}
    \label{shape-function-general}
    \Sigma(k;\, x_2,\, x_3) \equiv \mathcal{N} (k_1 k_2 k_3)^2 \mathscr{B}_{\zeta}(k_1,\,k_2,\,k_3)\, ,
\end{equation}
where $x_2=k_2/k$, $x_3 = k_3/k$, and the normalization factor $\mathcal{N}$ depends on the convention. For example, it can be chosen in such a way that the shape function equals unity in the equilateral configuration, $\Sigma(k;\,1,\,1)=1$. Note that the choice of the overall momentum scale $k$, and, therefore, the arguments of the shape function, is not unique, and one may find different conventions in the literature~\cite{Babich:2004gb,Fergusson:2008ra}; however, the right-hand side of the definition \eqref{shape-function-general} is always the same. In this work, we stick to the convention of Ref.~\cite{Babich:2004gb}. In the simplest cases, when the dependence on the overall momentum is very weak or absent (as in the case considered in Sec.~\ref{sec:application}), the shape function depends only on the ratios of momenta, $x_2$ and $x_3$.

Observational constraints on PNG follow from the CMB observations~\cite{Planck:2019kim} and are only known for a few particular patterns (shapes) of the bispectrum, such as the local~\cite{Gangui:1993tt,Komatsu:2001rj}, equilateral~\cite{Creminelli:2005hu}, or orthogonal~\cite{Senatore:2009gt} shapes. The corresponding shape functions (normalized to unity in the equilateral configuration) take the following form:
\begin{align}
    \Sigma_{\mathrm{local}}(x_2,\,x_3) &= \frac{1+x_2^3 + x_3^3}{3 x_2 x_3}\, , \label{pattern-local}\\
    \Sigma_{\mathrm{equil}}(x_2,\,x_3) &= \frac{(1\!+\!x_2\!-\!x_3)(1\!-\!x_2\!+\!x_3)(x_2\!+\!x_3\!-\!1)}{x_2 x_3}\, , \label{pattern-equil}\\
    \Sigma_{\mathrm{ortho}}(x_2,\,x_3) &= 3\Sigma_{\mathrm{equil}}(x_2,\,x_3)-2\, . \label{pattern-ortho}
\end{align}
To compare the predictions of any particular model to the CMB constraints, one needs to compare the shape function obtained for this model to each of the reference patterns. Usually, to quantify the ``similarity'' of two shapes $\Sigma$ and $\Sigma'$, one defines a  scalar product on the space of shape functions and then computes the cosine of the ``angle'' between the two shape functions. Here, again, there are different conventions in the literature~\cite{Babich:2004gb,Fergusson:2008ra}. For simplicity, in this article, we stick to the definition given in Ref.~\cite{Babich:2004gb}. For the scalar product, we simply use
\begin{equation}
    \label{scalar-product-shapes}
    (\Sigma\, , \Sigma') = \int dx_2 dx_3\,\Sigma(x_2,x_3) \Sigma'(x_2,x_3)\, ,
\end{equation}
where the integration is performed over the region constrained by the inequalities 
\begin{equation}
    \label{limits-shape}
    x_2+x_3\geq 1\, ,\qquad x_3\leq x_2\leq 1,
\end{equation}
which follow from the triangle inequality and our ordering of the variables, $k_1\geq k_2\geq k_3$.
The cosine of the generalized angle $\Theta_{\Sigma,\Sigma'}$ between two shapes in the vector space of shape functions is then given by
\begin{equation}
    \label{cosine-angle-shapes}
    \cos\Theta_{\Sigma,\Sigma'} = \frac{(\Sigma\, , \Sigma')}{\sqrt{(\Sigma\, , \Sigma)(\Sigma'\, , \Sigma')}}\, .
\end{equation}
In order to validate our formalism and compare its outcome to the known results in the literature, in the next section, we apply these equations to a particular physical model: axion inflation with negligible backreaction.

%%%%%%%%%%%%%%%%%%%%%%%%%%%%%%%%%%%%%%%%%%%%%%%%%%%%%%%%%
%%%%%%%%%%%%%%%%%%%%%%%%%%%%%%%%%%%%%%%%%%%%%%%%%%%%%%%%%
\section{Application: Axion inflation without backreaction}
\label{sec:application}
%%%%%%%%%%%%%%%%%%%%%%%%%%%%%%%%%%%%%%%%%%%%%%%%%%%%%%%%%
%%%%%%%%%%%%%%%%%%%%%%%%%%%%%%%%%%%%%%%%%%%%%%%%%%%%%%%%%

We  apply our general formalism to a simple, particular case of axion inflation where we have a purely axial coupling to the gauge fields. For simplicity, we also neglect backreaction in this first study.

%%%%%%%%%%%%%%%%%%%%%%%%%%%%%%%%%%%%%%%%%%%%%%%%%%%%%%%%%
\subsection{Inventory}
\label{subsec:axion-basics}
%%%%%%%%%%%%%%%%%%%%%%%%%%%%%%%%%%%%%%%%%%%%%%%%%%%%%%%%%

\subsubsection{Formulation of the model}

In the case under consideration, the kinetic coupling is trivial,
\begin{equation}
    I_1(\phi_c)\equiv 1\, , \qquad I'_1(\phi_c)=I''_1(\phi_c)=0\, ;
\end{equation}
the axial coupling function is assumed to be linear in the inflaton field
\begin{equation}
    I_2(\phi_c)=\frac{\alpha}{f}\phi_c\, , \ \  I'_2(\phi_c)=\frac{\alpha}{f}=\text{const}\, , \ \  I''_2(\phi_c)=0\, ;
\end{equation}
all gauge-field zero-mode contributions are neglected, i.e.,
\begin{equation}
    \langle \boldsymbol{E}^2\rangle \approx 0\, , \quad \langle \boldsymbol{B}^2\rangle \approx 0\, , \quad \langle\syml\boldsymbol{E}\cdot\boldsymbol{B}\symr\rangle \approx 0\, ,
\end{equation}
in all equations for the perturbations. This implies that the left-hand side (LHS) of Eq.~\eqref{zeta-prime-prime-2} takes the same form as in the Mukhanov--Sasaki equation,
\begin{equation}
 \mathrm{LHS} = \zeta''_{\boldsymbol{k}} + 2\frac{z'}{z}\zeta'_{\boldsymbol{k}} +k^2\zeta_{\boldsymbol{k}}\, ,
\end{equation}
where $z=a\phi'_c/\mathcal{H}$. The coefficients $r^{(L)}$ in the source terms on the right-hand side of Eq.~\eqref{zeta-prime-prime-2} for this particular case are listed in Appendix~\ref{subapp:axion-noBR-details}.

For further convenience, we introduce the dimensionless parameter
\begin{equation}
\label{xi-def}
    \xi= \frac{I'_2(\phi_c) \phi'_c}{2\mathcal{H}}=\frac{\alpha \dot{\phi}_c}{2Hf}\, ,
\end{equation}
which characterizes the efficiency of gauge-field production during axion inflation. The range of $\xi$ values for which the gauge-field backreaction can be neglected is strongly model-dependent. As we discuss in Appendix~\ref{app:no-BR}, the onset of backreaction may happen already for $\xi\sim 1$ in models with ultra-slow-roll stages while it may only occur for $\xi> 10$ in low-scale inflationary models. Therefore, we do not specify any constraints on the value of $\xi$ in what follows and perform the analysis assuming the absence of backreaction.

We determine the contribution to $\zeta$ coming from the gauge fields by neglecting slow-roll corrections. This corresponds to the assumption that the Hubble parameter $H$ and the inflaton velocity $\dot{\phi}_c$ are constant, implying that the parameter $\xi$ is constant as well.%
\footnote{Note that, while we neglect the contribution from $\dot\phi_c$ in the Friedmann equation by setting $H=\mathrm{const}$, we implicitly retain it in $\xi\neq 0$. It is, of course, also retained in the homogeneous $\zeta$ power spectrum, which diverges in the limit $\dot{\phi}_c\to 0$.}
In this approximation, the scale factor is given by  the de Sitter solution,
\begin{equation}
\label{a-of-eta-dS}
    a(\eta) = -\frac{1}{H\eta}\, .
\end{equation}

\subsubsection{The Green function}

The Green function for $\zeta$ satisfies the equation
\begin{equation}
    \Big(\frac{\partial^2}{\partial\eta^2}+2\frac{z'}{z}\frac{\partial}{\partial\eta} + k^2\Big) G_{\boldsymbol{k}}(\eta,\eta')=\delta(\eta-\eta')\, ,
\end{equation}
where again $z=a \dot{\phi}_{c}/H$. To find $G_{\boldsymbol{k}}(\eta,\eta')$, we follow the procedure described at the end of Sec.~\ref{sec:eqs-for-perturbations}; i.e., we look for two independent solutions to the corresponding homogeneous equation.

The Mukhanov--Sasaki variable $z$ behaves as $z\propto a \propto \eta^{-1}$ leading to the following homogeneous equation of motion:
\begin{equation}
    \zeta''_{\boldsymbol{k}} - \frac{2}{\eta} \zeta'_{\boldsymbol{k}} + k^2 \zeta_{\boldsymbol{k}} = 0\, .
\end{equation}
Its positive- (negative-) frequency solutions can be expressed in terms of Hankel functions,
%
%%ONE-COLUMN MODE
\begin{equation}
    \zeta_{1(2),\boldsymbol{k}}(\eta) = C \, (-k\eta)^{3/2} H^{(1,2)}_{3/2}(-k\eta)
    = -C \sqrt{\frac{2}{\pi}} (-k\eta \pm i) e^{\mp ik\eta}\, ,
\end{equation}
%
%%TWO-COLUMN MODE
% \begin{align}
%     \zeta_{1(2),\boldsymbol{k}}(\eta) &= C \, (-k\eta)^{3/2} H^{(1,2)}_{3/2}(-k\eta)
%     \nonumber\\& 
%     = -C \sqrt{\frac{2}{\pi}} (-k\eta \pm i) e^{\mp ik\eta}\, ,
% \end{align}
%
where $C$ is a normalization constant, which is not relevant for the Green function, and the upper (lower) sign corresponds to the first (second) solution. Substituting these solutions into Eq.~\eqref{Green-function-zeta12}, we obtain the following expression for the Green function:
%
%%ONE-COLUMN MODE
\begin{equation}
\label{Green-function-axion-full}
    G_{\boldsymbol{k}}(\eta,\eta') = \theta(y'-y)\frac{1}{k y^{\prime 2}} \big[(y-y') \cos(y-y') - (1+yy')\sin(y-y')\big]\, ,
\end{equation}
%
%%TWO-COLUMN MODE
% \begin{align}
% \label{Green-function-axion-full}
%     G_{\boldsymbol{k}}(\eta,\eta') &= \theta(y'-y)\frac{1}{k y^{\prime 2}} \\
%     &\times \big[(y-y') \cos(y-y') - (1+yy')\sin(y-y')\big]\, ,\nonumber
% \end{align}
%
where $y=-k\eta$ and $y'=-k\eta'$. If we further assume that the mode under consideration is well beyond the Hubble horizon at the moment of time $\eta$, i.e., $y=-k\eta \ll 1$, the expression for the Green function can be further simplified to
\begin{equation}
\label{Green-function-axion-approx}
    G_{\boldsymbol{k}}(\eta,\eta') \simeq \theta(y'-y)\frac{1}{k y^{\prime 2}}\big(\sin y' - y' \cos y'\big)\, .
\end{equation}
In what follows, we use this approximation to compute the scalar power spectrum and bispectrum.

\subsubsection{Gauge-field mode functions}

In the case of $\xi=\mathrm{const}$ in a de Sitter universe, the equation of motion for the gauge-field mode function $\mathcal{A}_{\lambda,\boldsymbol{p}}(\eta)$ [see Eq.~\eqref{A-decomposition}] has the form~\cite{Anber:2009ua}
\begin{equation}
\label{mode-eq-A}
    \mathcal{A}''_{\lambda,\boldsymbol{p}}(\eta) + \Big[p^2 + \lambda p \frac{2\xi}{\eta}\Big]\mathcal{A}_{\lambda,\boldsymbol{p}}(\eta)=0\, .
\end{equation}
Since the sign of the second term in brackets depends on the polarization of the mode, only one type of helicity mode becomes tachyonically unstable and exponentially enhanced. This is a manifestation of the spontaneous parity violation introduced by the axial coupling.

The solution to Eq.~\eqref{mode-eq-A} must satisfy the Bunch--Davies initial condition deep inside the horizon,
\begin{equation}
\label{BD-condition}
    \mathcal{A}_{\lambda,\boldsymbol{p}}(\eta) \simeq \frac{1}{\sqrt{2p}} e^{-ip\eta}\, , \quad |p\eta|\gg 1,
\end{equation}
which corresponds to the positive-frequency solution in the asymptotic past.%
\footnote{In this section, we use symbols $p$ and $q$ to denote absolute values of the momenta. This should not lead to a confusion with the coefficients $p$ and $q$ in the equation of motion for $\zeta$, which never show up in this section.}

By introducing the dimensionless variable $\rho=-p\eta$
and the function $\mathcal{W}_{\lambda}(\rho)$ via the relation%
\footnote{Here, we use the same symbol $\rho$ for the new dimensionless variable as for the energy density. However, this should not lead to confusion since this variable is never used outside the present subsection and Appendix~\ref{app:large-xi}, while the energy density does not occur there.}
\begin{equation}
    \mathcal{A}_{\lambda,\boldsymbol{p}}(\eta) = \frac{1}{\sqrt{2p}}\mathcal{W}_{\lambda}(\rho),
\end{equation}
we eliminate the momentum dependence both from the mode equation \eqref{mode-eq-A} and from the initial condition \eqref{BD-condition}:
\begin{align}
    \mathcal{W}''_{\lambda}(\rho) &+ \Big[1 - \lambda \frac{2\xi}{\rho}\Big]\mathcal{W}_{\lambda}(\rho) = 0\, ,\\
    \mathcal{W}_{\lambda}(\rho) &\simeq e^{i\rho}\, , \quad \rho\gg 1\, .
\end{align}
The exact solution to these equations is expressed in terms of special functions as
\begin{align}
\label{sol-Whittaker}
    \mathcal{W}_{\lambda}(\rho) &= e^{\lambda\pi\xi/2}W_{-i\lambda\xi,1/2}(-2i\rho)\, \\
\label{sol-Coulomb}
    &= G_0(\lambda\xi,\rho) + i F_0(\lambda\xi,\rho)\, ,
\end{align}
where $W_{\kappa,\mu}(z)$ is the Whittaker function, while $F_{0}(\varkappa,\rho)$ and $G_{0}(\varkappa,\rho)$ are the regular and irregular Coulomb wave functions with zero angular momentum, respectively.

For further convenience, we introduce an auxiliary function $\mathcal{U}_{\lambda}(\rho)$ by the relation
\begin{equation}
    \mathcal{A}'_{\lambda,\boldsymbol{p}}(\eta) = \sqrt{\frac{p}{2}}\mathcal{U}_{\lambda}(\rho)\, .
\end{equation}
The exact mode function, $\mathcal{U}_{\lambda}(\rho) = -\mathcal{W}'_{\lambda}(\rho)$, then takes the form
\begin{align}
\label{sol-Whittaker-U}
    \mathcal{U}_{\lambda}(\rho) &= e^{\lambda\pi\xi/2}\frac{1}{\rho}\big[i(\rho-\lambda\xi)W_{-i\lambda\xi,1/2}(-2i\rho) 
   % \nonumber\\& \hspace{2.6cm}     %%COMMENT IN 1-COLUMN
    + W_{1-i\lambda\xi,1/2}(-2i\rho)\big]\\
\label{sol-Coulomb-U}
    &= -\big[G'_0(\lambda\xi,\rho) + i F'_0(\lambda\xi,\rho)\big]\, ,
\end{align}
where we used the expression for the derivative of the Whittaker function given in Eq.~(13.4.33) in Ref.~\cite{AbramowitzStegun:1964handbook}. We show the mode functions $\mathcal{W}_{+}(\rho)$ and $\mathcal{U}_{+}(\rho)$ for a positive helicity in Fig.~\ref{fig:mode-functions} in Appendix~\ref{app:large-xi}.

Finally, the mode functions for the electric- and mag\-ne\-tic-field operators are expressed in terms of the dimensionless functions $\mathcal{W}_{\lambda}(\rho)$ and $\mathcal{U}_{\lambda}(\rho)$ as follows:
\begin{equation}
\label{E-B-functions-U-W}
    \hspace{-2mm}\mathcal{E}_{\lambda,\boldsymbol{p}}(\eta) \!=\! -\frac{1}{a^2} \sqrt{\frac{p}{2}} \mathcal{U}_{\lambda}(\rho)\, ,\ 
    \mathcal{B}_{\lambda,\boldsymbol{p}}(\eta)\! =\! \frac{\lambda}{a^2} \sqrt{\frac{p}{2}} \mathcal{W}_{\lambda}(\rho)\, .
\end{equation}

\subsubsection{Functions \texorpdfstring{$K_1$}{} and \texorpdfstring{$K_3$}{}}

For the following calculations, we will use the explicit expressions for the coefficients  $r^{(L)}$ derived in Appendix~\ref{subapp:axion-noBR-details} and compute the functions $K_1$ and $K_3$ in this simple case.
All six terms of the source term in Eq.~\eqref{zeta-prime-prime-2} can be combined, yielding
%
%%ONE-COLUMN MODE
\begin{multline}
\label{K1-axion-noBR}
    K_1(\lambda,\boldsymbol{p};\lambda',\boldsymbol{p}';\eta, \boldsymbol{k})
    = \big(\boldsymbol{\epsilon}_{\lambda}(\boldsymbol{p})\cdot\boldsymbol{\epsilon}_{\lambda'}(\boldsymbol{p}')\big) \frac{a^2\mathcal{H}^2}{\phi^{\prime 2}_c}
    \bigg\{\xi\big[1 + h_1(\eta)\big]\Big[\mathcal{E}_{\lambda,\boldsymbol{p}}(\eta) \mathcal{B}_{\lambda',\boldsymbol{p}'}(\eta)+\mathcal{B}_{\lambda,\boldsymbol{p}}(\eta) \mathcal{E}_{\lambda',\boldsymbol{p}'}(\eta)\Big]\\
    +h_2(\eta)\Big[\mathcal{E}_{\lambda,\boldsymbol{p}}(\eta) \mathcal{E}_{\lambda',\boldsymbol{p}'}(\eta)+\mathcal{B}_{\lambda,\boldsymbol{p}}(\eta) \mathcal{B}_{\lambda',\boldsymbol{p}'}(\eta)\Big]
    +h_3(\eta) \frac{k}{\mathcal{H}}\Big[\mathcal{E}_{\lambda,\boldsymbol{p}}(\eta) \mathcal{B}_{\lambda',\boldsymbol{p}'}(\eta)-\mathcal{B}_{\lambda,\boldsymbol{p}}(\eta) \mathcal{E}_{\lambda',\boldsymbol{p}'}(\eta)\Big] \bigg\}\, ,
\end{multline}
\begin{multline}
\label{K3-axion-noBR}
    K_3(\lambda,-\boldsymbol{p};\lambda',\boldsymbol{p}';\eta, \boldsymbol{k}) =
    \big(\boldsymbol{\epsilon}_{\lambda}(\boldsymbol{p})\cdot\boldsymbol{\epsilon}_{\lambda'}(\boldsymbol{p}')\big) \frac{a^2\mathcal{H}^2}{\phi^{\prime 2}_c} \bigg\{\xi\big[1 + h_1(\eta)]\Big[\mathcal{E}^{\ast}_{\lambda,-\boldsymbol{p}}(\eta) \mathcal{B}_{\lambda',\boldsymbol{p}'}(\eta)+\mathcal{B}^{\ast}_{\lambda,-\boldsymbol{p}}(\eta) \mathcal{E}_{\lambda',\boldsymbol{p}'}(\eta)\Big]\\
    +h_2(\eta)\Big[\mathcal{E}^{\ast}_{\lambda,-\boldsymbol{p}}(\eta) \mathcal{E}_{\lambda',\boldsymbol{p}'}(\eta)+\mathcal{B}^{\ast}_{\lambda,-\boldsymbol{p}}(\eta) \mathcal{B}_{\lambda',\boldsymbol{p}'}(\eta)\Big]    
    +h_3(\eta) \frac{k}{\mathcal{H}}\Big[\mathcal{E}^{\ast}_{\lambda,-\boldsymbol{p}}(\eta) \mathcal{B}_{\lambda',\boldsymbol{p}'}(\eta)-\mathcal{B}^{\ast}_{\lambda,-\boldsymbol{p}}(\eta) \mathcal{E}_{\lambda',\boldsymbol{p}'}(\eta)\Big] \bigg\}\, ,
\end{multline}
where $\boldsymbol{p}+\boldsymbol{p}'=\boldsymbol{k}$ and the dimensionless factors $h_i(\eta)$ are
\begin{align}
    h_1(\eta)  &= \frac{2\big(\boldsymbol{p}\cdot\boldsymbol{p}' 
    +\lambda\lambda' p p'\big)}{k^2}-1 =  - \frac{(\lambda p - \lambda' p')^2}{k^2} \vphantom{\Bigg[}\, , \label{h1-axion}\\
    h_2(\eta) &= -\Big(1+\frac{\phi''_c}{\mathcal{H}\phi'_c}\Big) + \frac{\big(\boldsymbol{p}\cdot\boldsymbol{p}' 
    +\lambda\lambda' p p'\big)}{k^2} %\vphantom{\Bigg[}\nonumber\\&\qquad \times    %%COMMENT IN 1-COLUMN
    \Big(1+\frac{\mathcal{H}'}{\mathcal{H}^2}
    +2\frac{\phi''_c}{\mathcal{H}\phi'_c}\Big) \vphantom{\Bigg[}\, , \label{h2-axion}\\
    h_3(\eta) & = \bigg[\frac{a^2\mathcal{H} V'(\phi_c)}{k^2 \phi'_c}\Big(2-\frac{\mathcal{H}'}{\mathcal{H}^2}-\frac{\phi''_c}{\mathcal{H}\phi'_c} + \frac{V''(\phi_c)\phi'_c}{\mathcal{H}V'(\phi_c)}\Big)
    %\vphantom{\Bigg[}\nonumber\\& \qquad    %%COMMENT IN 1-COLUMN
    + \frac{\big(\boldsymbol{p}\cdot\boldsymbol{p}' + \lambda\lambda' p p'\big)}{k^2}\bigg] \frac{\big(\lambda p-\lambda' p'\big)}{k} \vphantom{\Bigg[}\, . \label{h3-axion}
\end{align}
To the leading order in the slow-roll parameters $\epsilon_H$ and $\eta_\phi$, we have 
\begin{align}
\label{h1-axion-SR}
    h_1(\eta) & = -\frac{(\lambda p - \lambda' p')^2}{k^2}\, , \vphantom{\Bigg[}\\
\label{h2-axion-SR}
    h_2(\eta) & \simeq -2\frac{(\lambda p - \lambda' p')^2}{k^2} + \mathcal{O}(\epsilon_H)\, , \vphantom{\Bigg[}\\
\label{h3-axion-SR}
    h_3(\eta) & \simeq \frac{1}{2}\bigg[1-\frac{(\lambda p - \lambda' p')^2}{k^2}\bigg] \frac{\big(\lambda p-\lambda' p'\big)}{k}+\mathcal{O}(\epsilon_H, \eta_\phi)\, . \vphantom{\Bigg[}
\end{align}
Under the assumptions of constant $H$ and $\dot{\phi}_c$, the slow-roll parameters identically vanish and the above-mentioned expressions are exact. Note that, in this case, all three functions $h_i(\eta)$ do not depend on time, and we will omit this argument in what follows. They depend only on polarizations and ratios of the momenta.

Note that, in the first term in the curly brackets in Eqs.~\eqref{K1-axion-noBR} and \eqref{K3-axion-noBR}, we have explicitly separated unity from the function $h_1(\eta)$. This term is the only one remaining in the functions $K_1$ and $K_3$ if metric perturbations are neglected. This approximation was used in all the 
previous studies of scalar perturbations from axion inflation (see, e.g., Refs.~\cite{Barnaby:2011vw,Barnaby:2010vf}). 
In what follows, we use a tilde to denote the quantities computed in the case where the metric perturbations are neglected. Therefore, with the definitions above, we obtain the corresponding expressions for the functions
$\tilde{K}_1$ and $\tilde{K}_3$ simply by setting $h_1=h_2=h_3=0$.

For further convenience, we introduce the dimensionless momenta
\begin{equation}
    \boldsymbol{p}_{\ast} \equiv \frac{1}{q}\boldsymbol{p}\, ,\quad \boldsymbol{p}^{\prime}_{\ast} \equiv \frac{1}{q}\boldsymbol{p}^{\prime}\, , \quad x \equiv \frac{|\boldsymbol{k}|}{q}\, ,
\end{equation}
where $q$ is an overall momentum scale, which we keep arbitrary for now. Using this rescaling together with Eqs.~\eqref{Green-function-axion-approx}, \eqref{K1-axion-noBR} and \eqref{K3-axion-noBR}, the integral over conformal time that occurs in Eq.~\eqref{scalar-power-spectrum} for the scalar power spectrum and in Eq.~\eqref{zeta-bispectrum} for the bispectrum can be presented in the following form:
%
%%ONE-COLUMN MODE
\begin{equation}
    \int\limits_{-\infty}^{\eta} d\eta' G_{\boldsymbol{k}}(\eta,\eta') K_{1}(\lambda,\boldsymbol{p}; \lambda', \boldsymbol{p}'; \eta', \boldsymbol{k}) = -\frac{H^4}{2q^3\dot{\phi}_c^2} \big(\epsilon_{\lambda}(\boldsymbol{p}_{\ast})\cdot \epsilon_{\lambda'}(\boldsymbol{p}'_{\ast})\big) \frac{\sqrt{p_{\ast}p'_{\ast}}}{x^4} \mathcal{I}(\lambda, p_{\ast}; \lambda', p'_{\ast};x)\, . \label{integral-G-K1}
\end{equation}
%
%%TWO-COLUMN MODE
% \begin{multline}
%     \int\limits_{-\infty}^{\eta} d\eta' G_{\boldsymbol{k}}(\eta,\eta') K_{1}(\lambda,\boldsymbol{p}; \lambda', \boldsymbol{p}'; \eta', \boldsymbol{k}) = -\frac{H^4}{2q^3\dot{\phi}_c^2} \\ \times \big(\epsilon_{\lambda}(\boldsymbol{p}_{\ast})\cdot \epsilon_{\lambda'}(\boldsymbol{p}'_{\ast})\big) \frac{\sqrt{p_{\ast}p'_{\ast}}}{x^4} \mathcal{I}(\lambda, p_{\ast}; \lambda', p'_{\ast};x)\, .\label{integral-G-K1}
% \end{multline}
%
Rewriting the integral over $\eta'$  in terms of $y'=-k\eta'$ (then, the prime is omitted for brevity) $\mathcal{I}$ takes the following dimensionless form:
%
%%ONE-COLUMN MODE
\begin{align}
    \mathcal{I}\big(\lambda, p_{\ast}; \lambda', p'_{\ast};x\big) &= \int\limits_{0}^{y_{_\mathrm{UV}}}dy\,(\sin y - y\cos y) \bigg\{(\xi+\xi h_1 + y h_3)\lambda' \mathcal{U}_{\lambda}\Big(\frac{p_{\ast}}{x}y\Big)\mathcal{W}_{\lambda'}\Big(\frac{p'_{\ast}}{x}y\Big) \nonumber\\
    &\hspace*{-1cm} + (\xi+\xi h_1 - y h_3)\lambda \mathcal{W}_{\lambda}\Big(\frac{p_{\ast}}{x}y\Big)\mathcal{U}_{\lambda'}\Big(\frac{p'_{\ast}}{x}y\Big) - h_2 \Big[\mathcal{U}_{\lambda}\Big(\frac{p_{\ast}}{x}y\Big)\mathcal{U}_{\lambda'}\Big(\frac{p'_{\ast}}{x}y\Big)+\lambda\lambda'\mathcal{W}_{\lambda}\Big(\frac{p_{\ast}}{x}y\Big)\mathcal{W}_{\lambda'}\Big(\frac{p'_{\ast}}{x}y\Big)\Big]\bigg\}\, . \label{integral-I}
\end{align}
%
%%TWO-COLUMN MODE
% \begin{align}
%     &\mathcal{I}\big(\lambda, p_{\ast}; \lambda', p'_{\ast};x\big) = \int\limits_{0}^{y_{_\mathrm{UV}}}dy\, (\sin y - y\cos y) \nonumber\\
%     &\times\! \bigg\{(\xi+\xi h_1 + y h_3)\lambda' \mathcal{U}_{\lambda}\Big(\frac{p_{\ast}}{x}y\Big)\mathcal{W}_{\lambda'}\Big(\frac{p'_{\ast}}{x}y\Big) \nonumber\\
%     &\ \, + (\xi+\xi h_1 - y h_3)\lambda \mathcal{W}_{\lambda}\Big(\frac{p_{\ast}}{x}y\Big)\mathcal{U}_{\lambda'}\Big(\frac{p'_{\ast}}{x}y\Big)
% \label{integral-I}\\
%     & \,-\! h_2 \Big[\mathcal{U}_{\lambda}\Big(\frac{p_{\ast}}{x}y\Big)\mathcal{U}_{\lambda'}\Big(\frac{p'_{\ast}}{x}y\Big)\!+\!\lambda\lambda'\mathcal{W}_{\lambda}\Big(\frac{p_{\ast}}{x}y\Big)\mathcal{W}_{\lambda'}\Big(\frac{p'_{\ast}}{x}y\Big)\Big]\bigg\}\, . \nonumber
% \end{align}
%
Here, the factors $h_1$, $h_2$, and $h_3$ depend on the same arguments as the integral $\mathcal{I}$ itself, i.e., on the polarizations and rescaled momenta; we omit these arguments for brevity. The integrals with the functions $K_3$ or $K_1$ have exactly the same form with the only difference being in the mode functions $\mathcal{W}_{\lambda}(p_{\ast}y/x)$ and $\mathcal{U}_{\lambda}(p_{\ast}y/x)$, which have to be replaced by their complex conjugates. As a simple way to denote this fact while keeping the same notation $\mathcal{I}$ for the integral, we add a complex conjugation acting on the corresponding pair of arguments:
%
%%ONE-COLUMN MODE
\begin{equation}
    \int\limits_{-\infty}^{\eta} d\eta' G_{\boldsymbol{k}}(\eta,\eta') K_{3}(\lambda,-\boldsymbol{p}; \lambda', \boldsymbol{p}'; \eta', \boldsymbol{k}) = -\frac{H^4}{2q^3\dot{\phi}_c^2} \big(\epsilon_{\lambda}(\boldsymbol{p}_{\ast})\cdot \epsilon_{\lambda'}(\boldsymbol{p}'_{\ast})\big) \frac{\sqrt{p_{\ast}p'_{\ast}}}{x^4} \mathcal{I}\big((\lambda, p_{\ast})^{\ast}; \lambda', p'_{\ast};x\big)\, .\label{integral-G-K3}
\end{equation}
%
%%TWO-COLUMN MODE
% \begin{multline}
%     \int\limits_{-\infty}^{\eta} d\eta' G_{\boldsymbol{k}}(\eta,\eta') K_{3}(\lambda,-\boldsymbol{p}; \lambda', \boldsymbol{p}'; \eta', \boldsymbol{k}) = -\frac{H^4}{2q^3\dot{\phi}_c^2} \\\times \big(\epsilon_{\lambda}(\boldsymbol{p}_{\ast})\cdot \epsilon_{\lambda'}(\boldsymbol{p}'_{\ast})\big) \frac{\sqrt{p_{\ast}p'_{\ast}}}{x^4} \mathcal{I}\big((\lambda, p_{\ast})^{\ast}; \lambda', p'_{\ast};x\big)\, .\label{integral-G-K3}
% \end{multline}

The integration variable in Eq.~\eqref{integral-I}, $y=-k\eta'$, in principle should run from $-k\eta$ to infinity. However, since we are interested in computing the spectra for modes well outside the Hubble horizon, $k/(aH)\ll 1$, the lower integration boundary in $\mathcal{I}$ was set to zero. Keeping the infinite upper integration boundary leads to an ultraviolet divergence in the integral $\mathcal{I}$. This is the well-known UV divergence that arises in loop computations in quantum field theory (QFT). It could be regularized with one of the conventional methods such as, e.g., dimensional regularization, and then renormalized by introducing the counterterms in the Lagrangian. However, we use an approximation commonly adopted in the cosmology literature, and cut off the integral at a finite $y_{_\mathrm{UV}}$. The reasoning behind this approximation is the following. The gauge-field modes undergo amplification only when their momentum crosses the instability scale that, for axion inflation, is $k_{\mathrm{inst}}=2\xi aH$; see Ref.~\cite{Anber:2009ua}. The modes with higher momenta are not amplified and correspond to vacuum fluctuations of the gauge field. The common prescription is to exclude the latter modes from all observables. However, it is not clear whether one should set the cutoff directly at $k_{\mathrm{inst}}$ or somewhere else.%
\footnote{See, e.g., Refs.~\cite{Anber:2009ua,Durrer:2010mq,Jimenez:2017cdr,Peloso:2022ovc} for the different choices. In Ref.~\cite{Ballardini:2019rqh,Animali:2022lig}, the proper QFT renormalization was applied to the gauge-field energy densities and Chern--Pontryagin density in a simple case of constant $H$ and $\xi$. For large values of $\xi$, the result is in good agreement with the naive cutoff at the instability scale. See also Ref.~\cite{Seery:2008ms}, where the QFT renormalization was applied in the study of scalar perturbations in a simple case of the kinetic-coupling model in de Sitter space. However, application of remormalization techniques to the full time-dependent cases is a very challenging task.}
We, therefore, require that the momenta of all mode functions inside the integral in Eq.~\eqref{integral-I} are smaller than the cutoff scale of the same order as the instability threshold. This gives the following expression for the upper integration limit:
\begin{equation}
    y_{_\mathrm{UV}} = \gamma_{_\mathrm{UV}}\, \operatorname{min}\Big\{\frac{x}{p_{\ast}},\frac{x}{p'_{\ast}} \Big\}\, ,
\end{equation}
where we introduced an additional parameter $\gamma_{_\mathrm{UV}}$ that accounts for the possible theoretical uncertainties in the choice of the cutoff. Its reference value is $\gamma_{_\mathrm{UV}}=2\xi$, and we study the impact of its variation on the numerical results for the scalar spectrum and bispectrum in the next subsections.

%%%%%%%%%%%%%%%%%%%%%%%%%%%%%%%%%%%%%%%%%%%%%%%%%%%%%%%%%
\subsection{Numerical results}
\label{subsec:numerical-results}
%%%%%%%%%%%%%%%%%%%%%%%%%%%%%%%%%%%%%%%%%%%%%%%%%%%%%%%%%

\subsubsection{Scalar power spectrum}

We now compute the scalar power spectrum
for inflation  using Eq.~\eqref{scalar-power-spectrum}. The second term in this equation, which describes the contribution induced by the presence of gauge fields, i.e., the inverse-decay contribution, can be parametrized in the following form:
\begin{equation}
\label{f2-definition}
    P_{\zeta}^{(\text{inv.\ dec.})}(k) = e^{4\pi\xi} \Big(\frac{H^2}{2\pi |\dot{\phi}_c|}\Big)^{4} f_2(\xi)\, ,
\end{equation}
where $f_2(\xi)$ is a dimensionless function introduced in Refs.~\cite{Barnaby:2010vf,Barnaby:2011vw}. Using Eq.~\eqref{integral-G-K1} and choosing $k$ (the only momentum scale) to be the overall momentum scale $q$, we find the following expression for the function $f_2(\xi)$:
%
%%ONE-COLUMN MODE
\begin{equation}
\label{f2-general}
    f_2(\xi) = \frac{e^{-4\pi\xi}}{8\pi}\int d^3\boldsymbol{p}_{\ast}\sum\limits_{\lambda,\lambda'=\pm}  \Big(1-\lambda\lambda'\frac{\boldsymbol{p}_{\ast}\cdot\boldsymbol{p}'_{\ast}}{p_{\ast}p'_{\ast}}\Big)^2 p_{\ast}p'_{\ast} \big|\mathcal{I}\big(\lambda,p_{\ast};\lambda',p'_{\ast};1\big)\big|^2\, ,
\end{equation}
%
%%TWO-COLUMN MODE
% \begin{multline}
% \label{f2-general}
%     f_2(\xi) = \frac{e^{-4\pi\xi}}{8\pi}\int d^3\boldsymbol{p}_{\ast}\sum\limits_{\lambda,\lambda'=\pm}  \Big(1-\lambda\lambda'\frac{\boldsymbol{p}_{\ast}\cdot\boldsymbol{p}'_{\ast}}{p_{\ast}p'_{\ast}}\Big)^2 p_{\ast}p'_{\ast}
%     \\ \times \big|\mathcal{I}\big(\lambda,p_{\ast};\lambda',p'_{\ast};1\big)\big|^2\, ,
% \end{multline}
%
where $\boldsymbol{p}'_{\ast}=\hat{\boldsymbol{k}}-\boldsymbol{p}_{\ast}$,  $\hat{\boldsymbol{k}}=\boldsymbol{k}/k$, the integral $\mathcal{I}$ is given by Eq.~\eqref{integral-I}, and we used Eq.~\eqref{eps-scalar-product-particular} for the scalar product of the polarization vectors. Note that the power spectrum in Eq.~\eqref{f2-definition} is independent of $k$. This is a consequence of the fact that we are considering a de Sitter universe with $H=\text{const}$ and that we also set $\xi=\text{const}$. This implies that any scalar-perturbation mode appears to be in exactly the same condition as any other. The same result was also obtained in the seminal work~\cite{Barnaby:2011vw}. However, in any realistic inflationary model with a time-dependent background, the power spectrum will depend on $k$, which is captured by the general expression \eqref{scalar-power-spectrum}.

\begin{figure}
\centering
\includegraphics[width=0.95\textwidth]{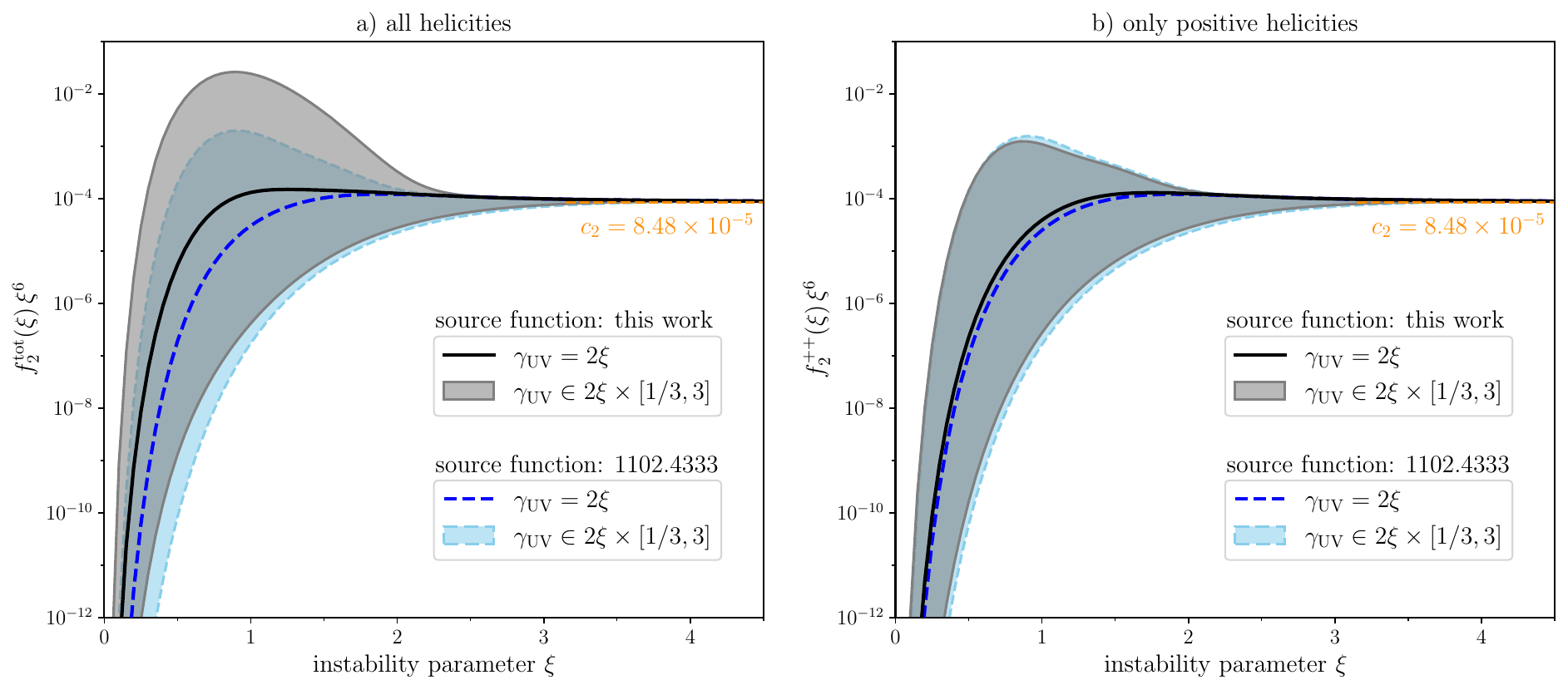}
\caption{The function $f_2(\xi)$ parametrizing the gauge-field-induced part of the scalar power spectrum, rescaled by a factor $\xi^{6}$, including contributions (a)~from both circular polarizations of the gauge field, (b)~only from the positive circular polarization, which is exponentially enhanced with respect to the negative one. The black solid lines and gray shaded regions represent the results of the full computation, which takes into account the metric perturbations, while the blue dashed line and light-blue shaded region correspond to the function $\tilde{f}_2(\xi)$ in the case when the metric perturbations are neglected. The thick black and blue lines correspond to the UV cutoff in the integral~\eqref{integral-I} at the instability scale, $\gamma_{_{\mathrm{UV}}}=2\xi$, while the shaded bands of the corresponding color show the variation of the final result under an increase or a decrease of the UV cutoff by a factor of 3. The orange dashed lines starting around $\xi=3$ show the leading term in the expansion of $f_2(\xi)$ in the large-$\xi$ approximation, see Eq.~\eqref{f2-large-xi-asymptote} and Appendix~\ref{app:large-xi}. \label{fig:f2}}
\end{figure}

\begin{figure}
    \centering
    \includegraphics[width=0.95\textwidth]{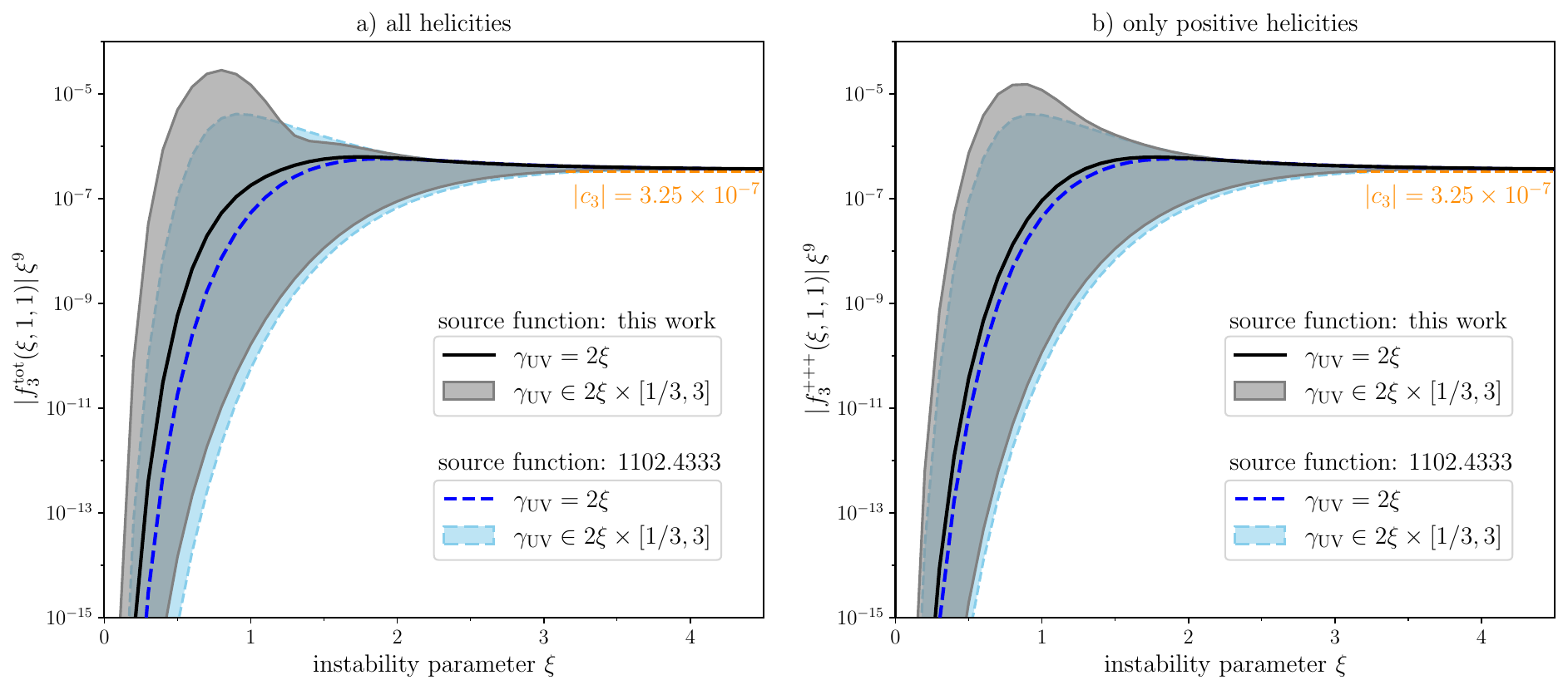}
    \caption{The function $f_3(\xi, 1, 1)$ parametrizing the equilateral scalar bispectrum rescaled by a factor $\xi^9$, including contributions (a)~from both circular polarizations of the gauge field, (b)~only from the positive circular polarization, which is exponentially enhanced with respect to  the negative one. The black solid lines and gray shaded regions represent the results of the full computation, which takes into account the metric perturbations, while the blue dashed line and light-blue shaded region correspond to the function $\tilde{f}_3(\xi,1,1)$ in the case when the metric perturbations are neglected. The thick black and blue lines correspond to the UV cutoff in the integral~\eqref{integral-I} at the instability scale, $\gamma_{_{\mathrm{UV}}}=2\xi$, while the shaded bands of the corresponding color show the variation of the final result under an increase or a decrease of the UV cutoff by a factor of 3. The orange dashed lines show the leading term in the expansion of $f_3(\xi,1,1)$ in the large-$\xi$ approximation [see Eq.~\eqref{f3-large-xi-asymptote} and Appendix~\ref{app:large-xi}]. The coefficient $c_3$ on the plot denotes $c_3(1,1)$ in Eq.~\eqref{f3-large-xi-asymptote}.\label{fig:f3}}
\end{figure}

To get some intuition about the analytic form of $f_2(\xi)$ and to compare with the well-known results in the literature (e.g., Ref.~\cite{Barnaby:2011vw}), we first compute this function in the large-$\xi$ approximation in Appendix~\ref{app:large-xi}. We call it an approximation because it is based on using the approximate expressions \eqref{W-funct-approx} and\eqref{U-funct-approx} for the gauge-field mode functions. These exponential functions are rather close to the exact solutions in Eqs.~\eqref{sol-Whittaker}--\eqref{sol-Coulomb-U} but still never reproduce them exactly for arbitrarily large $\xi$. Moreover, only one circular polarization of the gauge field (the one that is amplified due to the axial coupling to the inflaton) is taken into account. Nevertheless, this approximation allows us to compute the integral in Eq.~\eqref{f2-general} analytically in the form of a series in inverse powers of $\xi$. The leading term in this expansion behaves as
\begin{equation}
\label{f2-large-xi-asymptote}
    f_2(\xi)\simeq \frac{c_2}{\xi^{6}}
\end{equation}
with the coefficient $c_2\approx \num{8.48e-5}$ (see Appendix~\ref{app:large-xi} for more details). Note that this value of $c_2$ is about \qty{12}{\percent} larger than $\tilde{c}_2 \approx \num{7.47e-5}$ reported in Ref.~\cite{Barnaby:2011vw}, which has been calculated neglecting the metric perturbations.

We also compute the function $f_2(\xi)$ numerically for different values of $\xi$ and plot it rescaled by the factor of $\xi^6$ in Fig.~\ref{fig:f2}. The rescaling compensates for the decrease of the function $f_2(\xi)$ in the region of large $\xi$. The black solid lines and the gray shaded regions around them represent the results of the computations that self-consistently include the metric perturbations, while the blue dashed lines and the light-blue shaded regions show the results of the computations performed with the source term of Ref.~\cite{Barnaby:2011vw}, where the metric perturbations are neglected. Both source terms are evaluated using the exact mode functions reported in Eqs.~\eqref{sol-Whittaker}--\eqref{sol-Coulomb-U}. The shaded bands show the variation of the corresponding results under a change of the UV cutoff in the integral in Eq.~\eqref{integral-I}: the thick line in the middle corresponds to the reference value $\gamma_{_\mathrm{UV}}=2\xi$, which is the cutoff at the largest mode  experiencing the tachyonic instability; the upper (lower) boundary of the shaded region corresponds to $\gamma_{_\mathrm{UV}}$ increased (decreased) by a factor of 3. Figure~\ref{fig:f2}(a) shows the result including both circular polarizations of the gauge field, while \ref{fig:f2}(b) shows the contribution of the positive polarization only. The orange dashed lines correspond to the leading term in the expansion over the inverse powers of $\xi$ derived in the large-$\xi$ approximation in Appendix~\ref{app:large-xi}. Note that the number $c_2=\num{8.48e-5}$ does not exactly coincide with the true asymptotic behavior of $\xi^6 f_2(\xi)$ computed using the exact mode functions. But still, numerically, it turns out to be pretty close.

\subsubsection{Scalar bispectrum}

Let us now focus on the scalar bispectrum generated in the axion inflation model. Following Ref.~\cite{Barnaby:2011vw}, we introduce the function $f_3(\xi,\, x_2,\, x_3)$ defined by
%
%%ONE-COLUMN MODE
\begin{equation}
    \mathscr{B}_{\zeta}(k_1, k_2, k_3) = \frac{3}{10}(2\pi)^{5/2}
    \Big(\frac{H^2}{2\pi |\dot{\phi}_c|}\Big)^{\!6}\, \frac{e^{6\pi\xi}}{k^6} \frac{1+x_2^3 + x_3^3}{x_2^3 x_3^3} f_3(\xi,\, x_2,\, x_3)\, .
\end{equation}
%
%%TWO-COLUMN MODE
% \begin{align}
%     \mathscr{B}_{\zeta}(k_1, k_2, k_3) &=
%     \frac{3}{10}(2\pi)^{5/2}
%     \Big(\frac{H^2}{2\pi |\dot{\phi}_c|}\Big)^{6} \vphantom{\Biggl[}\nonumber \\ \vphantom{\Biggl[}
%     &\times \frac{e^{6\pi\xi}}{k^6} \frac{1+x_2^3 + x_3^3}{x_2^3 x_3^3}f_3(\xi,\, x_2,\, x_3)\, .
% \end{align}
%
Here, $k=k_1$ is the largest of the three momenta $k_1$, $k_2$, and $k_3$; the variables $x_2=k_2/k$ and $x_3=k_3/k$ satisfy the conditions detailed in Eq.~\eqref{limits-shape}.

Using Eq.~\eqref{zeta-bispectrum} for the bispectrum, and substituting  Eqs.~\eqref{integral-G-K1} and \eqref{integral-G-K3}, we arrive at the following expression for the function $f_3$:
%
%%ONE-COLUMN MODE
\begin{align}
    f_3(\xi, x_2, x_3) &= -\frac{5}{3\pi} \frac{e^{-6\pi\xi}}{x_2 x_3 (1+x_2^3 +x_3^3)} \int d^3\boldsymbol{p}_{\ast} \sum\limits_{\lambda_1,\lambda_2,\lambda_3 = \pm} p_{\ast} |\boldsymbol{p}_{\ast}-\hat{\boldsymbol{k}}_1| |\hat{\boldsymbol{k}}_3 x_3 +\boldsymbol{p}_{\ast}|\nonumber\\
    &\hspace{-16mm}\times \big(\boldsymbol{\epsilon}_{\lambda_1}(\boldsymbol{p}_{\ast})\cdot \boldsymbol{\epsilon}_{\lambda_2}(\hat{\boldsymbol{k}}_1-\boldsymbol{p}_{\ast})\big) \big(\boldsymbol{\epsilon}_{\lambda_2}(\boldsymbol{p}_{\ast}-\hat{\boldsymbol{k}}_1)\cdot \boldsymbol{\epsilon}_{\lambda_3}(-\hat{\boldsymbol{k}}_3 x_3 -\boldsymbol{p}_{\ast})\big) \big(\boldsymbol{\epsilon}_{\lambda_3}(\hat{\boldsymbol{k}}_3 x_3 +\boldsymbol{p}_{\ast})\cdot \boldsymbol{\epsilon}_{\lambda_1}(-\boldsymbol{p}_{\ast})\big) \nonumber\\
    &\hspace{-16mm}\times \mathcal{I}\big( \lambda_1, p_{\ast}; \lambda_2, |\boldsymbol{p}_{\ast}-\hat{\boldsymbol{k}}_1|; 1\big) \mathcal{I}\big((\lambda_2, |\boldsymbol{p}_{\ast}-\hat{\boldsymbol{k}}_1|)^{\ast}; \lambda_3, |\hat{\boldsymbol{k}}_3 x_3 +\boldsymbol{p}_{\ast}|; x_2\big) \mathcal{I}\big((\lambda_3, |\hat{\boldsymbol{k}}_3 x_3 +\boldsymbol{p}_{\ast}|)^{\ast}; (\lambda_1, p_{\ast})^{\ast}; x_3\big)\, , \label{f3-general}
\end{align}
%
%%TWo-COLUMN MODE
% \begin{align}
%     &f_3(\xi, x_2, x_3) = -\frac{5}{3\pi} \frac{e^{-6\pi\xi}}{x_2 x_3 (1+x_2^3 +x_3^3)}\! \int\!\! d^3\boldsymbol{p}_{\ast}\! \sum\limits_{\{\lambda\}}p_{\ast}\nonumber\\
%     &\ \ \times |\boldsymbol{p}_{\ast}-\hat{\boldsymbol{k}}_1| |\hat{\boldsymbol{k}}_3 x_3 +\boldsymbol{p}_{\ast}| \big(\boldsymbol{\epsilon}_{\lambda_1}(\boldsymbol{p}_{\ast})\cdot \boldsymbol{\epsilon}_{\lambda_2}(\hat{\boldsymbol{k}}_1-\boldsymbol{p}_{\ast})\big)\nonumber\\
%     &\ \ \times \big(\boldsymbol{\epsilon}_{\lambda_2}(\boldsymbol{p}_{\ast}-\hat{\boldsymbol{k}}_1)\cdot \boldsymbol{\epsilon}_{\lambda_3}(-\hat{\boldsymbol{k}}_3 x_3 -\boldsymbol{p}_{\ast})\big)\nonumber\\
%     &\ \ \times
%     \big(\boldsymbol{\epsilon}_{\lambda_3}(\hat{\boldsymbol{k}}_3 x_3 +\boldsymbol{p}_{\ast})\cdot \boldsymbol{\epsilon}_{\lambda_1}(-\boldsymbol{p}_{\ast})\big) \nonumber\\
%     &\ \ \times \mathcal{I}\big( \lambda_1, p_{\ast}; \lambda_2, |\boldsymbol{p}_{\ast}-\hat{\boldsymbol{k}}_1|; 1\big)\nonumber\\
%     &\ \ \times \mathcal{I}\big((\lambda_2, |\boldsymbol{p}_{\ast}-\hat{\boldsymbol{k}}_1|)^{\ast}; \lambda_3, |\hat{\boldsymbol{k}}_3 x_3 +\boldsymbol{p}_{\ast}|; x_2\big)\nonumber\\
%     &\ \ \times \mathcal{I}\big((\lambda_3, |\hat{\boldsymbol{k}}_3 x_3 +\boldsymbol{p}_{\ast}|)^{\ast}; (\lambda_1, p_{\ast})^{\ast}; x_3\big)\, , \label{f3-general}
% \end{align}
%
where $\hat{\boldsymbol{k}}_i=\boldsymbol{k}_i/k_i$. We also used the relation \eqref{rel-K1-K4} to express the function $K_4$ through $K_1$.

We first compute this function for the equilateral configuration, $x_2=x_3=1$. In the large-$\xi$ approximation, it behaves as
\begin{equation}
\label{f3-large-xi-asymptote}
    f_3(\xi,1,1)\simeq \frac{c_3(1,1)}{\xi^9}
\end{equation}
with $c_3(1,1)\approx \num{-3.25e-7}$; for more details, see Appendix~\ref{app:large-xi}. As for the power spectrum, this number is about \qty{15}{\percent} larger in absolute value than $\tilde{c}_3(1,1)\approx \num{-2.77e-7}$ reported in Ref.~\cite{Barnaby:2011vw}, where the metric perturbations are neglected. Note the difference of the overall sign of the bispectrum compared to Refs.~\cite{Barnaby:2010vf,Barnaby:2011vw} due to a different sign in the definition of the scalar perturbation variable $\zeta$; see also footnote~\ref{footnote-sign} on page~\pageref{footnote-sign}.

Further, we perform the numerical computations of the function $f_3(\xi,1,1)$ using the exact gauge-field mode functions. The results are shown in Fig.~\ref{fig:f3}, where we rescaled $f_3(\xi,1,1)$ by a factor of $\xi^9$ to compensate for its decrease in the limit of large $\xi$ values. The color scheme and notations are exactly the same as in Fig.~\ref{fig:f2}.

\begin{figure}
    \centering
    \includegraphics[width=0.95\textwidth]{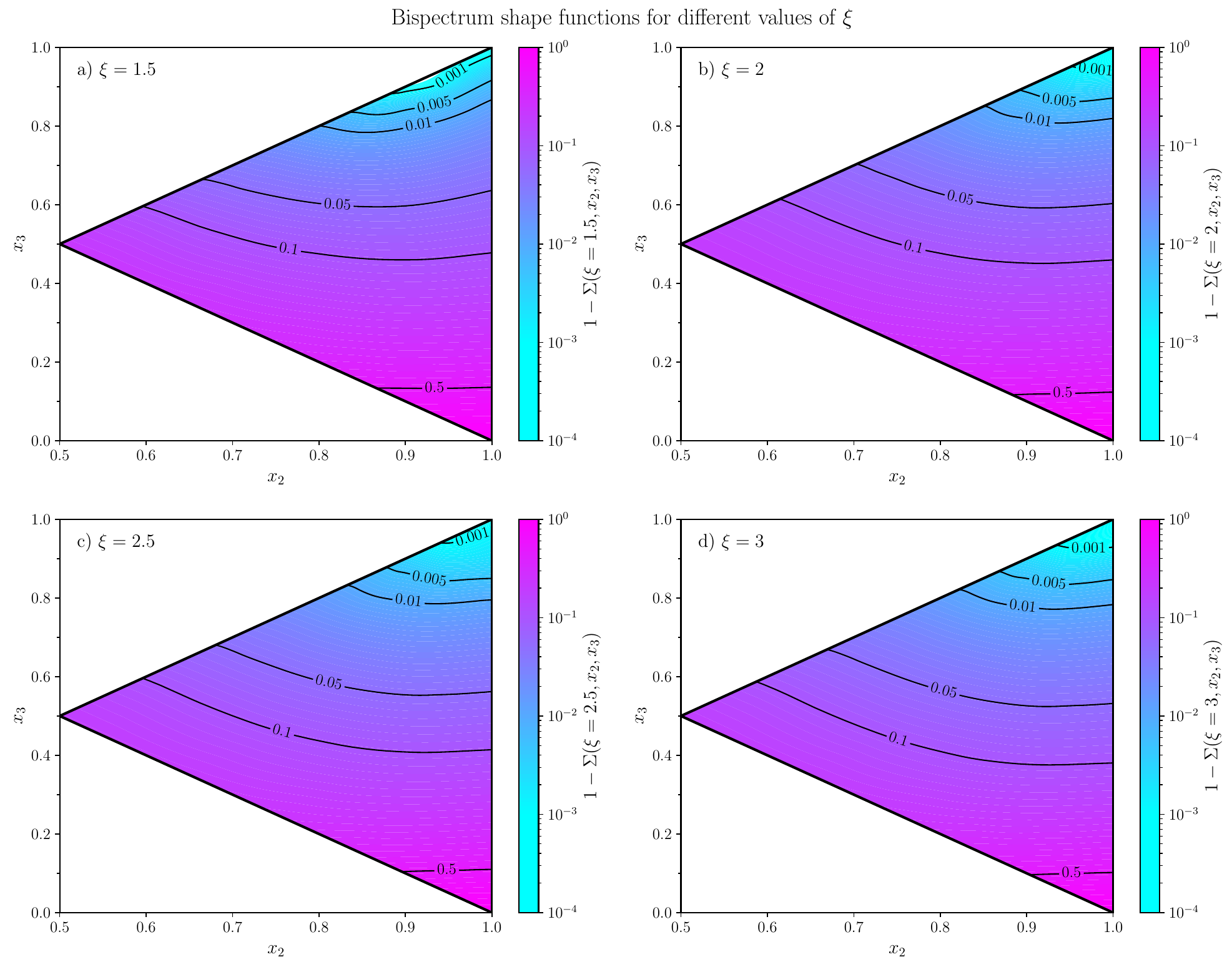}
    \caption{The bispectrum shape function $\Sigma(\xi, x_2, x_3)$ in the part of the $(x_2, x_3)$ plane limited by the inequalities \eqref{limits-shape} for four fixed values of $\xi$: (a)~$\xi=1.5$, (b)~$\xi=2$, (c)~$\xi=2.5$, and (d)~$\xi=3$. Since in all cases the shape function exhibits a broad maximum at the equilateral configuration, the color scheme and contour labels correspond to the deviation of the shape function from unity, $1-\Sigma(\xi, x_2, x_3)$ in a logarithmic scale. \label{fig:shapes}}
\end{figure}

\begin{table}
\renewcommand{\arraystretch}{1.4}
\centering
    \begin{tabular}{c|c|c|c}
     \hline
     \hline
     \multirow{2}{1.5cm}{\centering $\xi$} & \multicolumn{3}{c}{$\cos\Theta$ in Eq.~\eqref{cosine-angle-shapes} with respect to patterns:} \\
     \cline{2-4}
      & \multirow{1}{3cm}{\centering Local \eqref{pattern-local}} & \multirow{1}{3cm}{\centering Equilateral \eqref{pattern-equil}} & \multirow{1}{3cm}{\centering Orthogonal \eqref{pattern-ortho}} \\
     \hline
     \hline
     $1.5$ & $0.47$ & $0.97$ & $-0.10$ \\
     $2.0$ & $0.47$ & $0.97$ & $-0.10$ \\
     $2.5$ & $0.47$ & $0.94$ & $-0.10$ \\
     $3.0$ & $0.49$ & $0.94$ & $-0.10$ \\
     Large $\xi$ & $0.51$ & $0.92$ & $-0.18$\\
     \hline
     \hline
    \end{tabular}
\caption{Cosines of the angles \eqref{cosine-angle-shapes} between the shape functions $\Sigma(\xi,x_2,x_3)$ and the standard (local, equilateral, and orthogonal) patterns computed for four values of $\xi=1.5$, $2$, $2.5$, $3$ and in the large-$\xi$ approximation (see Appendix~\ref{app:large-xi}). Note that, for the local case, $(\Sigma_{\mathrm{local}}\, ,\Sigma_{\mathrm{local}})$ diverges and $\cos\Theta\to 0$ for $k_{\min}\to 0$. The value given in the table corresponds to the choice described in footnote~\ref{foot5}. \label{tab:cosines}
}
\end{table}

The relative value of the bispectrum for arbitrary configurations of momenta compared to its value for the equilateral configuration is characterized by the shape function $\Sigma$ given by Eq.~\eqref{shape-function-general}. It can be expressed in terms of the function $f_3$ as follows:
\begin{equation}
    \Sigma(\xi,x_2,x_3) = \frac{1 + x_2^3 + x_3^3}{3x_2 x_3}\frac{f_3(\xi,x_2,x_3)}{f_3(\xi, 1, 1)}\, .
\end{equation}
We plot the shape functions for fixed values of $\xi=1.5$, $2$, $2.5$, and $3$ in the contour and density plots in Fig.~\ref{fig:shapes}. The shape functions exhibit a broad peak around the equilateral configuration; therefore, the numbers on the contours and the color scheme of the density plots correspond to the deviations of the shape functions from unity. The shape function is smallest in the local configuration, $x_3\to 0$. In these calculations, we used the reference value $\gamma_{_\mathrm{UV}}=2\xi$ for the UV cutoff. The shape function in the large-$\xi$ approximation is shown in Fig.~\ref{fig:shapes-large-xi} in Appendix~\ref{app:large-xi}.

Finally, to see how similar the actual shape of non-Gaussianity is to the standard patterns \eqref{pattern-local}--\eqref{pattern-ortho}, we compute the cosines of the angles between the obtained shape functions and the standard patterns using Eq.~\eqref{cosine-angle-shapes}. They are listed in Table~\ref{tab:cosines}.%
\footnote{Note that the norm of the local shape function \eqref{pattern-local} with respect to the scalar product in Eq.~\eqref{scalar-product-shapes} is logarithmically divergent. This divergence was regularized by cutting the integral over $x_3$ at $x_{3,\mathrm{min}}=k_{\mathrm{min}}/k_{\mathrm{max}}=1/1000$ as it is usually done in the literature; see, e.g., Ref.~\cite{Fergusson:2008ra}. \label{foot5}}
The numbers show that the shape appears to be rather close to the equilateral pattern with the corresponding $\cos\Theta$ varying in the range from 0.92 in the large-$\xi$ approximation to 0.97 for small values of $\xi$.

%%%%%%%%%%%%%%%%%%%%%%%%%%%%%%%%%%%%%%%%%%%%%%%%%%%%%%%%%
\subsection{Discussion}
\label{subsec:discussion}
%%%%%%%%%%%%%%%%%%%%%%%%%%%%%%%%%%%%%%%%%%%%%%%%%%%%%%%%%

Even though in Ref.~\cite{Barnaby:2011vw}, only inflaton perturbations and the coupling of the inflaton to the gauge field were considered, and gravitational contributions to $\zeta$ were neglected, the final result agrees reasonably well with our more comprehensive treatment. We observe especially good agreement for $\xi\gtrsim 2$, where the black solid and blue dashed lines in Figs.~\ref{fig:f2} and \ref{fig:f3} coincide almost perfectly (note that this accuracy is much better than the \qty{12}{\percent}--\qty{15}{\percent} disagreement in the large-$\xi$ approximation mentioned above). This remarkable agreement can be explained as follows. For moderate and large values of $\xi$, the dominant contribution to the functions $f_2$ and $f_3$, as well as to any other quantity of interest in this paper, stems from gauge-field modes with positive circular polarization; compare panels (a) and (b) in Figs.~\ref{fig:f2} and \ref{fig:f3}. The corresponding mode functions are highly peaked and take the largest values for $\rho\lesssim 1/(8\xi)$; see Fig.~\ref{fig:mode-functions} in Appendix~\ref{app:large-xi}. Therefore, the maximal overlap of these peaked functions in the integral~\eqref{integral-I} occurs for the case $p_{\ast}\simeq p'_{\ast}$ and quickly decreases when $p_{\ast}$ and $p'_{\ast}$ become different. However, all differences between the new source term derived in the present work and the old one in Ref.~\cite{Barnaby:2011vw} are proportional to the functions $h_1$, $h_2$, and $h_3$, which, in the slow-roll approximation, are given by Eqs.~\eqref{h1-axion-SR}--\eqref{h3-axion-SR}. Note that all three of them are proportional to $(\lambda p_{\ast} - \lambda' p'_{\ast})=(p_{\ast}-p'_{\ast})$ or the same combination squared and, thus, vanish in the region $p_{\ast}\simeq p'_{\ast}$, which is required for the maximal overlap of the mode functions.

Note that the above-mentioned agreement is obtained in the simple case of axion inflation without backreaction where the slow-roll regime is realized. We expect that in other, physically more realistic and interesting cases, the effect of a self-consistent treatment of the metric perturbation may be important and may lead to $\mathcal{O}(1)$ corrections to the power spectrum and bispectrum. Especially interesting in this regard is the regime of strong gauge-field backreaction, whose background dynamics was studied, e.g., in Refs.~\cite{Domcke:2020zez,Gorbar:2021rlt,vonEckardstein:2023gwk,Durrer:2023rhc}. In this regime, at least two of the above-mentioned arguments are modified. Indeed, in the strong-backreaction regime, the inflaton and the gauge-field background quantities exhibit oscillations during which the slow-roll conditions are violated, e.g., $\ddot{\phi}_c$ is as important as other terms in the Klein--Gordon equation. This means that the functions $h_1$, $h_2$, and $h_3$ will have other contributions that do not vanish for $p_{\ast}\simeq p'_{\ast}$. Also, because of the oscillations, the peaks in the gauge-field mode functions may become broader, which will allow for a non-negligible overlap in the integral \eqref{integral-I} even for $p_{\ast}\neq p'_{\ast}$. Furthermore, even the left-hand side of the differential equation for $\zeta$ will be strongly modified by the backreaction (modified in a different way than in the absence of metric perturbations). This will also lead to modifications in the vacuum contribution to the scalar power spectrum and consequently to a different expression for the Green function \eqref{Green-function-zeta12} that enters the power spectrum and bispectrum. The numerical study of perturbations in the strong-backreaction regime is much more involved compared to the case considered in the present article. We  plan to address this problem in a future work.

The impact of the metric perturbations appears to be much more important for small values of $\xi$. Physically, this corresponds to early times during slow roll, when the cosmologically relevant scales exit the horizon. Here, one typically expects $\dot\phi_c/H$ to be very small so that $\xi$ may also be less than unity. In this case, the gauge-field amplification due to axial coupling is rather weak and the mode function is not as highly peaked as in the case of large $\xi$; see Fig.~\ref{fig:mode-functions} in Appendix~\ref{app:large-xi}. This means that the configurations of momenta with $p_{\ast}\neq p'_{\ast}$ also make a significant contribution to the integral \eqref{integral-I} and the new terms in the source proportional to the functions $h_1$, $h_2$, and $h_3$ become as important as the old source term. Note, however, that, in the region of small $\xi$, the numerical results presented in Figs.~\ref{fig:f2} and \ref{fig:f3} are subject to considerable theoretical uncertainties for at least two different reasons. The first one is related to the UV cutoff in the integral $\mathcal{I}$ to which the functions $f_2$ and $f_3$ appear to be very sensitive. Although our reference value $\gamma_{_{\mathrm{UV}}}=2\xi$ is well motivated, the fact that even a mild change of this parameter within 1 order of magnitude leads to a large change in the numerical results spanning over 4--6 orders of magnitude points toward the need for a more rigorous and accurate treatment of the UV divergence present in the problem. The second reason is related to the question of whether one has to take into account the contribution coming from the gauge-field mode functions of negative helicity, which appears to be comparable to that of the positive helicity in the region of small $\xi$. If we include it, it is unclear what the UV cutoff for these modes should be since they do not experience the tachyonic instability at all. In the present work, for definiteness, we took it to be the same as for the positive-helicity modes. Both issues can be resolved by employing a systematic QFT approach including conventional regularization and renormalization techniques. To the best of our knowledge, such an approach was only applied to the problem of gauge-field production during axion inflation in the absence of backreaction in Refs.~\cite{Ballardini:2019rqh,Animali:2022lig}. Undoubtedly, application of the proper renormalization techniques to the computation of scalar power spectra is very important for comparison with observations~\cite{Pla:2024renorm}. Talking about the gauge-field-induced spectra, the first step in this direction was recently made in Ref.~\cite{Galanti:2024jhw} where the case of constant $H$ and $\xi$ was considered. Application of renormalization methods to the fully time-dependent case, possibly including the gauge-field backreaction, is still missing in the literature and requires a separate, thorough investigation. We are planning to go back to this problem in a future work.

%%%%%%%%%%%%%%%%%%%%%%%%%%%%%%%%%%%%%%%%%%%%%%%%%%%%%%%%%
%%%%%%%%%%%%%%%%%%%%%%%%%%%%%%%%%%%%%%%%%%%%%%%%%%%%%%%%%
\section{Conclusion and outlook}
\label{sec:conclusion}
%%%%%%%%%%%%%%%%%%%%%%%%%%%%%%%%%%%%%%%%%%%%%%%%%%%%%%%%%
%%%%%%%%%%%%%%%%%%%%%%%%%%%%%%%%%%%%%%%%%%%%%%%%%%%%%%%%%

The production of gauge fields during inflation leads to new sources of cosmological perturbations. The slow roll of the inflaton amplifies the gauge fields, which in turn produce scalar fluctuations.
These additional cosmological perturbations are statistically independent from the vacuum perturbations of the inflaton and can be highly non-Gaussian. Hence, for any scenario of inflation where gauge fields  are generated, it is very important to keep track of the effect of these new source terms on the generation of scalar perturbations. Potentially, they can yield an additional contribution of perturbations that are both, non-Gaussian and, in general, not scale invariant. These properties are observable and can be used to constrain inflationary models that involve the production of gauge fields.

To the best of our knowledge, the majority of the current literature devoted to the investigation of this problem is based on the assumption that all relevant effects arise from the coupling between the inflaton and the gauge fields, while scalar metric perturbations are neglected (i.e., working with the unperturbed FLRW metric).

In the present work, we consider a generic model including the kinetic and axial coupling between the inflaton and an Abelian gauge field and self-consistently take into account all scalar perturbations existing in the problem. Indeed, in the presence of arbitrary sources, the curvature perturbation variable is given by Eq.~\eqref{zeta-def-2} and contains inflaton, metric, and  source perturbations. In the absence of sources, this variable describes the curvature of the spatial sections in unitary gauge.
We derived a second-order differential equation for the variable $\zeta$, Eq.~\eqref{zeta-prime-prime-2}, which is sourced by gauge-field perturbations. Standard inflationary predictions are obtained by simply setting the right-hand side of this equation to zero.

Equation~\eqref{zeta-prime-prime-2} for the variable $\zeta$ is an inhomogeneous linear differential equation; therefore, its general solution consists of two parts. The first part represents the general solution of the corresponding homogeneous equation and describes the vacuum scalar perturbations arising on a homogeneous background, which includes the inflaton as well as the gauge-field zero modes. 
The second part is induced by the source terms and corresponds to the scalar perturbations sourced by inhomogeneities in the gauge field.
Note also that the sources $F_\alpha$ themselves evolve according to equations of motion that include $\zeta$. Hence, one may try to solve this coupled system of equations treating all perturbations on the same footing as, e.g., it is done with the adiabatic and the entropy modes in the multifield inflationary models; see, e.g., Refs.~\cite{Nakamura:1996da,Gordon:2000hv,Gong:2011uw}.
However, since the gauge fields are already produced by the background inflaton motion, one may expect that the impact of scalar perturbations on the gauge-field evolution is small (provided that the former are small compared to the corresponding background quantities). 
In this work, we assume this and, as a first step, we neglect the impact of the inflaton and metric perturbations on the gauge-field mode functions. This allows us to derive closed expressions for the induced contributions to the scalar power spectrum and bispectrum
given in Eqs.~\eqref{scalar-power-spectrum} and \eqref{zeta-bispectrum}, respectively.

Although the main goal of this paper is to develop the formalism and to derive all equations needed to compute the scalar perturbations during inflation in the presence of the gauge field, in Sec.~\ref{sec:application}, we apply our formalism to the well-known case of axion inflation neglecting backreaction. This case, which has been previously studied extensively, allows a comparison with  the literature. 
For simplicity, we assume that the Hubble parameter $H$ as well as the gauge-field production parameter $\xi$ are both constant during inflation. This assumption allows us to solve the equations for the gauge-field mode functions and the Green function of the curvature perturbation analytically. We then use these to compute the induced scalar power spectrum and bispectrum. Comparing to Ref.~\cite{Barnaby:2011vw}, where metric perturbations are neglected, the source term in the equation for $\zeta$ \eqref{zeta-prime-prime-2} contains additional contributions. 
However, in the region of $\xi\gg 3$, the impact of extra terms is small and the power spectrum as well as the bispectrum (computed on the exact mode functions) arising from the sources in Ref.~\cite{Barnaby:2011vw} and in the present work differ little; see Figs.~\ref{fig:f2} and \ref{fig:f3}. Nevertheless, we argue that this result strongly depends on the gauge-field mode functions. Indeed, if we use the approximate exponential form for the gauge-field mode function, we obtain a $\sim 10\,\%$ deviation between our results and the ones of Ref.~\cite{Barnaby:2011vw}. 
A larger disagreement is observed in the region of small $\xi\lesssim 2$. These deviations lead us to the conclusion that there is no \textit{fundamental} reason for the impact of metric perturbations to be small, providing the main motivation for the revision of the formalism performed in this work.

A more interesting case, which potentially has many phenomenological applications, is the regime of strong backreaction of the gauge field on the background evolution. It is well known that the inflaton field and the gauge-field energy density both exhibit oscillatory behavior, which is due to the delay between the moment of time when the gauge-field modes are enhanced and the moment of time they start to backreact on the expansion of the  universe~\cite{Domcke:2018eki,Gorbar:2021rlt,Durrer:2023rhc,vonEckardstein:2023gwk}.
It was shown that these oscillations lead to oscillatory features in the tensor power spectrum~\cite{Garcia-Bellido:2023ser}, which are potentially observable by space interferometers such as LISA and by pulsar timing arrays.
Similar features are also expected to appear in the scalar power spectrum. Moreover, the periodic decrease of the inflaton velocity (sometimes even crossing zero) in the backreaction regime may lead to a strong enhancement of the vacuum scalar power spectrum, which can potentially lead to the creation of primordial black holes during the radiation-dominated stage of the early universe. The formalism developed in our paper allows us to investigate these phenomena, which requires, as a first step, the computation of the scalar power spectrum and bispectrum of scalar fluctuations in the presence of backreaction.
We intend to address these questions in more realistic physical models, including backreaction, with our new formalism in future studies.

\begin{acknowledgments}
R.~D. and S.~V. are supported by the Swiss National Science Foundation.
The work of S.~V. is sustained by a University of Geneva grant for Researchers at Risk.
The work of O.~S. is sustained by a Philipp--Schwartz fellowship of the University of M\"{u}nster.
The work of R.\,v.\,E.\ and K.\,S.\ is supported by the Deutsche Forschungsgemeinschaft (DFG) through the Research Training Group, GRK 2149: Strong and Weak Interactions\,---\,from Hadrons to Dark Matter.
\end{acknowledgments}

\vspace{2cm}
\appendix

%%%%%%%%%%%%%%%%%%%%%%%%%%%%%%%%%%%%%%%%%%%%%%%%%%%%%%%%%
%%%%%%%%%%%%%%%%%%%%%%%%%%%%%%%%%%%%%%%%%%%%%%%%%%%%%%%%%
\section{Details of the derivation of the perturbation equations}
\label{app:Einstein}
%%%%%%%%%%%%%%%%%%%%%%%%%%%%%%%%%%%%%%%%%%%%%%%%%%%%%%%%%
%%%%%%%%%%%%%%%%%%%%%%%%%%%%%%%%%%%%%%%%%%%%%%%%%%%%%%%%%

The components of the field-strength tensor read
\begin{align}
    F_{0i} &= a^2 E^{i}\, ,\label{EM-tensor-0i-low}\\
    F_{ij} &= -a^{2}\varepsilon_{ijk}B^{k}\, ,\label{EM-tensor-ij-low}\\
    F^{0i} &= -\frac{1}{a^2}(1-2\Psi+2\Phi)E^i\, ,\label{EM-tensor-0i-up}\\
    F^{ij} &= -\frac{1}{a^2}(1+4\Phi)\varepsilon^{ijk}B^k\, . \label{EM-tensor-ij-up}
\end{align}
The components of the dual tensor are given by
\begin{align}
    \tilde{F}^{0i} &= -\frac{1}{a^2}(1-\Psi+3\Phi)B^{i}\, , \label{EM-tensor-dual-0i-up}\\
    \tilde{F}^{ij} &= \frac{1}{a^{2}}(1-\Psi+3\Phi)\varepsilon^{ijk}E^{k}\, . \label{EM-tensor-dual-ij-up}\\
     \tilde{F}_{0i} &= a^2 (1+\Psi+\Phi) B^{i}\, , \label{EM-tensor-dual-0i-low}\\
    \tilde{F}_{ij} &= a^{2}(1-\Psi-\Phi)\varepsilon_{ijk}E^{k}\, . \label{EM-tensor-dual-ij-low}
   \end{align}
The gauge-field invariants are of the form
\begin{align}
    \syml F_{\mu\nu}F^{\mu\nu} \symr \!&= 2[(1+4\Phi)\boldsymbol{B}^2\!-\!(1-2\Psi+2\Phi)\boldsymbol{E}^2]\, , \label{F2-inv}\\ 
    \syml F_{\mu\nu}\tilde{F}^{\mu\nu} \symr \!&= -4(1-\Psi+3\Phi) \syml\boldsymbol{E}\cdot\boldsymbol{B}\symr\, . \label{FtF-inv}
\end{align}

%%%%%%%%%%%%%%%%%%%%%%%%%%%%%%%%%%%%%%%%%%%%%%%%%%%%%%%%%
\subsection{The Einstein tensor}
\label{subapp:Einstein-tensor}
%%%%%%%%%%%%%%%%%%%%%%%%%%%%%%%%%%%%%%%%%%%%%%%%%%%%%%%%%

The Einstein tensor in a linearly perturbed spatially flat FLRW universe in longitudinal gauge is given by
\begin{align}
    \langle G^{\ 0}_{0}\rangle &=3\frac{a^{\prime 2}}{a^4}=3\frac{\mathcal{H}^2}{a^2}\, ,\\
    \delta  G^{\ 0}_0&=\frac{2}{a^2}\big[\triangle\Phi
    -3\mathcal{H}\Phi' -3\mathcal{H}^2\Psi\big]\, , \label{dG00}\\ 
    \langle G^{\ 0}_{i}\rangle&= 0\, , \\
    \delta  G^{\ 0}_i&=\frac{2}{a^2}\partial_i
    \big[\Phi' +\mathcal{H}\Psi  \big]\, , \label{dG0i}\\
     \langle G_{\ j}^i\rangle&=\delta_j^i\frac{1}{a^2}\left[2\frac{a''}{a}-\left(\frac{a'}{a}\right)^2\right]  
     %\nonumber\\&    %%COMMENT IN 1-COLUMN
     =\delta_j^i\frac{1}{a^2}\big[2\mathcal{H}'+\mathcal{H}^2\big]\, ,\\
    \delta  G^{\ j}_{i}&=-\frac{1}{a^2}\partial_i\partial_j(\Phi-\Psi)  
    -\frac{2}{a^2}\delta_j^i\bigg\{\Phi''  + \frac{1}{2}\triangle(\Psi\!-\!\Phi)
    %\nonumber\\&\quad    %%COMMENT IN 1-COLUMN
    +\mathcal{H}(\Psi'\!+\!2\Phi')
    +(2\mathcal{H}'\!+\!\mathcal{H}^2)\Psi \bigg\}\, , \label{dGij}
\end{align}
where $\langle G^{\mu}_{\nu}\rangle$ are the background quantities.

%%%%%%%%%%%%%%%%%%%%%%%%%%%%%%%%%%%%%%%%%%%%%%%%%%%%%%%%%
\subsection{The energy--momentum tensor}
\label{subapp:EMT}
%%%%%%%%%%%%%%%%%%%%%%%%%%%%%%%%%%%%%%%%%%%%%%%%%%%%%%%%%

The energy--momentum tensor (\ref{EMT}) 
can be represented as $T_{\mu}^{\ \nu}=\langle T_{\mu}^{\ \nu}\rangle + \delta T_{\mu}^{\ \nu}$, where
%
%%ONE-COLUMN MODE
\begin{align}
    \langle T^0_{\ 0}\rangle&=\frac{\phi_{c}^{\prime 2}}{2a^2} +  V(\phi_c)
    +\frac{1}{2}I_1(\phi_c) \big(\langle\boldsymbol{E}^2\rangle+\langle\boldsymbol{B}^2\rangle\big)\, ,\\
    \delta T^{0}_{\ 0}&=\frac{\phi'_c}{a^2}\delta \varphi'
    + \Big[V'(\phi_{c})+\frac{1}{2}I'_1(\phi_{c})\big(\langle\boldsymbol{E}^2\rangle+\langle\boldsymbol{B}^2\rangle\big)\Big]\delta \varphi  -
    \frac{\phi_{c}^{\prime 2}}{a^2}\Psi \nonumber\\& \quad
    + I_1(\phi_{c})\big[ (\Phi-\Psi)\langle\boldsymbol{E}^2\rangle +2\Phi\langle\boldsymbol{B}^2\rangle \big] +
    \frac{1}{2} I_1(\phi_{c})\big[\delta_{\boldsymbol{E}^2}+\delta_{\boldsymbol{B}^2}\big]\, ,\\
    \langle T_i^{\ 0}\rangle&=\langle T_0^{\ i}\rangle=0\, ,\\
    \delta T_i^{\ 0}&=-\delta T_0^{\ i}=\frac{\phi'_c}{a^2}\partial_i\delta \varphi- I_1(\phi_{c}) \varepsilon_{ijk} \delta_{E_j B_k}\, ,\\
    \langle T_i^{\ j}\rangle&=
    -\delta_i^{j}\Big[\frac{\phi_{c}^{\prime 2}}{2 a^2} - V(\phi_c)+ \frac{1}{6}I_1(\phi_c) \big(\langle\boldsymbol{E}^2\rangle+\langle\boldsymbol{B}^2\rangle\big)\Big]\, ,\\
    \delta T_i^{\ j}&=
    - \delta_i^{j}\Big\{ 
    \frac{\phi'_c}{a^2} \delta \varphi'
    - \Big[V'(\phi_{c})-\frac{1}{6} I'_1(\phi_{c})\big(\langle\boldsymbol{E}^2\rangle+\langle\boldsymbol{B}^2\rangle\big)\Big]\delta \varphi -\frac{\phi_c^{\prime 2}}{a^2}\Psi
    \nonumber \\
    & \quad + \frac{1}{3} I_1(\phi_{c})\big[ (\Phi-\Psi) \langle\boldsymbol{E}^2\rangle + 2\Phi \langle \boldsymbol{B}^2\rangle\big]
    + \frac{1}{2} I_1(\phi_{c})\big[\delta_{\boldsymbol{E}^2}+\delta_{\boldsymbol{B}^2}\big] \Big\} + I_1(\phi_{c}) \big[\delta_{E_i E_j} + \delta_{B_i B_j}\big]\, .
\end{align}

With the  aim to extract only the scalar contribution to the energy--momentum tensor, we  introduce the scalar gauge-field sources $F_\rho,$ $F_v,$  $F_p$, and $F_\pi$  in Eqs.~\eqref{f-rho-def}--\eqref{f-pi-def}, where $\delta T^{\ \nu \,\mathrm{(EM)}}_{\mu}$ is the part of the energy--momentum tensor perturbation containing gauge-field perturbations, e.g., $\delta_{E_i E_j}$. Applying the corresponding projectors, we eventually get the following expressions:
%
%\begin{widetext}    %%COMMENT IN 1-COLUMN
\begin{align}
    \delta T^{\ 0 \,\mathrm{(s)}}_{0} &=\delta T^{\ 0}_{0}=\frac{\phi'_c}{a^2}\delta \varphi'+\Big[V'(\phi_{c})+\frac{1}{2}I'_1(\phi_{c})\Big(\langle\boldsymbol{E}^2\rangle+\langle\boldsymbol{B}^2\rangle\Big)\Big] \delta \varphi
	-\Big[\frac{\phi^{\prime 2}_c}{a^2}+ I_1(\phi_{c})\langle\boldsymbol{E}^2\rangle\Big] \Psi
	\nonumber \\& \hspace{9cm}
    + I_1(\phi_{c})\Big[\langle\boldsymbol{E}^2\rangle +2\langle\boldsymbol{B}^2\rangle \Big] \Phi + \frac{2M_{\mathrm{P}}^2}{a^2}F_{\rho}\, , \label{dT00-s}
    \\
	\delta T^{\ 0 \,\mathrm{(s)}}_{i}& =\frac{\partial_i \partial_j}{\triangle}\delta T^{\ 0}_{j}=\partial_i\Big[\frac{\phi'_c}{a^2}\delta \varphi +\frac{2M_{\mathrm{P}}^2}{a^2}F_v\Big]\, , \label{dT0i-s}
    \\
	\delta T^{\ j \,\mathrm{(s)}}_{i} &=-\delta^{j}_{i}\bigg\{\frac{\phi'_c}{a^2}\delta \varphi'+\Big[-V'(\phi_{c})+\frac{1}{6}I'_1(\phi_{c})\Big(\langle\boldsymbol{E}^2\rangle+\langle\boldsymbol{B}^2\rangle\Big)\Big] \delta \varphi
	-\Big[\frac{\phi^{\prime 2}_c}{a^2}+ \frac{1}{3}I_1(\phi_{c})\langle\boldsymbol{E}^2\rangle\Big] \Psi
	\nonumber \\ & \hspace{5.5cm}
    + \frac{1}{3}I_1(\phi_{c})\Big[\langle\boldsymbol{E}^2\rangle +2\langle\boldsymbol{B}^2\rangle \Big] \Phi + \frac{2M_{\mathrm{P}}^2}{a^2}F_{p}\bigg\}-\frac{2M_{\mathrm{P}}^2}{a^2}\Big(\partial_i \partial_j -\frac{1}{3}\delta^{j}_{i}\triangle \Big) F_{\pi}\, . \label{dTij-s}
\end{align}
Combining Eqs.~\eqref{dG00}, \eqref{dG0i}, and \eqref{dGij} with Eqs.~\eqref{dT00-s}--\eqref{dTij-s} into the perturbed Einstein equations $\delta  G_{\mu}^{\ \nu}=M_{\mathrm{P}}^{-2}\delta  T_{\mu}^{\ \nu}$, we obtain  Eqs.~\eqref{Einstein-laplace-Phi-1}--\eqref{Einstein-Psi-Phi} in the main text.

%%%%%%%%%%%%%%%%%%%%%%%%%%%%%%%%%%%%%%%%%%%%%%%%%%%%%%%%%
%%%%%%%%%%%%%%%%%%%%%%%%%%%%%%%%%%%%%%%%%%%%%%%%%%%%%%%%%
\section{Details of the derivation of the  equation for the curvature perturbation \texorpdfstring{\boldmath{$\zeta$}}{} variable}
\label{app:eq-for-zeta}
%%%%%%%%%%%%%%%%%%%%%%%%%%%%%%%%%%%%%%%%%%%%%%%%%%%%%%%%%
%%%%%%%%%%%%%%%%%%%%%%%%%%%%%%%%%%%%%%%%%%%%%%%%%%%%%%%%%

%%%%%%%%%%%%%%%%%%%%%%%%%%%%%%%%%%%%%%%%%%%%%%%%%%%%%%%%%
\subsection{The \texorpdfstring{\boldmath{$\zeta$}}{} equation}
\label{subapp:steps-derivation}
%%%%%%%%%%%%%%%%%%%%%%%%%%%%%%%%%%%%%%%%%%%%%%%%%%%%%%%%%

To derive the equation  for the variable $\zeta$, one should follow these steps:
\begin{enumerate}
    \item[(i)] Substituting $\Psi=\Phi-2F_\pi$ into Eq.~\eqref{Einstein-laplace-Phi} in the Fourier space (i.e., replacing $\triangle \to -k^2$), we solve it with respect to $\delta \varphi'$ and obtain the following result:
    \begin{multline}
        \label{delta-phi-prime-1}
        \ \ \ \ \,\delta \varphi'=\frac{2M_{\mathrm{P}}^2}{\phi'_c}\bigg\{
        \Big[-k^2 +\frac{\phi^{\prime 2}_c}{2M_{\mathrm{P}}^2}-\frac{a^2}{M_{\mathrm{P}}^2}I_1(\phi_c)\langle\boldsymbol{B}^2\rangle\Big]\Phi
        -\frac{1}{2M_{\mathrm{P}}^2}\Big[3\mathcal{H}\phi'_c+a^2 V'(\phi_c)+\frac{1}{2}a^2 I'_1(\phi_c)\big(\langle\boldsymbol{E}^2\rangle+\langle\boldsymbol{B}^2\rangle\big)\Big]\delta \varphi \\
        -F_\rho - 3\mathcal{H}F_v - \Big[\frac{\phi^{\prime 2}_c}{M_{\mathrm{P}}^2}+\frac{a^2}{M_{\mathrm{P}}^2}I_1(\phi_c)\langle\boldsymbol{E}^2\rangle\Big] F_\pi
        \bigg\}\, .
    \end{multline}

    \item[(ii)] Substituting $\Psi=\Phi-2F_\pi$ into Eq.~\eqref{Einstein-Phi-prime}, we solve it with respect to $\Phi'$ and obtain the following result:
    \begin{equation}
        \label{Phi-prime-1}
        \Phi'= - \mathcal{H}\Phi + \frac{\phi^{\prime}_c}{2M_{\mathrm{P}}^2} \delta \varphi + F_v + 2\mathcal{H} F_\pi\, .
    \end{equation}

    \item[(iii)] Using Eq.~\eqref{Einstein-Phi-prime-prime}, we determine the combination $(\Phi'+\mathcal{H}\Psi)'=\Phi'' + \mathcal{H}\Psi' + \mathcal{H}'\Psi$:
    \begin{multline}
    \ \ \ \ \,\Phi'' + \mathcal{H}\Psi' + \mathcal{H}'\Psi= -2\mathcal{H}\Phi' -(\mathcal{H}'+\mathcal{H}^2)\Psi
        +\frac{a^2}{2M_{\mathrm{P}}^2} \bigg\{ 
    \frac{\phi'_c}{a^2}\delta \varphi'+\Big[-V'(\phi_{c})+\frac{1}{6}I'_1(\phi_{c})\Big(\langle\boldsymbol{E}^2\rangle+\langle\boldsymbol{B}^2\rangle\Big)\Big] \delta \varphi \vphantom{\Bigg[}\\ \vphantom{\Bigg[}
    -\Big[\frac{\phi^{\prime 2}_c}{a^2}+ \frac{1}{3}I_1(\phi_{c})\langle\boldsymbol{E}^2\rangle\Big] \Psi
    +
    \frac{1}{3}I_1(\phi_{c})\Big[\langle\boldsymbol{E}^2\rangle +2\langle\boldsymbol{B}^2\rangle \Big] \Phi\bigg\} + \frac{1}{3}F_{p}-\frac{2}{3}k^2 F_\pi \, .
    \end{multline}
    Substituting also here $\Psi=\Phi-2F_\pi$ and Eq.~\eqref{Phi-prime-1}, we obtain the following result:
    \begin{multline}
        \label{Phi-prime-H-Psi-prime}
        \ \ \ \ \,(\Phi'\!+\!\mathcal{H}\Psi)'\!=\! \Big[\!-\!k^2\! +\!\frac{\phi^{\prime 2}_c}{2M_{\mathrm{P}}^2}\!+\!\frac{a^2}{3M_{\mathrm{P}}^2}I_1(\phi_c)\big(\langle\boldsymbol{E}^2\rangle\!-\!\langle\boldsymbol{B}^2\rangle\big)\Big]\Phi
        -\frac{1}{2M_{\mathrm{P}}^2}\Big[5\mathcal{H}\phi'_c+\! 2a^2 V'(\phi_c)\!+\!\frac{a^2}{3} I'_1(\phi_c)\big(\langle\boldsymbol{E}^2\rangle\!+\!\langle\boldsymbol{B}^2\rangle\big)\Big]\delta \varphi \vphantom{\Bigg[}\\\vphantom{\Bigg[}
        -\frac{2}{3} F_\rho - 5\mathcal{H}F_v - \Big[\frac{2}{3} k^2 +\frac{\phi^{\prime 2}_c}{M_{\mathrm{P}}^2}+\frac{2a^2}{3M_{\mathrm{P}}^2}I_1(\phi_c)\big(2\langle\boldsymbol{E}^2\rangle+\langle\boldsymbol{B}^2\rangle\big)\Big] F_\pi\, .
    \end{multline}
    The same expression can also be obtained by direct differentiation of Eq.~\eqref{Einstein-Phi-prime} and substituting $F'_v$ from Eq.~\eqref{eq-F-v}, $\delta \varphi'$ from Eq.~\eqref{delta-phi-prime-1}, and $\phi''_c$ from Eq.~\eqref{KGF-c}.

    \item[(iv)] Equation~\eqref{zeta-def-2} can be represented in the following form:
    \begin{empheq}{equation}
        \label{zeta-matrix}
        \zeta=M_{11} \Phi + M_{12} \delta \varphi + c_1^{(\rho)} F_\rho +c_1^{(v)} F_v + c_1^{(\pi)} F_\pi\, ,
    \end{empheq}
    where
    \begin{equation}
        M_{11}=1,\qquad M_{12}=\frac{\mathcal{H}}{\mathcal{H}^2-\mathcal{H}^\prime}\frac{\phi'_c}{2M_{\mathrm{P}}^2},  
       \qquad c_1^{(\rho)}=c_1^{(\pi)}=0, \qquad c_1^{(v)}=\frac{\mathcal{H}}{\mathcal{H}^2-\mathcal{H}^\prime}\, .
    \end{equation}
    We want to represent $\zeta'$ in a  similar form. For this, we take the time derivative of the right-hand side of Eq.~\eqref{zeta-def-1} and use Eqs.~\eqref{Phi-prime-1} and \eqref{Phi-prime-H-Psi-prime} to get rid of the derivatives of the perturbation variables. We obtain the following result:
    \begin{empheq}{equation}
        \label{zeta-prime-matrix}
        \zeta'=M_{21} \Phi + M_{22} \delta \varphi + c_2^{(\rho)} F_\rho +c_2^{(v)} F_v + c_2^{(\pi)} F_\pi\, ,
    \end{empheq}
    with
    \begin{align}
        M_{21} &=
        -\frac{\mathcal{H}}{\mathcal{H}^2-\mathcal{H}^\prime}\Big[k^2 +\frac{2a^2}{3M_{\mathrm{P}}^2}I_1(\phi_c)\langle\boldsymbol{B}^2\rangle\Big]\, ,\vphantom{\Bigg[}\\
        M_{22} &=
        -\frac{\mathcal{H}}{M_{\mathrm{P}}^2(\mathcal{H}^2-\mathcal{H}^\prime)^2}\bigg\{
        \frac{\phi^{\prime 2}_c a^2}{6M_{\mathrm{P}}^2}\big[I'_1(\phi_c) \langle\boldsymbol{E}^2\rangle
        + I'_2(\phi_c) \langle\syml\boldsymbol{E}\cdot\boldsymbol{B}\symr\rangle \big]
        \vphantom{\Bigg[}\nonumber \\ & \hspace{3cm}
        + \frac{a^2}{3M_{\mathrm{P}}^2}I_1(\phi_c)\big(\langle\boldsymbol{E}^2\rangle
        +\langle\boldsymbol{B}^2\rangle\big)
        \,\Big[\mathcal{H}\phi'_c +a^2 V'(\phi_c)+\frac{a^2}{6}I'_1(\phi_c)\big(\langle\boldsymbol{E}^2\rangle+\langle\boldsymbol{B}^2\rangle\big)\Big]
        \bigg\}\, ,
        \\
        c_2^{(\rho)} &=-\frac{2}{3}\frac{\mathcal{H}}{\mathcal{H}^2-\mathcal{H}^\prime}\, ,
        \\
        c_2^{(\pi)} &=-\frac{2}{3} \frac{\mathcal{H}}{\mathcal{H}^2-\mathcal{H}^\prime} \Big[k^2 +\frac{a^2}{M_{\mathrm{P}}^2}I_1(\phi_c)\langle\boldsymbol{E}^2\rangle\Big]\, ,
        \\
        c_2^{(v)} &=1+\Big(\frac{\mathcal{H}}{\mathcal{H}^2-\mathcal{H}^\prime}\Big)^\prime-5\frac{\mathcal{H}^2}{\mathcal{H}^2-\mathcal{H}^\prime}
        = \frac{\mathcal{H}a^2}{M_{\mathrm{P}}^2(\mathcal{H}^2-\mathcal{H}^\prime)^2}\bigg\{
         \phi'_c\Big[V'(\phi_c)-
         \frac{1}{6}I'_1(\phi_c) \big(\langle\boldsymbol{E}^2\rangle-\langle\boldsymbol{B}^2\rangle\big)
         \nonumber \\ & \hspace{7cm}
         -\frac{1}{3}I'_2(\phi_c) \langle\syml\boldsymbol{E}\cdot\boldsymbol{B}\symr\rangle\Big] -\frac{2}{3}\mathcal{H}I_1(\phi_c) \big(\langle\boldsymbol{E}^2\rangle+\langle\boldsymbol{B}^2\rangle\big)
        \bigg\}\, .
    \end{align}

    \item[(v)] Solving Eqs.~\eqref{zeta-matrix} and \eqref{zeta-prime-matrix} for $\Phi$ and $\delta \varphi$, we find the following results:
    \begin{align}
        \label{Phi-matrix}
        \Phi &=\big(M^{-1}\big)_{11} \zeta + \big(M^{-1}\big)_{12} \zeta'
         -\sum_{\alpha=\rho,\pi,v}\Big[\big(M^{-1}\big)_{11}c_1^{(\alpha)} + \big(M^{-1}\big)_{12}c_2^{(\alpha)}\Big] F_\alpha\, ,
        \\
        \label{delta-varphi-matrix}
        \delta \varphi &=\big(M^{-1}\big)_{21} \zeta + \big(M^{-1}\big)_{22} \zeta'
        -\sum_{\alpha=\rho,\pi,v}\Big[\big(M^{-1}\big)_{21}c_1^{(\alpha)} + \big(M^{-1}\big)_{22}c_2^{(\alpha)}\Big] F_\alpha\, ,
    \end{align}
    where $M^{-1}$ denotes the inverse matrix of the matrix $M$.

    \item[(vi)] We now take the time derivative of Eq.~\eqref{zeta-prime-matrix}:
    \begin{equation}\label{zeta-pprime}
        \zeta''=M_{21}\Phi' + M'_{21}\Phi + M_{22}\delta \varphi' + M'_{22}\delta \varphi
        +\sum_{\alpha=\rho,\pi,v}\big[c_2^{(\alpha)} F'_\alpha+(c_2^{(\alpha)})' F_\alpha\big]\, .
    \end{equation}
    Inserting Eqs.~\eqref{Phi-prime-1}, \eqref{delta-phi-prime-1}, and the equations of motion for the sources Eqs.~\eqref{eq-F-rho} and \eqref{eq-F-v} in \eqref{zeta-pprime}, we can express $\zeta''$ in terms of $\Phi$, $\varphi$, and the sources $F_\rho$, $F_v$, $F_\pi$, $F'_\pi$, $\delta_{\boldsymbol{E}^2}$, and $\delta_{\boldsymbol{E}\cdot\boldsymbol{B}}$ as follows:
    \begin{equation}
        \label{zeta-prime-prime-1}
        \zeta''=d^{(\Phi)} \Phi + d^{(\delta \varphi)}\delta \varphi + d^{(v)} F_v +
         d^{(\rho)} F_\rho + d^{(\pi)}_1 F_\pi
         + d^{(\pi)}_2 F'_\pi + d^{(\boldsymbol{E}^2)} \delta_{\boldsymbol{E}^2} + d^{(\boldsymbol{E}\cdot\boldsymbol{B})} \delta_{\boldsymbol{E}\cdot\boldsymbol{B}}\, ,
    \end{equation}
    where
   \begin{align}
    d^{(\Phi)} &=M'_{21}-\mathcal{H} M_{21} 
    +\frac{2M_{\mathrm{P}}^2}{\phi'_c}
    \Big[-k^2 +\frac{\phi^{\prime 2}_c}{2M_{\mathrm{P}}^2}-\frac{a^2}{M_{\mathrm{P}}^2}I_1(\phi_c)\langle\boldsymbol{B}^2\rangle\Big] M_{22}
    + \frac{a^2}{3M_{\mathrm{P}}^2}I_1(\phi_c)\big(\langle\boldsymbol{E}^2\rangle+2\langle\boldsymbol{B}^2\rangle\big) c_2^{(v)}
    \nonumber \\ &
    + \bigg\{
    \frac{a^2}{\phi'_c}\big[I'_1(\phi_c) \langle\boldsymbol{E}^2\rangle + I'_2(\phi_c) \langle\syml\boldsymbol{E}\cdot\boldsymbol{B}\symr\rangle \big]\Big[k^2 + \frac{a^2}{M_{\mathrm{P}}^2}I_1(\phi_c)\langle\boldsymbol{B}^2\rangle\Big]
    -\frac{a^2 \phi'_c}{2M_{\mathrm{P}}^2}\big[I'_1(\phi_c) \langle\boldsymbol{E}^2\rangle + 3I'_2(\phi_c) \langle\syml\boldsymbol{E}\cdot\boldsymbol{B}\symr\rangle \big]
    \nonumber \\ & \hspace{8cm}
    -\frac{a^2}{M_{\mathrm{P}}^2}I_1(\phi_c)\big[\mathcal{H} \langle\boldsymbol{E}^2\rangle 
    - 2\langle\syml \boldsymbol{E}\cdot\operatorname{rot}\boldsymbol{B}\symr \rangle\big]
    \bigg\}c_2^{(\rho)}\, ,
    \\
    d^{(\delta \varphi)} &=M'_{22}+\frac{\phi'_c}{2M_{\mathrm{P}}^2} M_{21} - \Big[3\mathcal{H}+\frac{a^2}{\phi'_c} V'(\phi_c)+\frac{a^2}{2\phi'_c} I'_1(\phi_c)\big(\langle\boldsymbol{E}^2\rangle+\langle\boldsymbol{B}^2\rangle\big)\Big] M_{22}
    \nonumber \\ & 
    - \frac{a^2}{6M_{\mathrm{P}}^2}\big[I'_1(\phi_c)\big(\langle\boldsymbol{E}^2\rangle-2\langle\boldsymbol{B}^2\rangle\big)+3I'_2(\phi_c) \langle\syml\boldsymbol{E}\cdot\boldsymbol{B}\symr\rangle\big] c_2^{(v)}
    \nonumber \\ &
    +\frac{a^2}{2M_{\mathrm{P}}^2}\bigg\{
    \big[I'_1(\phi_c) \langle\boldsymbol{E}^2\rangle + I'_2(\phi_c) \langle\syml\boldsymbol{E}\cdot\boldsymbol{B}\symr\rangle \big]\Big[3\mathcal{H}+\frac{a^2}{\phi'_c} V'(\phi_c)+\frac{a^2}{2\phi'_c} I'_1(\phi_c)\big(\langle\boldsymbol{E}^2\rangle+\langle\boldsymbol{B}^2\rangle\big)\Big]
    \nonumber \\ &
    -\phi'_c\Big[\Big(I''_1(\phi_c)-\frac{I^{\prime 2}_1(\phi_c)}{I_1(\phi_c)}\Big) \langle\boldsymbol{E}^2\rangle + \Big(I''_2(\phi_c)-\frac{I'_1(\phi_c)I'_2(\phi_c)}{I_1(\phi_c)}\Big) \langle\syml\boldsymbol{E}\cdot\boldsymbol{B}\symr\rangle \Big] + \frac{\phi' _c}{M_{\mathrm{P}}^2}I_1(\phi_c)\langle\boldsymbol{E}^2\rangle
    \bigg\}c_2^{(\rho)}\, ,
    \\
    d^{(v)} &=c_2^{(v)\prime}-2\mathcal{H}c_2^{(v)}+ M_{21} - \frac{6M_{\mathrm{P}}^2\mathcal{H}}{\phi'_c} M_{22}
    + \Big\{-k^2 + \frac{a^2}{M_{\mathrm{P}}^2}I_1(\phi_c)\langle\boldsymbol{E}^2\rangle + \frac{3\mathcal{H}a^2}{\phi'_c}\big[I'_1(\phi_c) \langle\boldsymbol{E}^2\rangle
    \nonumber \\ & \hspace{10.5cm}
    + I'_2(\phi_c) \langle\syml\boldsymbol{E}\cdot\boldsymbol{B}\symr\rangle \big]\Big\} c_2^{(\rho)}\, ,
    \\
    d^{(\rho)} &=c_2^{(\rho)\prime}+        \Big\{-2\mathcal{H} + \frac{I'_1(\phi_c)}{I_1(\phi_c)}\phi'_c + \frac{a^2}{\phi'_c}\big[I'_1(\phi_c) \langle\boldsymbol{E}^2\rangle + I'_2(\phi_c) \langle\syml\boldsymbol{E}\cdot\boldsymbol{B}\symr\rangle \big]\Big\}c_2^{(\rho)}+\frac{1}{3}c_2^{(v)}- \frac{2M_{\mathrm{P}}^2}{\phi'_c} M_{22}\, ,
    \\
    d^{(\pi)}_1 &=c_2^{(\pi)\prime}+ 2\mathcal{H}M_{21} - \Big[2\phi^{\prime}_c+\frac{2a^2}{\phi'_c}I_1(\phi_c)\langle\boldsymbol{E}^2\rangle\Big] M_{22}
    -\Big[\frac{2}{3}k^2 +\frac{a^2}{3M_{\mathrm{P}}^2}I_1(\phi_c)\big(\langle\boldsymbol{E}^2\rangle+2\langle\boldsymbol{B}^2\rangle\big)\Big] c_2^{(v)}
    \nonumber \\ &
    +
    \frac{a^2}{M_{\mathrm{P}}^2}\Big\{
    \phi'_c\big[I'_1(\phi_c) \langle\boldsymbol{E}^2\rangle + 2I'_2(\phi_c) \langle\syml\boldsymbol{E}\cdot\boldsymbol{B}\symr\rangle \big] + \frac{a^2}{\phi'_c}I_1(\phi_c) \langle\boldsymbol{E}^2\rangle\big[I'_1(\phi_c) \langle\boldsymbol{E}^2\rangle + I'_2(\phi_c) \langle\syml\boldsymbol{E}\cdot\boldsymbol{B}\symr\rangle \big]
    \nonumber \\ & \hspace{8.5cm}
    + 2I_1(\phi_c) \big[\mathcal{H} \langle\boldsymbol{E}^2\rangle - \langle\syml\boldsymbol{E}\cdot\operatorname{rot}\boldsymbol{B}\symr\rangle\big]\Big\}c_2^{(\rho)}\, ,
    \\
    d^{(\pi)}_2 &=c_2^{(\pi)} -\frac{a^2}{M_{\mathrm{P}}^2}I_1(\phi_c)\langle\boldsymbol{E}^2\rangle c_2^{(\rho)} = -\frac{2k^2}{3} \frac{\mathcal{H}}{\mathcal{H}^2-\mathcal{H}^\prime}\, ,
    \\
    d^{(\boldsymbol{E}^2)} &=-\frac{a^2\phi'_c}{2M_{\mathrm{P}}^2}I'_1(\phi_c) c_2^{(\rho)}=\frac{a^2 I'_1(\phi_c)\phi'_c}{3M_{\mathrm{P}}^2}\frac{\mathcal{H}}{\mathcal{H}^2-\mathcal{H}^\prime}\, ,
    \\
    d^{(\boldsymbol{E}\cdot\boldsymbol{B})} &=-\frac{a^2\phi'_c}{2M_{\mathrm{P}}^2}I'_2(\phi_c) c_2^{(\rho)}=\frac{a^2 I'_2(\phi_c)\phi'_c}{3M_{\mathrm{P}}^2}\frac{\mathcal{H}}{\mathcal{H}^2-\mathcal{H}^\prime}\, .
    \end{align}

    Finally, we use Eqs.~\eqref{Phi-matrix} and\eqref{delta-varphi-matrix} to write an equation of motion for the $\zeta$ variable alone in the following form:
    \begin{equation}
        \label{zeta-prime-prime-2a}
        \zeta''_{\boldsymbol{k}}+p \zeta'_{\boldsymbol{k}} + q\zeta_{\boldsymbol{k}}=r^{(v)} F_v + r^{(\rho)} F_\rho +\\+ r^{(\pi)}_1 F_\pi + r^{(\pi)}_2 F'_\pi + r^{(\boldsymbol{E}^2)} \delta_{\boldsymbol{E}^2} + r^{(\boldsymbol{E}\cdot\boldsymbol{B})} \delta_{\boldsymbol{E}\cdot\boldsymbol{B}}\, ,
    \end{equation}
    where we added a lower index $\boldsymbol{k}$ representing the momentum of the Fourier mode. The coefficients read
    \begin{align}
        p &=-\big(M^{-1}\big)_{12} d^{(\Phi)} -\big(M^{-1}\big)_{22} d^{(\delta \varphi)}\, ,
        \\
        q &=-\big(M^{-1}\big)_{11} d^{(\Phi)} -\big(M^{-1}\big)_{21} d^{(\delta \varphi)}\, ,
        \\
        r^{(v)} &=d^{(v)}-\Big[\big(M^{-1}\big)_{11}c_1^{(v)} + \big(M^{-1}\big)_{12}c_2^{(v)}\Big]d^{(\Phi)}- \Big[\big(M^{-1}\big)_{21}c_1^{(v)} + \big(M^{-1}\big)_{22}c_2^{(v)}\Big]d^{(\delta \varphi)}\, ,
        \\
        r^{(\rho)} & =d^{(\rho)}-\Big[\big(M^{-1}\big)_{11}c_1^{(\rho)} + \big(M^{-1}\big)_{12}c_2^{(\rho)}\Big]d^{(\Phi)}
        -\Big[\big(M^{-1}\big)_{21}c_1^{(\rho)} + \big(M^{-1}\big)_{22}c_2^{(\rho)}\Big]d^{(\delta \varphi)}\, ,
        \\
        r^{(\pi)}_1 &=d^{(\pi)}_1-\Big[\big(M^{-1}\big)_{11}c_1^{(\pi)} + \big(M^{-1}\big)_{12}c_2^{(\pi)}\Big]d^{(\Phi)}
        -\Big[\big(M^{-1}\big)_{21}c_1^{(\pi)} + \big(M^{-1}\big)_{22}c_2^{(\pi)}\Big]d^{(\delta \varphi)}\, ,
        \\
       r^{(\pi)}_2 &=d^{(\pi)}_2\, ,\qquad
         r^{(\boldsymbol{E}^2)}=d^{(\boldsymbol{E}^2)}\, ,\qquad
         r^{(\boldsymbol{E}\cdot\boldsymbol{B})}=d^{(\boldsymbol{E}\cdot\boldsymbol{B})}\, .
    \end{align} 
\end{enumerate}

%%%%%%%%%%%%%%%%%%%%%%%%%%%%%%%%%%%%%%%%%%%%%%%%%%%%%%%%%
\subsection{Particular case: No direct gauge-field coupling to the inflaton}
\label{subapp:no-direct-coupling}
%%%%%%%%%%%%%%%%%%%%%%%%%%%%%%%%%%%%%%%%%%%%%%%%%%%%%%%%%

With the aim to show that our formalism is  applicable even in the case with no direct coupling to the inflaton, let us consider a model where the kinetic and axial couplings are trivial, i.e., $I_1(\phi)=1$ and $I_2(\phi)=0$. This corresponds to the situation when the gauge field already exists at a given moment of time during inflation and we are not interested in its source (which is already switched off). In addition, let us assume that the backreaction is negligible at this stage. In fact, without a source for the gauge field, it will quickly decay due to the universe's expansion and become sufficiently weak. Therefore, it will not cause significant backreaction. Setting $\langle \boldsymbol{E}^2 \rangle\approx 0$, $\langle \boldsymbol{B}^2 \rangle\approx 0$, $\langle \syml\boldsymbol{E}\cdot\boldsymbol{B}\symr\rangle\approx 0$ etc., it is straightforward to show that the left-hand side of the $\zeta$-equation \eqref{zeta-prime-prime-2a} takes the form of the Mukhanov--Sasaki equation
\begin{equation}
\label{eq:LHS-Mukhanov-Sasaki}
 \mathrm{LHS} = \zeta''_{\boldsymbol{k}}+2\Big(\frac{\phi''_c}{\phi'_c}-\frac{\mathcal{H}'}{\mathcal{H}}+\mathcal{H}\Big) \zeta'_{\boldsymbol{k}} + k^2 \zeta_{\boldsymbol{k}}= \zeta''_{\boldsymbol{k}} + 2\frac{z'}{z}\zeta'_{\boldsymbol{k}} +k^2\zeta_{\boldsymbol{k}}\, ,
\end{equation}
where $z=a\phi'_c/\mathcal{H}$ is the Mukhanov--Sasaki variable. The coefficients on the right-hand side read
\begin{eqnarray}
    r^{(v)}&=&\frac{4M_p^2 \mathcal{H}}{\phi_c^{\prime 2}}\bigg[\frac{k
    ^2}{3}+\frac{a^2 V'(\phi_c)}{\phi'_c}\bigg(2\mathcal{H}-\frac{\mathcal{H}'}{\mathcal{H}}-\frac{\phi''_c}{\phi'_c}+\frac{V''(\phi_c)\phi'_c}{V'(\phi_c)}\bigg)\bigg]\approx \frac{2k^2}{3\epsilon_{H}\mathcal{H}}\,,\\
    r^{(\rho)}&=&\frac{4M_p^2 \mathcal{H}}{3\phi_c^{\prime 2}}\bigg(-2\mathcal{H}+\frac{\mathcal{H}'}{\mathcal{H}}-\frac{\phi''_c}{\phi'_c}\bigg)\approx -\frac{4}{3\epsilon_{H}}\,,\\
    r^{(\pi)}_1&=&\frac{4M_p^2 k^2 \mathcal{H}}{3\phi_c^{\prime 2}}\bigg(-\mathcal{H}+\frac{\mathcal{H}'}{\mathcal{H}}+2\frac{\phi''_c}{\phi'_c}\bigg)\approx\frac{4k^2}{3\epsilon_{H}}\, ,\\
    r^{(\pi)}_2&=&-\frac{4M_p^2 k^2 \mathcal{H}}{3\phi_c^{\prime 2}}\approx -\frac{2k^2}{3\epsilon_{H}\mathcal{H}}\,,\\
    r^{(\boldsymbol{E}^2)}&=&0\,,\\
    r^{(\boldsymbol{E}\cdot\boldsymbol{B})}&=&0\,.
\end{eqnarray}
In the expressions after the $\approx$ sign, we kept only the leading order in the slow-roll parameters. Thus, the source term has the form
\begin{equation}
    \mathcal{S}_{\boldsymbol{k}} \approx \frac{2}{3\epsilon_{H}\mathcal{H}}\Big[k^2 F_v - 2\mathcal{H}F_\rho +2 k^2\mathcal{H} F_\pi -k^2 F'_\pi\Big]\, ,
    \label{source-new-no-interaction-noBR}
\end{equation}
which exactly coincides with the source term reported in Ref.~\cite{Bonvin:2011dt}, where the case without direct gauge-field coupling to the inflaton and without backreaction was studied.

%%%%%%%%%%%%%%%%%%%%%%%%%%%%%%%%%%%%%%%%%%%%%%%%%%%%%%%%%
\subsection{Particular case: Axion inflation in the absence of backreaction}
\label{subapp:axion-noBR-details}
%%%%%%%%%%%%%%%%%%%%%%%%%%%%%%%%%%%%%%%%%%%%%%%%%%%%%%%%%

In Sec.~\ref{sec:application}, we apply our formalism to axion inflation, where $I_1(\phi)=1$, $I_2(\phi)=\tfrac{\alpha}{f}\phi$, neglecting backreaction. In this case the left-hand side of Eq.~\eqref{zeta-prime-prime-2a} again has the form of the Mukhanov--Sasaki equation \eqref{eq:LHS-Mukhanov-Sasaki} and the coefficients on the right-hand side read
\begin{eqnarray}
    r^{(v)}&=&\frac{4M_{\mathrm{P}}^2 \mathcal{H}}{\phi_c^{\prime 2}}\bigg[\frac{k
    ^2}{3}+\frac{a^2 V'(\phi_c)}{\phi'_c}\bigg(2\mathcal{H}-\frac{\mathcal{H}'}{\mathcal{H}}-\frac{\phi''_c}{\phi'_c}+\frac{V''(\phi_c)\phi'_c}{V'(\phi_c)}\bigg)\bigg]\, ,\\
    r^{(\rho)}&=&\frac{4M_{\mathrm{P}}^2 \mathcal{H}}{3\phi_c^{\prime 2}}\bigg(-2\mathcal{H}+\frac{\mathcal{H}'}{\mathcal{H}}-\frac{\phi''_c}{\phi'_c}\bigg)\, ,\\
    r^{(\pi)}_1&=&\frac{4M_{\mathrm{P}}^2 k^2 \mathcal{H}}{3\phi_c^{\prime 2}}\bigg(-\mathcal{H}+\frac{\mathcal{H}'}{\mathcal{H}}+2\frac{\phi''_c}{\phi'_c}\bigg)\, ,\\
    r^{(\pi)}_2&=&-\frac{4M_{\mathrm{P}}^2 k^2 \mathcal{H}}{3\phi_c^{\prime 2}},\\
    r^{(\boldsymbol{E}^2)}&=&0\, ,\\
    r^{(\boldsymbol{E}\cdot\boldsymbol{B})}&=&\frac{2}{3}a^2 I'_2(\phi_c)\frac{\mathcal{H}}{\phi'_c}\, .
\end{eqnarray}

We stress that the decomposition of the source into six terms is performed just for convenience and different terms do not correspond to different physical effects. In particular, one may guess that the last term $r^{(\boldsymbol{E}\cdot\boldsymbol{B})} \delta_{\boldsymbol{E}\cdot\boldsymbol{B}}$ is the only term that comes directly from the axial coupling and that one has to keep only this term to get the results of Ref.~\cite{Barnaby:2011vw}. This is not correct. In particular,  the axial coupling $I'_2$ is also present in the term $r^{(\pi)}_2 F'_\pi$; see Eq.~\eqref{eq-F-pi}. Moreover, other terms, although not directly proportional to $I'_2$, depend on it implicitly through the gauge-field mode functions. Therefore, the presence of the metric perturbations does not simply lead to new terms in the source but changes its entire 
structure. A detailed comparison with the earlier results in the literature is performed in Sec.~\ref{sec:application}.

%%%%%%%%%%%%%%%%%%%%%%%%%%%%%%%%%%%%%%%%%%%%%%%%%%%%%%%%%
%%%%%%%%%%%%%%%%%%%%%%%%%%%%%%%%%%%%%%%%%%%%%%%%%%%%%%%%%
\section{Conditions for the absence of backreaction during axion inflation}
\label{app:no-BR}
%%%%%%%%%%%%%%%%%%%%%%%%%%%%%%%%%%%%%%%%%%%%%%%%%%%%%%%%%
%%%%%%%%%%%%%%%%%%%%%%%%%%%%%%%%%%%%%%%%%%%%%%%%%%%%%%%%%

In this appendix, we discuss the conditions under which backreaction is negligible during axion inflation. This is crucial for understanding the applicability range of the results of Sec.~\ref{sec:application}.

Before turning to the quantitative analysis, we would like to emphasize that the relevance of backreaction at a given moment of time during inflation cannot be inferred just from the numerical value of the gauge-field production parameter $\xi$ in Eq.~\eqref{xi-def}. In fact, even in the strong-backreaction regime, this parameter can take small values (sometimes even crossing  zero) for a short time because of its oscillatory behavior in time~\cite{Domcke:2020zez,Gorbar:2021rlt,Figueroa:2023oxc,Durrer:2023rhc}. Moreover, even if we start in the absence of backreaction with $\xi=0$ and gradually increase it, the transition to the backreaction regime occurs at different threshold values $\xi_{\mathrm{thr}}$ for different inflationary models. Indeed, let us introduce the following parameters:
\begin{align}
    \delta_{\mathrm{F}\phantom{\mathrm{G}}}^{(1)} & =\,\frac{\frac{1}{2}(\langle \boldsymbol{E}^2\rangle+\langle \boldsymbol{B}^2\rangle)}{3H^2 M_{\mathrm{P}}^2}\,=\frac{1}{6}\left(\frac{H}{M_{\mathrm{P}}}\right)^{2}\left[e_0(\xi)+b_0(\xi)\right] \,,\\
    \delta_{\mathrm{F}\phantom{\mathrm{G}}}^{(2)} & =\,\frac{-\frac{1}{3}(\langle \boldsymbol{E}^2\rangle+\langle \boldsymbol{B}^2\rangle)}{\dot{H} M_{\mathrm{P}}^2}\,=\frac{1}{3\epsilon_{H}}\left(\frac{H}{M_{\mathrm{P}}}\right)^{2}\left[e_0(\xi)+b_0(\xi)\right] \,,\\
    \delta_{\mathrm{KG}} & = \frac{|(\alpha/f)\langle \boldsymbol{E}\cdot\boldsymbol{B}\rangle |}{|3H\dot{\phi}|}=\frac{1}{3\epsilon_{H}\big(1-\delta_{\mathrm{F}}^{(2)}\big)}\left(\frac{H}{M_{\mathrm{P}}}\right)^{2}\left|\xi g_0(\xi)\right| \,.
\end{align}
They quantify the relative gauge-field contributions to $H^2$ (first Friedmann equation), $\dot{H}$ (second Friedmann equation), and to the Klein--Gordon equation for the inflaton, respectively. Here $e_0(\xi)=\langle\boldsymbol{E}^2\rangle/H^4$, $b_0(\xi)=\langle\boldsymbol{B}^2\rangle/H^4$, and $g_0(\xi)=\langle\boldsymbol{E}\cdot\boldsymbol{B}\rangle/H^4$ are dimensionless functions, which, in the case of $\xi=\text{const}$ and $H=\text{const}$, depend only on $\xi$~\cite{Anber:2009ua,Sobol:2019xls,vonEckardstein:2023gwk}. The backreaction is negligible if all these parameters are much smaller than unity. Typically, due to the appearance of the slow-roll parameter $\epsilon_H$, the conditions $\delta_{\mathrm{KG}} \ll 1$ and $\delta_{\mathrm{F}}^{(2)}\ll 1$ are much stronger than $\delta_{\mathrm{F}}^{(1)} \ll 1$. Therefore, as a criterion for the backreaction to become relevant, we may use
\begin{equation}
\label{eq:conditions-BR}
    |\xi g_0(\xi)| \simeq \kappa\, ,\qquad e_0(\xi) + b_0(\xi) \simeq \kappa\, ,
\end{equation}
where
\begin{equation}
    \kappa \equiv 3\epsilon_H \left(\frac{M_{\mathrm{P}}}{H}\right)^{2}\, .
\end{equation}
The minimal value of $\xi$ for which at least one of the conditions in Eq.~\eqref{eq:conditions-BR} is satisfied is the threshold value $\xi_{\mathrm{thr}}$. $\xi_{\mathrm{thr}}$ therefore quanitifies when backreaction is expected to occur in a given model.

\begin{figure}[t]
\centering
\includegraphics[width=0.59\textwidth]{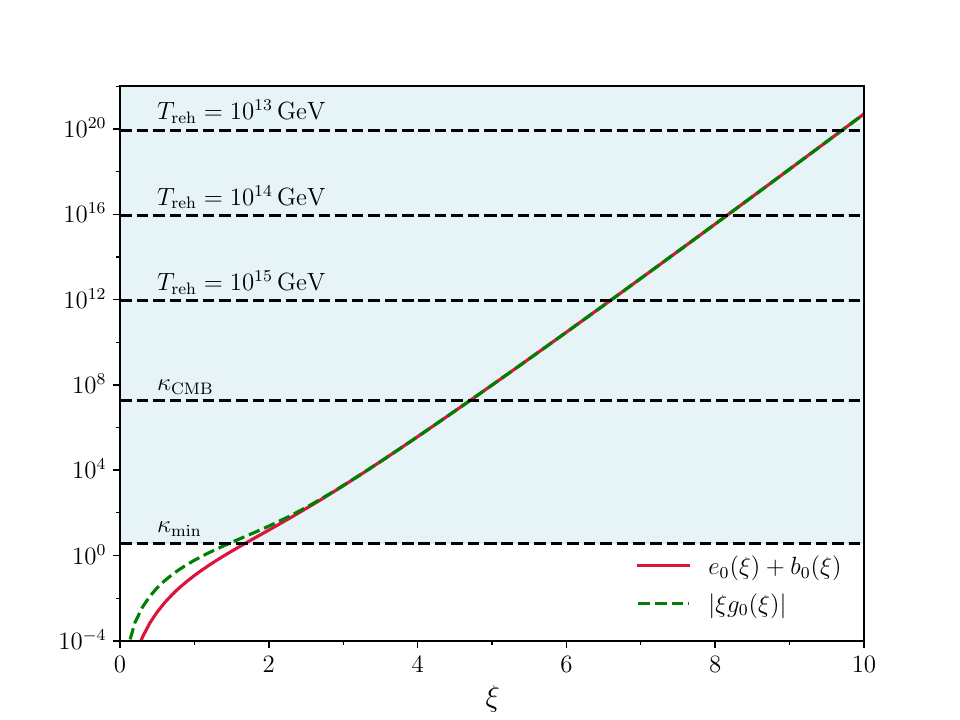}
\caption{The dimensionless functions $|\xi g_0(\xi)|$ (red solid line) and $e_0(\xi)+b_0(\xi)$ (green dashed line) together with five reference values for the parameter $\kappa$ (horizontal black dashed lines). The lowest black dashed line corresponds to the minimal estimate in Eq.~\eqref{eq:kappa-min}; the second-to-lowest line shows the value which follows from the CMB constraints, Eq.~\eqref{eq:kappa-CMB}; three upper black dashed lines represent maximal estimates \eqref{eq:kappa-max} for the parameter $\kappa$ for three values of the reheating temperature $T_{\mathrm{reh}}=10^{15}$, $10^{14}$, and $10^{13}\,\SI{}{\giga\eV}$. Lower reheating temperatures would correspond to even larger values of $\kappa_{\mathrm{max}}$. The blue shaded region covers the range of $\kappa$ values that may be accessible in different inflation models.  \label{fig:BR}}
\end{figure}

In the case of constant $\xi$ and $H$, the functions $e_0(\xi)$, $g_0(\xi)$, and $b_0(\xi)$ read~\cite{vonEckardstein:2023gwk}
\begin{align}
e_0\left(\xi\right) & = \int\limits_0^{2\xi} \frac{dx}{4\pi^2} \sum\limits_{\lambda=\pm} x\,e^{\lambda\pi\xi}\left|\left(x-\lambda\xi\right)W_{-i\lambda\xi,1/2}\left(-2ix\right)-iW_{1-i\lambda\xi,1/2}\left(-2ix\right)\right|^2 \,, \label{eq:e0}\\
g_0\left(\xi\right) & = -\int\limits_0^{2\xi}\frac{dx}{4\pi^2} \sum\limits_{\lambda=\pm}\lambda x^{2}e^{\lambda\pi\xi}\operatorname{Re}\left[W_{i\lambda\xi,1/2}\left(2ix\right)W_{1-i\lambda\xi,1/2}\left(-2ix\right)\right] \,, \label{eq:g0}\\
b_0\left(\xi\right) & = \int\limits_0^{2\xi}\frac{dx}{4\pi^2} \sum\limits_{\lambda=\pm} x^{3}e^{\lambda\pi\xi}\left|W_{-i\lambda\xi,1/2}\left(-2ix\right)\right|^2 \,, \label{eq:b0}
\end{align}
where $W_{i\lambda\xi,1/2}(z)$ is the Whittaker function. For any given value of $\xi$, they can be evaluated numerically. In the limit of large $\xi$, these functions can be approximated with simple analytical expressions~\cite{Sobol:2019xls} leading to
\begin{align}
\label{eq:e0-approx}
    e_0(\xi)&\approx \phantom{-}\frac{9}{\num{1120}\,\pi^3}\frac{e^{2\pi\xi}}{\xi^3}\,, \\
\label{eq:g0-approx}
    g_0(\xi)&\approx -\frac{9}{\num{1120}\,\pi^3}\frac{e^{2\pi\xi}}{\xi^4}\,, \\
\label{eq:b0-approx}
    b_0(\xi)&\approx \phantom{-}\frac{10}{\num{1120}\,\pi^3}\frac{e^{2\pi\xi}}{\xi^5}\,.
\end{align}
We plot the  functions $|\xi g_0(\xi)|$ and $e_0(\xi)+b_0(\xi)$ in Fig.~\ref{fig:BR} by the red solid and green dashed lines, respectively. In the limit of large $\xi$, the two lines coincide. This is in agreement with Eqs.~\eqref{eq:e0-approx}--\eqref{eq:b0-approx}.

Furthermore, let us estimate the bounds for the parameter $\kappa$ that may be achieved in a generic inflationary model. For this, we observe that, in the absence of backreaction,
\begin{equation}
    \kappa = \frac{3}{8\pi^2} \frac{1}{P_{\zeta}^{(0)}}\, ,
\end{equation}
where $P_{\zeta}^{(0)} = (H^2/2\pi \dot{\phi}_c)^2$ is the scalar power spectrum in single-field inflation. Then, the minimal possible value of $\kappa$ is given by the maximal allowed value of the scalar power spectrum. Taking the latter to be $P_{\zeta,\mathrm{max}}^{(0)}\sim 10^{-2}$, which follows from the constraints on the production of primordial black holes (for review, see, e.g., Ref.~\cite{Ozsoy:2023ryl}), we obtain
\begin{equation}
\label{eq:kappa-min}
    \kappa_{\mathrm{min}} = \frac{3}{8\pi^2} \frac{1}{P_{\zeta,\mathrm{max}}^{(0)}} \simeq 4\, .
\end{equation}
This value of $\kappa$ may be reached in inflationary models which include an ultra-slow-roll stage. Of course, this can only happen during horizon crossing of $k$ values which are much larger than those accessible to CMB observations.

Another benchmark value of $\kappa$ is found by inserting the CMB constraints for the scalar power spectrum:
\begin{equation}
\label{eq:kappa-CMB}
    \kappa_{\mathrm{CMB}} = \frac{3}{8\pi^2} \frac{1}{A_s} \simeq \num{2e7}\, .
\end{equation}
where $A_s=P_{\zeta,\mathrm{CMB}}^{(0)}\approx \num{2.1e-9}$ is the amplitude of the scalar power spectrum for the CMB pivot scale~\cite{Planck:2018vyg}. This value of $\kappa$ is significant if one wants to estimate the strength of the backreaction while the modes relevant for the CMB cross the horizon during inflation.

Finally, the maximal value of $\kappa$, can be estimated as follows:
\begin{equation}
\label{eq:kappa-max-1}
    \kappa_{\mathrm{max}}=3\epsilon_{H,\mathrm{max}} \frac{M_{\mathrm{P}}^2}{H_{\mathrm{min}}^2}\, ,
\end{equation}
where $\epsilon_{H,\mathrm{max}}$ is the maximal value of the slow-roll parameter which we will take to be $\epsilon_{H,\mathrm{max}}\sim 0.1$ (still to be in the slow-roll regime) while $H_{\mathrm{min}}$ is the minimal value of the Hubble rate. The latter value is achieved at the end of inflation since $\dot{H}<0$. We express it in terms of the reheating temperature assuming that reheating is instantaneous,
\begin{equation}
    H_{\mathrm{min}} = \sqrt{\frac{\pi^2 g_\ast}{90}}\frac{T_{\mathrm{reh}}^2}{M_{\mathrm{P}}}\, ,
\end{equation}
where $g_\ast$ is the effective number of relativistic degrees of freedom, which we take to be $g_\ast = 106.75$, corresponding to the full Standard Model. Inserting this expression into Eq.~\eqref{eq:kappa-max-1}, we obtain the maximal estimate
\begin{equation}
\label{eq:kappa-max}
    \kappa_{\mathrm{max}}=\frac{270\,\epsilon_{H,\mathrm{max}}}{\pi^2 g_{\ast}}\left( \frac{M_{\mathrm{P}}}{T_{\mathrm{reh}}}\right)^4 \simeq \num{9e11}\,\left( \frac{10^{15}\,\SI{}{\giga\eV}}{T_{\mathrm{reh}}}\right)^4 \, .
\end{equation}
Varying the reheating temperature from the maximal possible value allowed by the CMB constraints $T_{\mathrm{reh,max}}=M_{\mathrm{P}}\sqrt{45A_s r_{\mathrm{max}}/g_{\ast}}=\qty{5.6e15}{\giga\eV}$ (here $r_{\mathrm{max}}=0.032$ is the maximal value of the tensor-to-scalar ratio at \qty{95}{\percent} CL~\cite{Tristram:2021tvh}) to the lowest possible one that still allows for the electroweak phase transition, $T_{\mathrm{reh,min}}\sim \qty{100}{\giga\eV}$, we cover an extremely wide range of $\kappa_{\mathrm{max}}$ from $\num{9e8}$ to $\sim 10^{64}$.

We show the values of $\kappa_{\mathrm{min}}$, $\kappa_{\mathrm{CMB}}$, and $\kappa_{\mathrm{max}}$ (for three particular values of the reheating temperature) as black dashed lines in Fig.~\ref{fig:BR}. The blue shaded region shows the possible range of $\kappa$ values that can be achieved during inflation. A horizontal line with $\kappa=\text{const}$ corresponds to a specific moment of time in a given model of inflation. The crossing of the green dashed line with this horizontal line gives the threshold value of $\xi$ for which backreaction becomes relevant at this moment of time in this model of inflation. As one can see from the plot, $\xi_{\mathrm{thr}}$ may vary from values of order unity to well above 10.

%%%%%%%%%%%%%%%%%%%%%%%%%%%%%%%%%%%%%%%%%%%%%%%%%%%%%%%%%
%%%%%%%%%%%%%%%%%%%%%%%%%%%%%%%%%%%%%%%%%%%%%%%%%%%%%%%%%
\section{Expressions for the functions \texorpdfstring{\boldmath{$K_1$}}{} and \texorpdfstring{\boldmath{$K_3$}}{}}
\label{app:K-functions}
%%%%%%%%%%%%%%%%%%%%%%%%%%%%%%%%%%%%%%%%%%%%%%%%%%%%%%%%%
%%%%%%%%%%%%%%%%%%%%%%%%%%%%%%%%%%%%%%%%%%%%%%%%%%%%%%%%%

The source in Eq.~\eqref{zeta-prime-prime-2} was divided into six terms for convenience . Here, we introduce the capital Latin index $L$ running from 1 to 6 and denote each term as $\mathcal{S}^{(L)}$ with $L$ corresponding to its number; e.g., $\mathcal{S}^{(1)}=r^{(v)}F_v$ and $\mathcal{S}^{(4)}=r^{(\pi)}_2 F'_\pi$. In  Fourier space, each of the terms can be characterized by the functions $K_1^{(L)}$ to $K_4^{(L)}$, as it is shown in Eq.~\eqref{S-Fourier}. Because of the relations \eqref{K1-sym}--\eqref{rel-K1-K4}, not all of them are independent. One can choose, e.g., the functions $K^{(L)}_1$ and $K^{(L)}_3$ to form the minimal set of independent quantities. In this appendix, we give explicit expressions for these functions. They are listed below:
\begin{align}
    K^{(1)}_1(\lambda,\boldsymbol{p};\lambda',\boldsymbol{p}';\eta, \boldsymbol{k}) &=
  \frac{a^2 r^{(v)}}{4M_{\mathrm{P}}^2} \frac{i}{k^2}\big(\boldsymbol{k}\cdot[\boldsymbol{\epsilon}_{\lambda}(\boldsymbol{p})\times\boldsymbol{\epsilon}_{\lambda'}(\boldsymbol{p}')]\big)
  \Big[\mathcal{E}_{\lambda,\boldsymbol{p}}(\eta) \mathcal{B}_{\lambda',\boldsymbol{p}'}(\eta)-\mathcal{B}_{\lambda,\boldsymbol{p}}(\eta) \mathcal{E}_{\lambda',\boldsymbol{p}'}(\eta)\Big]\, , \label{K-1-v}
    \\
    K^{(1)}_3(\lambda,\boldsymbol{p};\lambda',\boldsymbol{p}';\eta, \boldsymbol{k}) &= \frac{a^2 r^{(v)}}{4M_{\mathrm{P}}^2} \frac{i}{k^2}\big(\boldsymbol{k}\cdot[\boldsymbol{\epsilon}^{\ast}_{\lambda}(\boldsymbol{p})\times\boldsymbol{\epsilon}_{\lambda'}(\boldsymbol{p}')]\big)
    \Big[\mathcal{E}^{\ast}_{\lambda,\boldsymbol{p}}(\eta) \mathcal{B}_{\lambda',\boldsymbol{p}'}(\eta)-
    \mathcal{B}^{\ast}_{\lambda,\boldsymbol{p}}(\eta) \mathcal{E}_{\lambda',\boldsymbol{p}'}(\eta)\Big]\, , \label{K-3-v}
    \\
    K^{(2)}_1(\lambda,\boldsymbol{p};\lambda',\boldsymbol{p}';\eta, \boldsymbol{k}) &= \frac{a^2 r^{(\rho)}}{4M_{\mathrm{P}}^2} \big(\boldsymbol{\epsilon}_{\lambda}(\boldsymbol{p})\cdot\boldsymbol{\epsilon}_{\lambda'}(\boldsymbol{p}')\big)\Big[\mathcal{E}_{\lambda,\boldsymbol{p}}(\eta) \mathcal{E}_{\lambda',\boldsymbol{p}'}(\eta)+
    \mathcal{B}_{\lambda,\boldsymbol{p}}(\eta) \mathcal{B}_{\lambda',\boldsymbol{p}'}(\eta)\Big]\, , \label{K-1-rho}
    \\
    K^{(2)}_3(\lambda,\boldsymbol{p};\lambda',\boldsymbol{p}';\eta, \boldsymbol{k}) &= \frac{a^2 r^{(\rho)}}{4M_{\mathrm{P}}^2} \big(\boldsymbol{\epsilon}^{\ast}_{\lambda}(\boldsymbol{p})\cdot\boldsymbol{\epsilon}_{\lambda'}(\boldsymbol{p}')\big) \Big[\mathcal{E}^{\ast}_{\lambda,\boldsymbol{p}}(\eta) \mathcal{E}_{\lambda',\boldsymbol{p}'}(\eta)+\mathcal{B}^{\ast}_{\lambda,\boldsymbol{p}}(\eta) \mathcal{B}_{\lambda',\boldsymbol{p}'}(\eta)\Big]\, , \label{K-3-rho}
    \\
    K^{(3)}_1(\lambda,\boldsymbol{p};\lambda',\boldsymbol{p}';\eta, \boldsymbol{k}) &= 
    -\frac{a^2 r^{(\pi)}_1}{4M_{\mathrm{P}}^2 k^2}\Big(\delta_{ij}-3\frac{k_i k_j}{k^2}\Big) \epsilon^{i}_{\lambda}(\boldsymbol{p})\epsilon^{j}_{\lambda'}(\boldsymbol{p}')
    \Big[\mathcal{E}_{\lambda,\boldsymbol{p}}(\eta) \mathcal{E}_{\lambda',\boldsymbol{p}'}(\eta)+\mathcal{B}_{\lambda,\boldsymbol{p}}(\eta) \mathcal{B}_{\lambda',\boldsymbol{p}'}(\eta)\Big]\, , \label{K-1-pi}
    \\
    K^{(3)}_3(\lambda,\boldsymbol{p};\lambda',\boldsymbol{p}';\eta, \boldsymbol{k}) &= -\frac{a^2 r^{(\pi)}_1}{4M_{\mathrm{P}}^2 k^2}\Big(\delta_{ij}-3\frac{k_i k_j}{k^2}\Big) \epsilon^{\ast i}_{\lambda}(\boldsymbol{p})\epsilon^{j}_{\lambda'}(\boldsymbol{p}')
    \Big[\mathcal{E}^{\ast}_{\lambda,\boldsymbol{p}}(\eta) \mathcal{E}_{\lambda',\boldsymbol{p}'}(\eta)+\mathcal{B}^{\ast}_{\lambda,\boldsymbol{p}}(\eta) \mathcal{B}_{\lambda',\boldsymbol{p}'}(\eta)\Big]\, , \label{K-3-pi}
    \\
    K^{(4)}_1(\lambda,\boldsymbol{p};\lambda',\boldsymbol{p}';\eta, \boldsymbol{k}) &= \frac{a^2 r^{(\pi)}_2}{4M_{\mathrm{P}}^2 k^2}\Big(\delta_{ij}-3\frac{k_i k_j}{k^2}\Big) \epsilon^{i}_{\lambda}(\boldsymbol{p})\epsilon^{j}_{\lambda'}(\boldsymbol{p}')\bigg\{\Big(2\mathcal{H}+\frac{I'_1(\phi_c)\phi'_c}{I_1(\phi_c)}\Big)\mathcal{E}_{\lambda,\boldsymbol{p}}(\eta) \mathcal{E}_{\lambda',\boldsymbol{p}'}(\eta)
    \nonumber \\& \qquad
    +\Big(2\mathcal{H}-\frac{I'_1(\phi_c)\phi'_c}{I_1(\phi_c)}\Big)\mathcal{B}_{\lambda,\boldsymbol{p}}(\eta) \mathcal{B}_{\lambda',\boldsymbol{p}'}(\eta)
    +\frac{I'_2(\phi_c)\phi'_c}{I_1(\phi_c)}\Big[\mathcal{E}_{\lambda,\boldsymbol{p}}(\eta) \mathcal{B}_{\lambda',\boldsymbol{p}'}(\eta)+\mathcal{B}_{\lambda,\boldsymbol{p}}(\eta) \mathcal{E}_{\lambda',\boldsymbol{p}'}(\eta)\Big]
    \nonumber \\& \hspace{5.5cm}
    +(\lambda p-\lambda' p') \Big[\mathcal{E}_{\lambda,\boldsymbol{p}}(\eta) \mathcal{B}_{\lambda',\boldsymbol{p}'}(\eta)-\mathcal{B}_{\lambda,\boldsymbol{p}}(\eta) \mathcal{E}_{\lambda',\boldsymbol{p}'}(\eta)\Big]\bigg\}\, , \label{K-1-pi-prime}
    \\
    K^{(4)}_3(\lambda,\boldsymbol{p};\lambda',\boldsymbol{p}';\eta, \boldsymbol{k}) &= 
    \frac{a^2 r^{(\pi)}_2}{4M_{\mathrm{P}}^2 k^2}\Big(\delta_{ij}-3\frac{k_i k_j}{k^2}\Big) \epsilon^{\ast i}_{\lambda}(\boldsymbol{p})\epsilon^{j}_{\lambda'}(\boldsymbol{p}') \bigg\{\Big(2\mathcal{H}+\frac{I'_1(\phi_c)\phi'_c}{I_1(\phi_c)}\Big)\mathcal{E}^{\ast}_{\lambda,\boldsymbol{p}}(\eta) \mathcal{E}_{\lambda',\boldsymbol{p}'}(\eta)
    \nonumber \\& \qquad
    +\Big(2\mathcal{H}-\frac{I'_1(\phi_c)\phi'_c}{I_1(\phi_c)}\Big)\mathcal{B}^{\ast}_{\lambda,\boldsymbol{p}}(\eta) \mathcal{B}_{\lambda',\boldsymbol{p}'}(\eta)
    +\frac{I'_2(\phi_c)\phi'_c}{I_1(\phi_c)}\Big[\mathcal{E}^{\ast}_{\lambda,\boldsymbol{p}}(\eta) \mathcal{B}_{\lambda',\boldsymbol{p}'}(\eta)+\mathcal{B}^{\ast}_{\lambda,\boldsymbol{p}}(\eta) \mathcal{E}_{\lambda',\boldsymbol{p}'}(\eta)\Big] 
    \nonumber \\& \hspace{5.5cm}
    +(\lambda p-\lambda' p') \Big[\mathcal{E}^{\ast}_{\lambda,\boldsymbol{p}}(\eta) \mathcal{B}_{\lambda',\boldsymbol{p}'}(\eta)-\mathcal{B}^{\ast}_{\lambda,\boldsymbol{p}}(\eta) \mathcal{E}_{\lambda',\boldsymbol{p}'}(\eta)\Big]\bigg\}\, ,  \label{K-3-pi-prime}
    \\
    K^{(5)}_1(\lambda,\boldsymbol{p};\lambda',\boldsymbol{p}';\eta, \boldsymbol{k}) &=\frac{r^{(\boldsymbol{E}^2)}}{I_1(\phi_c)} \big(\boldsymbol{\epsilon}_{\lambda}(\boldsymbol{p})\cdot\boldsymbol{\epsilon}_{\lambda'}(\boldsymbol{p}')\big)\mathcal{E}_{\lambda,\boldsymbol{p}}(\eta) \mathcal{E}_{\lambda',\boldsymbol{p}'}(\eta)\, , \label{K-1-E2}
    \\
    K^{(5)}_3(\lambda,\boldsymbol{p};\lambda',\boldsymbol{p}';\eta, \boldsymbol{k}) &= \frac{r^{(\boldsymbol{E}^2)}}{I_1(\phi_c)} \big(\boldsymbol{\epsilon}^{\ast}_{\lambda}(\boldsymbol{p})\cdot\boldsymbol{\epsilon}_{\lambda'}(\boldsymbol{p}')\big)\mathcal{E}^{\ast}_{\lambda,\boldsymbol{p}}(\eta) \mathcal{E}_{\lambda',\boldsymbol{p}'}(\eta)\, , \label{K-3-E2}
    \\
    K^{(6)}_1(\lambda,\boldsymbol{p};\lambda',\boldsymbol{p}';\eta, \boldsymbol{k})
    &= \frac{r^{(\boldsymbol{E}\cdot\boldsymbol{B})}}{2I_1(\phi_c)} \big(\boldsymbol{\epsilon}_{\lambda}(\boldsymbol{p})\cdot\boldsymbol{\epsilon}_{\lambda'}(\boldsymbol{p}')\big)\Big[\mathcal{E}_{\lambda,\boldsymbol{p}}(\eta) \mathcal{B}_{\lambda',\boldsymbol{p}'}(\eta)+\mathcal{B}_{\lambda,\boldsymbol{p}}(\eta) \mathcal{E}_{\lambda',\boldsymbol{p}'}(\eta)\Big]\, , \label{K-1-EB}
    \\
    K^{(6)}_3(\lambda,\boldsymbol{p};\lambda',\boldsymbol{p}';\eta, \boldsymbol{k}) &= \frac{r^{(\boldsymbol{E}\cdot\boldsymbol{B})}}{2I_1(\phi_c)} \big(\boldsymbol{\epsilon}^{\ast}_{\lambda}(\boldsymbol{p})\cdot\boldsymbol{\epsilon}_{\lambda'}(\boldsymbol{p}')\big)\Big[\mathcal{E}^{\ast}_{\lambda,\boldsymbol{p}}(\eta) \mathcal{B}_{\lambda',\boldsymbol{p}'}(\eta)+\mathcal{B}^{\ast}_{\lambda,\boldsymbol{p}}(\eta) \mathcal{E}_{\lambda',\boldsymbol{p}'}(\eta)\Big]\, . \label{K-3-EB}
\end{align}
Note that the functions entering the expressions for the scalar power spectrum in Eq.~\eqref{scalar-power-spectrum} and the scalar bispectrum in Eq.~\eqref{zeta-bispectrum} are just summations over the index $L$ of the above-mentioned expressions, i.e.,
\begin{align}
    K_1(\lambda,\boldsymbol{p};\lambda',\boldsymbol{p}';\eta, \boldsymbol{k}) &= \sum\limits_{L=1}^{6}K_1^{(L)}(\lambda,\boldsymbol{p};\lambda',\boldsymbol{p}';\eta, \boldsymbol{k})\, ,\\
    K_3(\lambda,\boldsymbol{p};\lambda',\boldsymbol{p}';\eta, \boldsymbol{k}) &= \sum\limits_{L=1}^{6}K_3^{(L)}(\lambda,\boldsymbol{p};\lambda',\boldsymbol{p}';\eta, \boldsymbol{k})\, .
\end{align}
%\end{widetext}     %%COMMENT IN 1-COLUMN

%%%%%%%%%%%%%%%%%%%%%%%%%%%%%%%%%%%%%%%%%%%%%%%%%%%%%%%%%
%%%%%%%%%%%%%%%%%%%%%%%%%%%%%%%%%%%%%%%%%%%%%%%%%%%%%%%%%
\section{Polarization vectors}
\label{app:polarization}
%%%%%%%%%%%%%%%%%%%%%%%%%%%%%%%%%%%%%%%%%%%%%%%%%%%%%%%%%
%%%%%%%%%%%%%%%%%%%%%%%%%%%%%%%%%%%%%%%%%%%%%%%%%%%%%%%%%

In this appendix,  expressions for the convolutions of polarization vectors with different tensors are derived. These expressions appear in Eqs.~\eqref{K-1-v}--\eqref{K-3-EB}.

First of all, we note that we always deal with expressions involving the product of two polarization vectors with momenta $\boldsymbol{p}$ and $\boldsymbol{p}'$ satisfying 
\begin{equation}
    \boldsymbol{p}+\boldsymbol{p}'=\boldsymbol{k}\, ,
\end{equation}
where $\boldsymbol{k}$ is one of the ``external'' momenta in the correlation function of perturbations, i.e., is fixed. In all cases, where the complex conjugate of the polarization appears, one may use the property
\begin{equation}
    \boldsymbol{\epsilon}^{\ast}_{\lambda}(\boldsymbol{p}) = \boldsymbol{\epsilon}_{\lambda}(-\boldsymbol{p})
\end{equation}
to bring the corresponding expression into the above-mentioned form.

One can then show straightforwardly that all three different types of convolutions appearing in the main text can be expressed in terms of a scalar product of the polarization vectors with, possibly, some prefactors:
%
%%ONE-COLUMN MODE
\begin{align}
\label{eps-scalar-product}
    \delta_{ij}\epsilon^{i}_{\lambda}(\boldsymbol{p})\epsilon^{j}_{\lambda'}(\boldsymbol{p}')&=\big(\boldsymbol{\epsilon}_{\lambda}(\boldsymbol{p})\cdot\boldsymbol{\epsilon}_{\lambda'}(\boldsymbol{p}')\big)\, ,\\
 \label{eps-ki-kj}
    k_i k_j \epsilon^{i}_{\lambda}(\boldsymbol{p})\epsilon^{j}_{\lambda'}(\boldsymbol{p}')&=(\boldsymbol{p}\cdot\boldsymbol{p}' + \lambda\lambda' pp')\big(\boldsymbol{\epsilon}_{\lambda}(\boldsymbol{p})\cdot\boldsymbol{\epsilon}_{\lambda'}(\boldsymbol{p}')\big)\, ,\\
\label{mixed-product}
    i\big(\boldsymbol{k}\cdot[\boldsymbol{\epsilon}_{\lambda}(\boldsymbol{p})\times\boldsymbol{\epsilon}_{\lambda'}(\boldsymbol{p}')]\big)&=(\lambda p -\lambda' p')\big(\boldsymbol{\epsilon}_{\lambda}(\boldsymbol{p})\cdot\boldsymbol{\epsilon}_{\lambda'}(\boldsymbol{p}')\big)\, .
\end{align}
%
%%TWO-COLUMN MODE
% \begin{equation}
% \label{eps-scalar-product}
%     \delta_{ij}\epsilon^{i}_{\lambda}(\boldsymbol{p})\epsilon^{j}_{\lambda'}(\boldsymbol{p}')=\big(\boldsymbol{\epsilon}_{\lambda}(\boldsymbol{p})\cdot\boldsymbol{\epsilon}_{\lambda'}(\boldsymbol{p}')\big)\, ,
% \end{equation}
% \begin{equation}
%  \label{eps-ki-kj}
%     k_i k_j\epsilon^{i}_{\lambda}(\boldsymbol{p})\epsilon^{j}_{\lambda'}(\boldsymbol{p}')\!=\!(\boldsymbol{p}\cdot\boldsymbol{p}'\! +\! \lambda\lambda' pp')\big(\boldsymbol{\epsilon}_{\lambda}(\boldsymbol{p})\cdot\boldsymbol{\epsilon}_{\lambda'}(\boldsymbol{p}')\big)\, ,
% \end{equation}   
% \begin{equation}
% \label{mixed-product}
%     i\big(\boldsymbol{k}\cdot[\boldsymbol{\epsilon}_{\lambda}(\boldsymbol{p})\!\times\!\boldsymbol{\epsilon}_{\lambda'}(\boldsymbol{p}')]\big)=(\lambda p\! -\!\lambda' p')\big(\boldsymbol{\epsilon}_{\lambda}(\boldsymbol{p})\cdot\boldsymbol{\epsilon}_{\lambda'}(\boldsymbol{p}')\big)\, .
% \end{equation}
%
Therefore, in what follows we discuss how to compute the scalar product of two polarization vectors.

Here, we note that the definition of polarization vectors corresponding to the mode with momentum $\boldsymbol{p}$ is not unique because one can rotate the coordinate system in the plane orthogonal to $\boldsymbol{p}$. In particular, this means that the circular polarization vectors are defined up to a complex phase. However, below we present an algorithm that allows one to unambiguously fix both polarization vectors, $\boldsymbol{\epsilon}_{\lambda}(\boldsymbol{p})$ and $\boldsymbol{\epsilon}_{\lambda'}(\boldsymbol{p}')$, and compute their convolutions for given values of $\boldsymbol{k}$ and $\boldsymbol{p}$.

Let us start from the definition of linear polarization vectors for the mode with momentum $p$. To this end, we need some vector $\boldsymbol{q}$ that is not collinear to $\boldsymbol{p}$. Then, one of the linear polarization vectors can be defined as, e.g.,
\begin{equation}
    \label{eps2-gen}
    \boldsymbol{\epsilon}_{2}(\boldsymbol{p})=\frac{\boldsymbol{q}\times\boldsymbol{p}}{|\boldsymbol{q}\times\boldsymbol{p}|}\, .
\end{equation}
Consequently, the other linear polarization vector reads as
\begin{equation}
    \label{eps1-gen}
    \boldsymbol{\epsilon}_{1}(\boldsymbol{p})=\frac{\boldsymbol{\epsilon}_2(\boldsymbol{p})\times \boldsymbol{p}}{p}=\frac{\boldsymbol{p}(\boldsymbol{p}\cdot\boldsymbol{q})-\boldsymbol{q} p^2}{p |\boldsymbol{p}\times\boldsymbol{q}|}\, .
\end{equation}
By construction, $\boldsymbol{p}$, $\boldsymbol{\epsilon}_{1}(\boldsymbol{p})$, and $\boldsymbol{\epsilon}_{2}(\boldsymbol{p})$ form a right-handed system of vectors, i.e.,
\begin{equation}
    [\boldsymbol{\epsilon}_{1}(\boldsymbol{p})\times \boldsymbol{\epsilon}_{2}(\boldsymbol{p})]\cdot \frac{\boldsymbol{p}}{p} = +1\, .
\end{equation}
The circular polarization vectors can be constructed as usual,
\begin{equation}
    \label{eps-lambda}
    \boldsymbol{\epsilon}_{\lambda}(\boldsymbol{p})=\frac{1}{\sqrt{2}}\Big[\boldsymbol{\epsilon}_{1}(\boldsymbol{p})+i\lambda \boldsymbol{\epsilon}_{2}(\boldsymbol{p})\Big]\, ,\qquad \lambda=\pm 1\, .
\end{equation}

Taking the same vector $\boldsymbol{q}$ for the mode with momentum $\boldsymbol{p}'$, one can derive similar expressions for the corresponding linear polarization vectors. In principle, one could express all the required convolutions of polarization vectors in terms of $\boldsymbol{p}$, $\boldsymbol{p}'$, and the arbitrary vector $\boldsymbol{q}$ not collinear with  $\boldsymbol{p}$ and $\boldsymbol{p}'$ ($\boldsymbol{q}$ should disappear from all physical quantities in the end). However, it is more convenient to take a specific form of $\boldsymbol{q}$ which strongly simplifies the computations.

Here, we will distinguish between two situations. In the case when we have not just one pair of polarization vectors but several different configurations, we need to choose the vector $\boldsymbol{q}$ globally for all of them (this is, e.g., the case of computing the bispectrum in Sec.~\ref{sec:application}). Then, one may just choose $\boldsymbol{q} = \boldsymbol{e}_z = (0,\,0,\,1)^{\mathrm{T}}$ (assuming $\boldsymbol{p}$ is not parallel to $\boldsymbol{e}_z $) . In this case, for any given $\boldsymbol{p}=p(\sin\theta \cos\varphi, \, \sin\theta \sin\varphi,\, \cos\theta)^{\mathrm{T}}$, the polarization vectors of circular polarization $\lambda$ will then have the form
%
%%ONE-COLUMN MODE
\begin{equation}
    \label{eps-lambda-ez}
    \boldsymbol{\epsilon}_{\lambda}(\boldsymbol{p}) = \frac{1}{\sqrt{2}}\big(\cos\theta \cos\varphi - i\lambda \sin\varphi,\, \cos\theta \sin\varphi + i\lambda \cos\varphi,\, -\sin\theta \big)^{\mathrm{T}}\, .
\end{equation}
%
%%TWO-COLUMN MODE
% \begin{multline}
%     \label{eps-lambda-ez}
%     \boldsymbol{\epsilon}_{\lambda}(\boldsymbol{p}) = \frac{1}{\sqrt{2}}\big(\cos\theta \cos\varphi - i\lambda \sin\varphi,\, \cos\theta \sin\varphi \\+ i\lambda \cos\varphi,\, -\sin\theta \big)^{\mathrm{T}}\, .
% \end{multline}
%
Then, the scalar product of two polarization vectors has the form
%
%%ONE-COLUMN MODE
\begin{equation}
\label{scalar-prod-general}
    \big(\boldsymbol{\epsilon}_{\lambda}(\boldsymbol{p})\cdot\boldsymbol{\epsilon}_{\lambda'}(\boldsymbol{p}')\big) = \frac{1}{2}\big[\sin\theta \sin\theta' 
    + \big(\cos\theta \cos\theta' - \lambda\lambda'\big)\cos(\varphi-\varphi')
    + i(\lambda' \cos\theta - \lambda \cos\theta')\sin(\varphi-\varphi')\big]\, .
\end{equation}
%
%%TWO-COLUMN MODE
% \begin{multline}
%     \label{scalar-prod-general}
%     \big(\boldsymbol{\epsilon}_{\lambda}(\boldsymbol{p})\cdot\boldsymbol{\epsilon}_{\lambda'}(\boldsymbol{p}')\big) = \frac{1}{2}\big[\sin\theta \sin\theta' 
%     \\+ \big(\cos\theta \cos\theta' - \lambda\lambda'\big)\cos(\varphi-\varphi')
%     \\+ i(\lambda' \cos\theta - \lambda \cos\theta')\sin(\varphi-\varphi')\big]\, .
% \end{multline}
%
One can show that the absolute value of this expression equals
\begin{align}
    \big|\big(\boldsymbol{\epsilon}_{\lambda}(\boldsymbol{p})\cdot\boldsymbol{\epsilon}_{\lambda'}(\boldsymbol{p}')\big)\big| &= \frac{1}{2}\big\{1-\lambda\lambda'[\cos\theta \cos\theta'
    %\nonumber \\&\qquad     %%COMMENT IN 1-COLUMN
    + \sin\theta \sin\theta' \cos(\varphi-\varphi')]\big\}
    \nonumber \\& = \frac{1}{2}\Big(1-\lambda\lambda'\frac{\boldsymbol{p}\cdot\boldsymbol{p}'}{pp'}\Big)\, ,
\end{align}
while the complex phase depends on the choice of $\boldsymbol{q}$; however, this dependence always disappears in all physical quantities such as the power spectrum of the bispectrum.

On the other hand, if only one pair of momenta $\boldsymbol{p}$ and $\boldsymbol{p}'$ appears in the computation as, e.g., in the case of the scalar power spectrum, one can choose the vector $\boldsymbol{q}$ more conveniently as $\boldsymbol{q}=\boldsymbol{p}\times \boldsymbol{p}'$. Then, the linear polarization vectors read as
\begin{align}
    \label{eps1}
    \boldsymbol{\epsilon}_{1}(\boldsymbol{p}) &=\boldsymbol{\epsilon}_{1}(\boldsymbol{p}')=-\frac{\boldsymbol{p}\times \boldsymbol{p}'}{|\boldsymbol{p}\times \boldsymbol{p}'|}\, ,
    \\
    \label{eps2-p}
    \boldsymbol{\epsilon}_{2}(\boldsymbol{p}) &=\frac{\boldsymbol{p}' p^2-\boldsymbol{p}(\boldsymbol{p}\cdot\boldsymbol{p}')}{p |\boldsymbol{p}\times \boldsymbol{p}'|}\, ,
    \\
    \label{eps2-p-prime}
    \boldsymbol{\epsilon}_{2}(\boldsymbol{p}') &=\frac{\boldsymbol{p}'(\boldsymbol{p}\cdot\boldsymbol{p}')-\boldsymbol{p} p^{\prime 2}}{p' |\boldsymbol{p}\times \boldsymbol{p}'|}\, ,
\end{align}
and the circular polarization vectors are constructed according to Eq.~\eqref{eps-lambda}. Then, the scalar product of two polarization vectors becomes
\begin{equation}
    \label{eps-scalar-product-particular}
    \big(\boldsymbol{\epsilon}_{\lambda}(\boldsymbol{p})\cdot\boldsymbol{\epsilon}_{\lambda'}(\boldsymbol{p}')\big)=\frac{1}{2}\Big(1-\lambda\lambda'\frac{\boldsymbol{p}\cdot\boldsymbol{p}'}{pp'}\Big)\, .
\end{equation}
We obtained Eq.~\eqref{scalar-prod-general} in the case when the complex phase vanishes. This is due to the convenient choice of the vector $\boldsymbol{q}$. However, we  emphasize once again that the above expression cannot be used in a general case when there are more than two different momenta of the gauge-field modes simultaneously in one expression. In this case, one should rather use Eq.~\eqref{scalar-prod-general}.

%%%%%%%%%%%%%%%%%%%%%%%%%%%%%%%%%%%%%%%%%%%%%%%%%%%%%%%%%
%%%%%%%%%%%%%%%%%%%%%%%%%%%%%%%%%%%%%%%%%%%%%%%%%%%%%%%%%
\section{Scalar power spectrum and bispectrum in the large-\texorpdfstring{\boldmath{$\xi$}}{xi} approximation during axion inflation}
\label{app:large-xi}
%%%%%%%%%%%%%%%%%%%%%%%%%%%%%%%%%%%%%%%%%%%%%%%%%%%%%%%%%
%%%%%%%%%%%%%%%%%%%%%%%%%%%%%%%%%%%%%%%%%%%%%%%%%%%%%%%%%

%%%%%%%%%%%%%%%%%%%%%%%%%%%%%%%%%%%%%%%%%%%%%%%%%%%%%%%%%
\subsection{Large-\texorpdfstring{\boldmath{$\xi$}}{xi} approximation}
\label{subapp:large-xi-approx}
%%%%%%%%%%%%%%%%%%%%%%%%%%%%%%%%%%%%%%%%%%%%%%%%%%%%%%%%%

\begin{figure}
\centering
\includegraphics[width=0.98\textwidth]{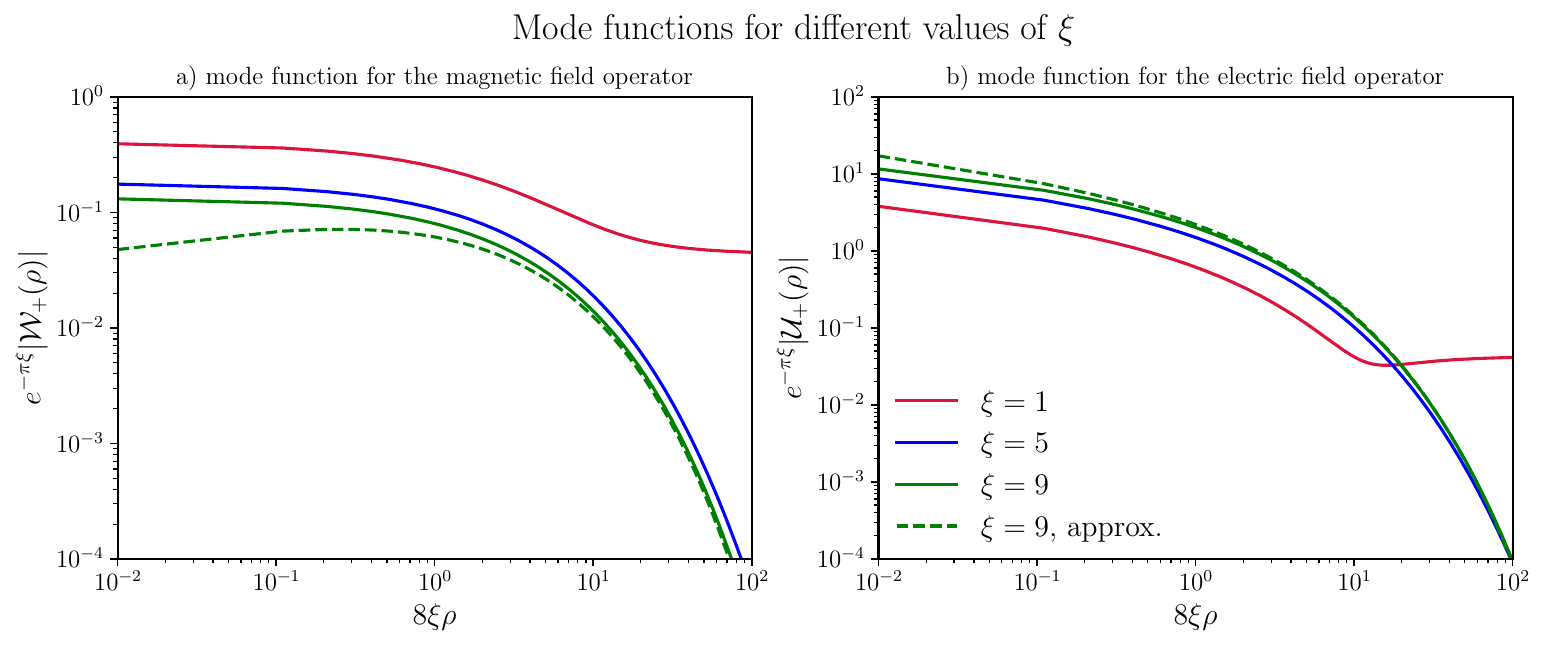}
\caption{The absolute values of the gauge-field mode functions with positive helicity multiplied by $e^{-\pi\xi}$: (a)~$\mathcal{W}_{+}(\rho)$ and (b)~$\mathcal{U}_{+}(\rho)$. The red, blue, and green solid curves show the exact mode functions in Eqs.~\eqref{sol-Whittaker}--\eqref{sol-Coulomb-U} for $\xi=1$, $5$, and $9$, respectively. The green dashed lines show the approximate mode functions in Eqs.~\eqref{W-funct-approx} and \eqref{U-funct-approx} for $\xi=9$. \label{fig:mode-functions}}
\end{figure}

To obtain a simple analytical estimate for the scalar power spectrum and bispectrum generated during axion inflation without gauge-field backreaction, one can follow Ref.~\cite{Barnaby:2011vw} and apply the following approximations (in addition to assumptions of a constant Hubble rate $H$ and gauge-field production parameter $\xi$ which were made in Sec.~\ref{sec:application}):
\begin{itemize}
    \item[(i)] take into account only the circular polarization of the gauge field that is amplified by the axion (for definiteness, $\lambda=+$);
    \item[(ii)] use the large-$\xi$ approximation for the mode functions derived in Ref.~\cite{Anber:2009ua},
    \begin{align}
        \mathcal{W}_{+}(\rho) &\simeq
         \Big(\frac{\rho}{2\xi}\Big)^{1/4} e^{\pi\xi - \sqrt{8\xi \rho}}\, , \label{W-funct-approx}\\ 
        \mathcal{U}_{+}(\rho) &\simeq\Big(\frac{\rho}{2\xi}\Big)^{-1/4} e^{\pi\xi - \sqrt{8\xi \rho}}\, , \label{U-funct-approx}
    \end{align}
    valid for $\xi\gg1$ and $(8\xi)^{-1}\ll \rho \ll 2\xi$ (we plot these expressions together with the exact mode functions in Fig.~\ref{fig:mode-functions})%
    \footnote{Note that for this approximate mode function the relation $\mathcal{U}_{\lambda}(\rho) = -\mathcal{W}'_{\lambda}(\rho)$ does not hold. Therefore, it is important to keep different notations for the two functions.}
    ;

    \item[(iii)] approximate the Green function in Eq.~\eqref{Green-function-axion-approx} by the first nontrivial term in its Taylor expansion, namely 
    \begin{equation}
        G_{\boldsymbol{k}}(\eta,\eta')\approx \theta(y'-y) \frac{y'}{3k}\, ,
    \end{equation}
    where $y=-k\eta$ and $y'=-k\eta'$.
\end{itemize}

With these approximations, the integral in Eq.~\eqref{integral-I} can be computed analytically. The upper integration limit can be set to infinity since the mode functions in Eqs.~\eqref{W-funct-approx} and \eqref{U-funct-approx} exponentially decay at large arguments and the integral does not diverge in the UV.%
\footnote{Note that the approximate expressions for the mode functions in Eqs.~\eqref{W-funct-approx} and \eqref{U-funct-approx} are not valid for $\rho \gtrsim 2\xi$ and they do not approximate the true mode functions in that region. However, their exponential damping is in some sense analogous to the cutoff at $y_{_\mathrm{UV}}$.}
Keeping only the leading term in the series over the inverse powers of $\xi$, we find the following result:
%
%%ONE-COLUMN MODE
\begin{equation}
\label{int-I-approx}
    \mathcal{I}(+,p_{\ast};+,p'_{\ast};x) = \frac{e^{2\pi \xi}}{\xi^3} \frac{15}{128}\frac{x^4 \big[7+Q(p_{\ast},p'_{\ast},x)\big]}{(p_{\ast}p'_{\ast})^{1/4}\big(\sqrt{p_{\ast}}+\sqrt{p'_{\ast}}\big)^{7}} 
    + \mathcal{O}\Big(\frac{e^{2\pi\xi}}{\xi^5}\Big)\, .
\end{equation}
%
%%TWO-COLUMN MODE
% \begin{multline}
% \label{int-I-approx}
%     \mathcal{I}(+,p_{\ast};+,p'_{\ast};x) = \frac{e^{2\pi \xi}}{\xi^3} \frac{15}{128}\frac{x^4 \big[7+Q(p_{\ast},p'_{\ast},x)\big]}{(p_{\ast}p'_{\ast})^{1/4}\big(\sqrt{p_{\ast}}+\sqrt{p'_{\ast}}\big)^{7}} 
%     \\+ \mathcal{O}\Big(\frac{e^{2\pi\xi}}{\xi^5}\Big)\, .
% \end{multline}
%
Here we have introduced the quantity $Q(p_{\ast},p'_{\ast},x)$ defined by
%
%%ONE-COLUMN MODE
\begin{equation}
   Q(p_{\ast},p'_{\ast},x) = 7h_1(+,p_{\ast};+,p'_{\ast};x) - 4h_2(+,p_{\ast};+,p'_{\ast};x) 
  = \frac{(p_{\ast}-p'_{\ast})^2}{x^2}
\end{equation}
%
%%TWO-COLUMN MODE
% \begin{align}
%     Q(p_{\ast},p'_{\ast},x) &= 7h_1(+,p_{\ast};+,p'_{\ast};x) - 4h_2(+,p_{\ast};+,p'_{\ast};x) 
%     \nonumber\\& = \frac{(p_{\ast}-p'_{\ast})^2}{x^2}
% \end{align}
%
in the case when we take into account the metric perturbations. $Q(p_{\ast},p'_{\ast},x)$ is equal to zero if metric perturbations are neglected.

%%%%%%%%%%%%%%%%%%%%%%%%%%%%%%%%%%%%%%%%%%%%%%%%%%%%%%%%%
\subsection{Scalar power spectrum}
\label{subapp:power-spectrum-large-xi}
%%%%%%%%%%%%%%%%%%%%%%%%%%%%%%%%%%%%%%%%%%%%%%%%%%%%%%%%%

Now, let us compute the scalar power spectrum, which is parametrized by the function $f_2(\xi)$ given by Eq.~\eqref{f2-general}. From Eq.~\eqref{int-I-approx}, it is obvious that the function $f_2(\xi)$ in the limit of large $\xi$ behaves as $f_2(\xi) = c_2/\xi^6 + \mathcal{O}(\xi^{-8})$, where the amplitude $c_2$ is given by the integral
%
%%ONE-COLUMN MODE
\begin{equation}
    c_2 = \frac{15^2}{2^{15} \pi}\int d^3\boldsymbol{p}_{\ast} 
    \sqrt{p_{\ast} |\boldsymbol{p}_{\ast}-\hat{\boldsymbol{k}}|}
    \frac{\big|\big(\boldsymbol{\epsilon}_{+}(\boldsymbol{p}_{\ast})\cdot \boldsymbol{\epsilon}_{+}(\hat{\boldsymbol{k}}-\boldsymbol{p}_{\ast})\big)\big|^2}{\Big(\sqrt{p_{\ast}}+\sqrt{|\boldsymbol{p}_{\ast}-\hat{\boldsymbol{k}}|}\Big)^{14}}
    \big[7+Q(p_{\ast},|\boldsymbol{p}_{\ast}-\hat{\boldsymbol{k}}|,1)\big]^2\, , \label{f2-spectrum}
\end{equation}
%
%%TWO-COLUMN MODE
% \begin{multline}
%     c_2 = \frac{15^2}{2^{15} \pi}\!\int\!\! d^3\boldsymbol{p}_{\ast} 
%    \sqrt{p_{\ast} |\boldsymbol{p}_{\ast}\!\!-\!\hat{\boldsymbol{k}}|}\,
%     \frac{\big|\big(\boldsymbol{\epsilon}_{+}(\boldsymbol{p}_{\ast})\cdot \boldsymbol{\epsilon}_{+}(\hat{\boldsymbol{k}}\!-\!\boldsymbol{p}_{\ast})\big)\big|^2}{\Big(\sqrt{p_{\ast}}+\sqrt{|\boldsymbol{p}_{\ast}\!\!-\!\hat{\boldsymbol{k}}|}\Big)^{14}}\\ \times
%     \big[7+Q(p_{\ast},|\boldsymbol{p}_{\ast}-\hat{\boldsymbol{k}}|,1)\big]^2\, , \label{f2-spectrum}
% \end{multline}
%
where $\hat{\boldsymbol{k}}=\boldsymbol{k}/|\boldsymbol{k}|$ and $\boldsymbol{p}_{\ast}=\boldsymbol{p}/|\boldsymbol{k}|$.

Numerically, we find
\begin{equation}
    c_2 = \frac{\num{97 637}}{\num{473 561 088}}-\frac{195\pi}{\num{5 046272}}\approx \num{8.48e-5}
\end{equation}
for the case when the metric perturbations are included and
\begin{equation}
    \tilde{c}_2 = \frac{172}{\num{2 342 912}}\approx \num{7.47e-5}
\end{equation}
for the case where the metric perturbations are neglected (the value reported in Ref.~\cite{Barnaby:2011vw}). Thus, taking into account the metric perturbations leads to a change of  approximately \qty{12}{\percent}. Note that this difference is present only if one uses the approximate mode functions \eqref{W-funct-approx} and \eqref{U-funct-approx} for the gauge field. If one takes the exact mode functions instead, both results converge to the same asymptotic value in the limit of large $\xi$; see Fig.~\ref{fig:f2} and the discussion in the main text.

%%%%%%%%%%%%%%%%%%%%%%%%%%%%%%%%%%%%%%%%%%%%%%%%%%%%%%%%%
\subsection{Scalar bispectrum}
\label{subapp:bispectrum-large-xi}
%%%%%%%%%%%%%%%%%%%%%%%%%%%%%%%%%%%%%%%%%%%%%%%%%%%%%%%%%

To calculate the function $f_3(\xi, x_2, x_3)$, which describes the scalar bispectrum, we substitute Eq.~\eqref{int-I-approx} into Eq.~\eqref{f3-general}. The large-$\xi$ expansion of this function has the form $f_3(\xi,\, x_2,\, x_3) = c_3(x_2,x_3)/\xi^{9} + \mathcal{O}(\xi^{-11})$ with
%
%%ONE-COLUMN MODE
\begin{align}
    c_3(x_2,x_3) &= 
    -\frac{3^2 5^4}{2^{21} \pi}\frac{x_2^3 x_3^3}{1+x_2^3 + x_3^3}
     \int d^3\boldsymbol{p}_{\ast} 
    \sqrt{p_{\ast} |\boldsymbol{p}_{\ast}-\hat{\boldsymbol{k}}_1| |\hat{\boldsymbol{k}}_3 x_3 
    +\boldsymbol{p}_{\ast}|} \nonumber \\
    &\times \frac{\big(\boldsymbol{\epsilon}_{+}(\boldsymbol{p}_{\ast})\cdot 
    \boldsymbol{\epsilon}_{+}
    (\hat{\boldsymbol{k}}_1-\boldsymbol{p}_{\ast})\big) 
    \big(\boldsymbol{\epsilon}_{+}
    (\boldsymbol{p}_{\ast}-\hat{\boldsymbol{k}}_1)\cdot 
    \boldsymbol{\epsilon}_{+}(-\hat{\boldsymbol{k}}_3 x_3 -\boldsymbol{p}_{\ast})\big) \big(\boldsymbol{\epsilon}_{+}(\hat{\boldsymbol{k}}_3 x_3 +\boldsymbol{p}_{\ast})\big)\cdot \boldsymbol{\epsilon}_{+}(-\boldsymbol{p}_{\ast})\big)}{\Big(\sqrt{p_{\ast}}+
    \sqrt{|\boldsymbol{p}_{\ast}-\hat{\boldsymbol{k}}_1|}\Big)^7 \Big(\sqrt{|\hat{\boldsymbol{k}}_3 x_3 +
    \boldsymbol{p}_{\ast}|}+\sqrt{|\boldsymbol{p}_{\ast}-\hat{\boldsymbol{k}}_1|}\Big)^7 \Big(\sqrt{p_{\ast}}+\sqrt{|\hat{\boldsymbol{k}}_3x_3 +\boldsymbol{p}_{\ast}|}\Big)^7} \nonumber\\
    &\times \big[7+Q(p_{\ast},|\boldsymbol{p}_{\ast}-\hat{\boldsymbol{k}}|,1)\big]\big[7+Q(|\boldsymbol{p}_{\ast}-\hat{\boldsymbol{k}}|,|\hat{\boldsymbol{k}}_3x_3 +\boldsymbol{p}_{\ast}|,x_2)\big]\big[7+Q(p_{\ast},|\hat{\boldsymbol{k}}_3x_3 +\boldsymbol{p}_{\ast}|,x_3)\big]
    \, , \label{c3-coefficient}
\end{align}
%
%%TWO-COLUMN MODE
% \begin{align}
%     &c_3(x_2,x_3) = 
%     -\frac{3^2 5^4}{2^{21} \pi}\frac{x_2^3 x_3^3}{1+x_2^3 + x_3^3} \int\!\! d^3\boldsymbol{p}_{\ast}\, \sqrt{p_{\ast}}\nonumber\\
%     &\times  
%     \sqrt{ |\boldsymbol{p}_{\ast}\!-\!\hat{\boldsymbol{k}}_1| |\hat{\boldsymbol{k}}_3 x_3 
%     \!+\!\boldsymbol{p}_{\ast}|} \ 
%     \frac{\big(\boldsymbol{\epsilon}_{+}
%     (\boldsymbol{p}_{\ast}\!\!-\!\hat{\boldsymbol{k}}_1)\cdot 
%     \boldsymbol{\epsilon}_{+}(-\hat{\boldsymbol{k}}_3 x_3\! -\!\boldsymbol{p}_{\ast})\big)}{\Big(\sqrt{|\hat{\boldsymbol{k}}_3 x_3\! +\!
%     \boldsymbol{p}_{\ast}|}+\sqrt{|\boldsymbol{p}_{\ast}\!\!-\!\hat{\boldsymbol{k}}_1|}\Big)^7}\nonumber \\
%     &\times \frac{\big(\boldsymbol{\epsilon}_{+}(\boldsymbol{p}_{\ast})\cdot 
%     \boldsymbol{\epsilon}_{+}
%     (\hat{\boldsymbol{k}}_1\!-\!\boldsymbol{p}_{\ast})\big)
%     \big(\boldsymbol{\epsilon}_{+}(\hat{\boldsymbol{k}}_3 x_3 \!+\!\boldsymbol{p}_{\ast})\cdot \boldsymbol{\epsilon}_{+}(-\boldsymbol{p}_{\ast})\big)}{\Big(\sqrt{p_{\ast}}+
%     \sqrt{|\boldsymbol{p}_{\ast}-\hat{\boldsymbol{k}}_1|}\Big)^7 \Big(\sqrt{p_{\ast}}\!+\!\sqrt{|\hat{\boldsymbol{k}}_3x_3\! +\!\boldsymbol{p}_{\ast}|}\Big)^7} \nonumber\\
%     &\times\big[7+Q(p_{\ast},|\boldsymbol{p}_{\ast}-\hat{\boldsymbol{k}}|,1)\big]\,\big[7+Q(p_{\ast},|\hat{\boldsymbol{k}}_3x_3 +\boldsymbol{p}_{\ast}|,x_3)\big]\nonumber\\
%     &\times\big[7+Q(|\boldsymbol{p}_{\ast}-\hat{\boldsymbol{k}}|,|\hat{\boldsymbol{k}}_3x_3 +\boldsymbol{p}_{\ast}|,x_2)\big]
%     \, , \label{c3-coefficient}
% \end{align}
%
where $\boldsymbol{p}_{\ast}=\boldsymbol{p}/ k$, $\hat{\boldsymbol{k}}_i = \boldsymbol{k}_i/|\boldsymbol{k}_i|$.

In the equilateral configuration $x_2=x_3=1$, numerically, we obtain
\begin{equation}
    c_3(1,1)\approx \num{-3.25e-7}\, ,
\end{equation}
if metric perturbations are included and
\begin{equation}
    \tilde{c}_3(1,1)\approx \num{-2.77e-7}\, ,
\end{equation}
when they are neglected, which (up to an overall sign) is the same value as reported in Ref.~\cite{Barnaby:2011vw}.%
\footnote{Note that the difference in the overall sign of the bispectrum is due to the sign difference in the definition of the scalar perturbation variable $\zeta$; compare Eq.~\eqref{zeta-def-2} in the present work and Eq.~(5.16) in Ref.~\cite{Barnaby:2011vw}.\label{footnote-sign}}
Again, considering metric perturbations on the same footing with those of the inflaton we obtain a result which is about \qty{15}{\percent} larger (in the absolute value). Note, however, that in the same manner as for the scalar power spectrum, this difference disappears if one uses the exact gauge-field mode functions instead of the approximate ones (see Fig.~\ref{fig:f3} in Sec.~\ref{sec:application}).

\begin{figure}
\centering
\includegraphics[width=0.98\textwidth]{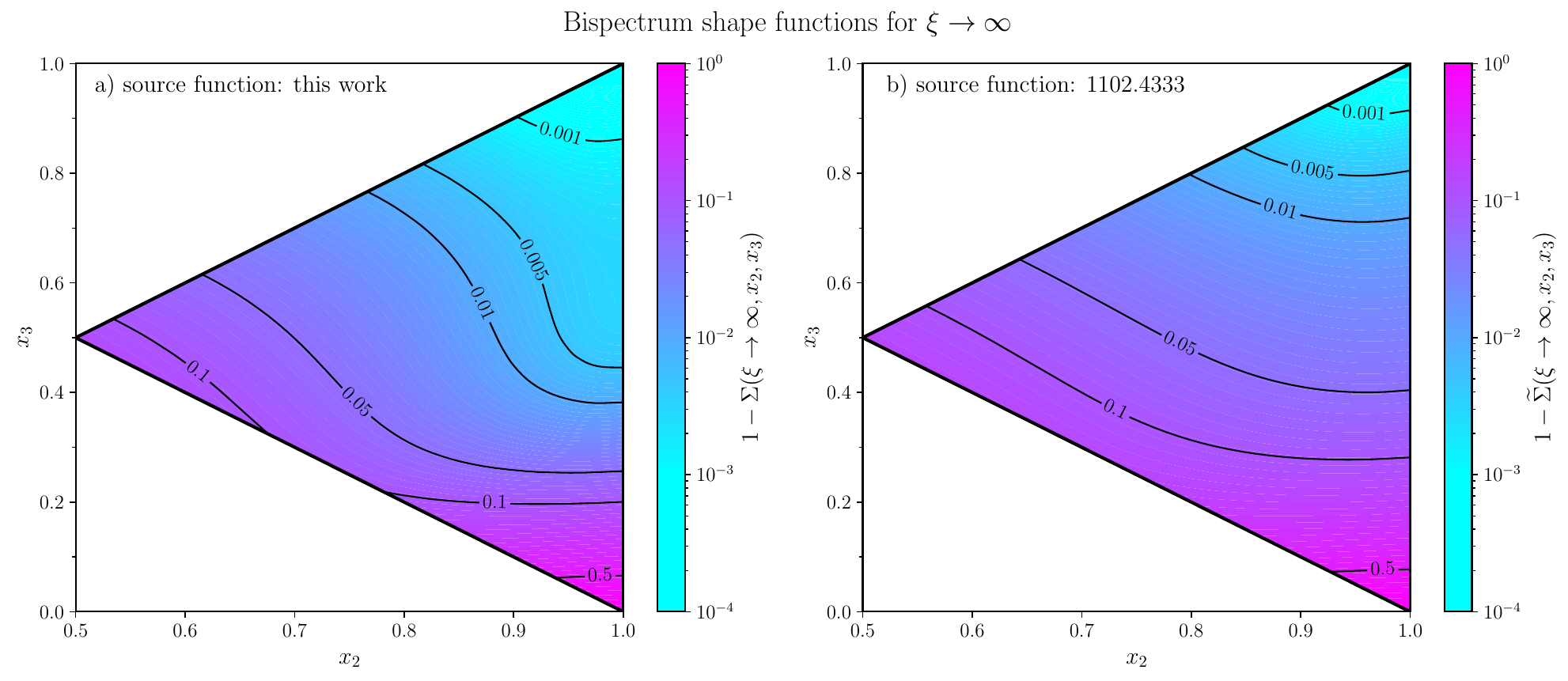}
\caption{The bispectrum shape function $\Sigma(\xi, x_2, x_3)$ in the region of the $(x_2, x_3)$ plane limited by the inequalities \eqref{limits-shape} in the large-$\xi$ approximation for the cases when the metric perturbations are (a)~included and (b)~excluded from the analysis. The latter one coincides with the result of Ref.~\cite{Barnaby:2011vw}. Since the shape function exhibits a broad maximum around the equilateral configuration, the color scheme and contour labels correspond to the deviation of the shape function from unity, $1-\Sigma(\xi, x_2, x_3)$, in logarithmic scale. \label{fig:shapes-large-xi}}
\end{figure}

The bispectrum in nonequilateral momentum configurations is conveniently represented by the shape function $\Sigma$. We show these functions in the large-$\xi$ approximation in Fig.~\ref{fig:shapes-large-xi}, where panel~(a) corresponds to the case with metric perturbations included, while panel~(b) represents the case with the metric perturbations neglected, which corresponds to  Fig.~6(a) in Ref.~\cite{Barnaby:2011vw}.

\vspace{-2mm} 
%
%\bibliographystyle{apsrev4-2.bst}
%\bibliography{references.bib}

%apsrev4-2.bst 2019-01-14 (MD) hand-edited version of apsrev4-1.bst
%Control: key (0)
%Control: author (72) initials jnrlst
%Control: editor formatted (1) identically to author
%Control: production of article title (0) allowed
%Control: page (0) single
%Control: year (1) truncated
%Control: production of eprint (0) enabled
%

\end{document}